%% file: info-measure.tex
\documentclass[useAMS,usenatbib,usegraphicx]{mn2e}
\usepackage[utf8]{inputenc}
\usepackage{amsmath}
\usepackage{amssymb}
\usepackage{color}
\usepackage{aas_macros}
\usepackage{tikz}
\usepackage{times}
\include{hacks}

\title{A Bayesian method for pulsar template generation}
\author[M. Imgrund, D.J. Champion, M. Kramer, H. Lesch]{M. Imgrund$^{1,2}$\thanks{E-mail:
mimgrund@mpifr-bonn.mpg.de}, D.J. Champion$^1$, M. Kramer$^{1,3}$, H. Lesch$^2$\\
$^{1}$Max-Planck-Institute for Radio Astronomy, Auf dem Hügel 69, 53121 Bonn, Germany\\
$^{2}$University Observatory Munich, Scheinerstraße 1, 81679 Munich, Germany\\
$^{3}$Jodrell Bank Centre for Astrophysics, University of Manchester, Alan Turing Building, MP13 9PL, UK }

\begin{document}
\date{Accepted x. Received x; in original form x.}
\maketitle

\begin{abstract}
Extracting Times of Arrival from pulsar radio signals depends on the knowledge of the pulsars pulse profile and how this template is generated. We examine pulsar template generation with Bayesian methods. We will contrast the classical generation mechanism of averaging intensity profiles with a new approach based on Bayesian inference. We introduce the Bayesian measurement model imposed and derive the algorithm to reconstruct a ``statistical template'' out of noisy data. The properties of these ``statistical templates'' are analysed both with simulated and real measurement data from PSR B1133+16. We explain how to put this new form of template to use in analysing secondary parameters of interest and give various examples:\\
We show how to reconstruct a detuned measurement's phase shifts and demonstrate how to discriminate between different modes of radiation by implementing a nulling detection. Combining elements of the former, we implement a nonlinear filter for determining ToAs of pulsars. Applying this method to data from PSR J1713+0747 we derive ToAs self consistently, meaning all epochs were timed and we used the same epochs for template generation.\\
The phase shift reconstruction is found to measure a shift in simulated data up to the estimated statistically possible accuracy out of noisy data.\\
Average templates as well as Bayesian templates are subject to uncertainties by fluctuations and noise. While the average template contains these as unavoidable artifacts, we find that the ``statistical template'' derived by Bayesian inference quantifies fluctuations and remaining uncertainty. This is why the algorithm suggested turns out to reconstruct templates of statistical significance from as few as ten to fifty single pulses.\\
A moving data window of fifty pulses, taking out one single pulse at the beginning and adding one at the end of the window unravels the characteristics of the methods to be compared. It shows that the change induced in the classical reconstruction is dominated by random fluctuations for the average template, while statistically significant changes drive the dynamics of the proposed method's reconstruction.\\
The analysis of phase shifts with simulated data reveals that the proposed nonlinear algorithm is able to reconstruct correct phase information along with an acceptable estimation of the remaining uncertainty.\\
\end{abstract}
\section{Introduction}
Pulsars are well known for their high precision measurements,
particularly in areas of fundamental physics such as general
relativity \citep{2004NewAR..48.1413C,2006Sci...314...97K}. These measurements make use of the exceptional
rotational stability of the pulsar as observed via its electromagnetic
pulses, most often observed in the radio. Most of these analyses rely on determining the arrival times of the pulses out of tens of thousands integrated single pulses.\\
The individual (or single-) pulses observed from a pulsar are
extremely variable, not only in flux but also in the shape of the
pulse. It is only when 10s of 1000s of pulses are averaged in an
integrated pulse profile that it becomes stable \citep {1975ApJ...198..661H,2012MNRAS.420..361L}. Some pulsars exhibit temporal variations in this pulse profile; showing no emission
for a period of time - nulling (see Sec. \ref{nulling}), or switching between
one or more additional stable profiles (see Sec. \ref{moding}), however for
the majority of pulsars the pulse profile remains stable, in some
cases over at least years).\\
There exist ready to use toolsets like PSRCHIVE\citep{2004PASA...21..302H} and TEMPO2 \citep{2006MNRAS.372.1549E,2006MNRAS.369..655H} to analyse raw observation data in order to assign times of arrival to integrated observations and constrain the parameters of the physical model under consideration. The remaining differences between the model's prediction of an arrival time and the actual time measured, also known as timing residuals, can be as low as a few tens or hundreds nanoseconds on average while the periods of observational data spans several years. This amounts to tracking the rotational phase of the pulsar at a certain observational epoch to an accuracy of $10^{-4}$ and well below. In the light of a single pulse being a highly variable object that apparently cannot be tracked down to that accuracy (see e.g. fig. \ref{fig:nulling}) it becomes clear that this very exact phase information is imprinted in and has to be recovered out of tens of thousands of very individual pulses.\\
The classical way of generating times of arrival (ToAs) is based on techniques comparing the shift of an epoch under consideration to a so called template. 
The template is often an integrated pulse profile of the entire
dataset for that pulsar (at the appropriate observing frequency and
bandwidth). Usually it is then fit with an analytic function to
produce a noiseless template. These templates are occasionally updated
and the ToAs reproduced.\\
Observations integrated over usually shorter times are then tested for a phase shift w.r.t. the template. The user of PSRCHIVE e.g. has the choice among five algorithms all comparing the data of an observation to the reference and yielding a relative shift and an error estimate on it for example by discrete cross-correlation and interpolation to yield sub-bin accuracy like \citet{2005MNRAS.362.1267H} or examining the phase gradient in the fourier space representation as in \citet{1992RSPTA.341..117T} to name just a few. The phase shift is then converted into a time of arrival using the timestamp and folding period of the observation. In a next step, the actual pulsar timing is carried out.\\
Pulsar timing is the process by which a rotational model of the pulsar is
produced (the ephemeris) and the value of its rotation phase is mapped to the ToA at the observatory of every pulse. This is then fit to the ToAs from
observations to produce a phase coherent solution that accounts for
every rotation of the pulsar from the first observation to the most
recent. It is the coherence of this solution that gives pulsar timing
its extraordinary precision.\\
This constitutes the classical procedure of pulsar timing. Both the template and the observed profile to be converted to a ToA rely on integrated and thus averaged data before correlating. Single pulses cannot be compared directly for they are often not bright enough to be distinguished from noise introduced on the signal's way to the telescope and by the antenna-receiver-amplifier system. Furthermore their distinct and individual shape makes it difficult to correlate them to the average profile as simply matching the shape of the pulse against the average is not possible. Thus timing individual pulses by comparing them to a template lacks the necessary precision.\\
For these reasons and yielding a such precise timing the classical way of integrating the single pulses to yield a stable and confident profile justifies itself. However in actual observations, the reduced $\chi^2$ value, a measure for the believed precision of the method compared to the actual residual error of the timing model, is much larger than unity (see e.g. \citet{2010IAUS..261..212B}) pointing to an underestimation of the error by the classical procedure and algorithms. This additional error is believed to have its root in additional noise sources like e.g. scattering by the interstellar medium. A comprehensive overview of the noise budgets can be found e.g. in  \citet{2010arXiv1010.3785C}. \\
By assessing and fitting the possibly introduced effects on the ToAs, mitigating errors after determining ToAs has been intensively investigated \citet{2011MNRAS.418..561C,2010arXiv1010.3785C}. As Bayesian methods already have been successfully used for correlating ToA-data from different pulsars, extending these to mitigate the timing errors by supposing a Bayesian timing model on the ToAs as described by \citet{2014MNRAS.441.2831L,2013arXiv1310.2606V,2014MNRAS.437.3004L} and others suggests itself.\\
The shape of single pulses is affected mainly by the short term noise contributions coming from current observational conditions one can aptly paraphrase as interstellar weather and the magnetosphere or radiation process intrinsic pulse jitter. Both influences affect mainly the single pulse statistics or influence the profile over short timescales, but can systematically influence timing and even dominate timing precision \citep{2011MNRAS.418.1258O}.\\
Increasing the gain of the observations with future systems like the Square Kilometre Array (SKA) aggravates this problem and demands for longer integration times if no solution is found to incorporate the single pulse fluctuations into the problem of determining ToAs rather than trying to average them out \citep{2012MNRAS.420..361L}.\\
Finding a generic way of dealing with single pulse indivituality can ameliorate the situation and fill the statistical gap between single pulses and an integrated profile without the need to take further modelling assumptions\footnote{Incorporating assumptions on e.g. scattering of the ISM or pulse jitter behaviour however may further improve the results presented}.\\
In this paper we will show a way to address the problem of generating pulsar ToAs with single pulse statistics and the benefits of having a more accurate statistical representation for single pulses' behaviour. By using more information than the classical method (which integrates this information) we hope to improve the precision of pulsar ToA generation and to allow ToAs from shorter observations to be unbiased by pulse jitter.\\ 
We introduce a model for statistics of the single pulses and substitute the classical template by a ``statistical template'' representing the single pulse statistics. We will then work out the steps of statistical template generation and phase shift detection replacing the classical counterparts.\\
Hereto we will derive a measurement model for squared and amplified receiver voltage in Sec. \ref{modelderiv} and argue how to handle the signals from pulsars best, splitting them in a radiation process and a phase coupled template part. Using the Bayesian formalism, we will deduce the posterior distribution for both the amplitude and phase model introduced and thus show a way to infere a statistical template from input data. Though focusing on the important application of generating ToAs, we will also demonstrate the benefit of using statistical templates on other applications such as determining nulling and analysing a moding pulsar. Sec. \ref{application} will evaluate the proposed algorithm and methods using both simulated and real pulsar data also giving insight on how a statistical template can enhance the view on the pulsars profile from a statistical perspective. Comparing the generation of ToAs to the classical results we will highlight improvements both in timing residuals as well as the $\chi^2$ value of the fit. We summarize our findings in Sec. \ref{conclusion}.
\section{Modelling pulsar measurements}\label{modelderiv}
\subsection{Terminology of the used statistical model for single pulses}\label{stmod}

\begin{figure}
\input{graphics/profile_exp}
\caption{\label{fig:profileexp}Scheme for single pulse statistics as described in Sec. \ref{stmod}}
\end{figure}
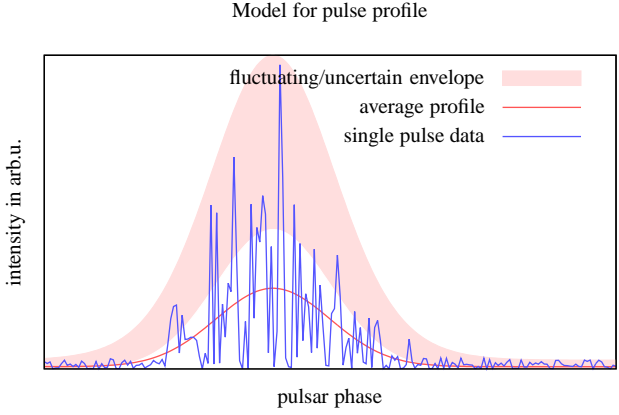
In order to describe single pulse statistics we use the notion of a fluctuating envelope besides to the classical set of terms of single pulses and integrated profiles.\\
Instead of modeling an average profile, we assume a single pulse to be the result of multiplying a stationary Gaussian random noise process, which amounts to the physical process that generates the radiation in the pulsar magnetosphere, with a possibly unstable envelope (see fig.\ref{fig:profileexp}). This envelope models the periodic influence of the magnetosphere. The uncertainty in the envelope may reduce with observation time when the envelope is not fluctuating. However if the magnetosphere conditions for some region mapping to the pulse phase are fluctuating, the envelope will stay fluctuating in that region.\\
Only in the limit of no uncertainty the envelope would equal the (scaled) average of the pulses and amount to the same thing as a classical integrated profile. As averaging single pulses generates a pulse profile, we emphasize that there are many possible fluctuating envelopes which are integrated to the same profile.\\
The empiric or analytic template that is used for generating times of arrival will consequently be substituted by a Bayesian statistical template describing the fluctuating envelope.

\subsection{Amplitude model}
Every Bayesian analysis is based on formulating a signal and noise model to describe the statistical imprint of a certain signal in the data. Thus, before inserting the outlined statistical model of the signal , we have to understand the influence of an arbitrary signal. Therefore we will introduce the basic measurement model, insert our signal model and then derive the formulae for signal reconstruction using Bayes' theorem.\\
Typical pulsar measurements at radio frequencies give the intensity of received radiation by suitable sampling of $\vv d$, the quadrature of the received voltage. The problem is to infer a unknown signal $\vv s$ (which in our case amounts to the noise-free series of single pulses as emitted from the pulsar) from the noisy data $\vv d$. The noise - leaving aside radio interference - is dominated by thermal fluctuations of both receiver and antenna and is accurately described by white noise of a usually known variance $\sigma_{\mathrm n}^2$ entering before quadrature of the signal. Thus, the baseline corrected data $\vv d$ measured, given a signal $\vv s$ and a random noise $\vv n$ amounts to 
\begin{align}
 \vv d &= ( \vv n + \vv s)^2\\
 \vv d - \vv n^2 &= \vv s^2 +  2\vv s \vv n\label{classum}\\\intertext{where}
 \cali P (\vv d|\vv s,\vv n) &= \delta(\vv d-(\vv n + \vv s)^2)\\
 \cali P(\vv n) &= \cali G_\vv n(0,\vv N)
 \end{align}
 where $\PP{x|y}$ is the short notation for the probability density of $X = x$ given $Y = y$, where $X,Y$ would be the corresponding random variables.
 We use boldface lower case letters to emphasize that data, signal and noise consist of many single values $\vv d = \{d_1 , d_2, \cdots d_t, \cdots\}$ for which the equations stated are valid component-wise. Thus operations like taking a root of a vector or multiplying two vectors are carried out only on the components of the vector. For the scalar product  of vectors $\vv a$ and $\vv b$ we use the notations $\vv a \dgr \vv b$ and $\sqrt{\vv a\dgr \vv a} = ||\vv a||$. We abbreviate $\det(\cdot)$ as $|\cdot|$ and define $\delta (\vv a) := \prod _k \delta (a_k)$. Putting a hat over a vector $\hat{\vv v} = \mathrm{diagmat}(\vv v) $ denotes constructing a diagonal matrix with a diagonal with the elements of$\vv v$. When dealing with sets, $\vv A\setminus{\vv g}$ denotes the set $\vv A$ without the elements $\vv g$.\\
 $\cali G_\vv x(\vv m,\vv V)$ denotes that $\vv x$ is distributed as the multivariate Gaussian distribution with mean $\vv m$ and covariance matrix $\vv V$. It is defined as 
 $$\cali G_\vv x(\vv m,\vv V):= \frac {1}{ \sqrt{2\pi|\vv V|}} \Exp{-\frac{1}{2} (\vv x - \vv m )\dgr \inv {\vv V} (\vv x - \vv m )}$$
 Integrating over a vector $\vv a$ is denoted by $\PP{\vv b} = \Dint \vv a \PP{\vv a, \vv b} = \iint \mathrm d a_1 \cdots \mathrm d a_k \PP{\vv a,\vv b}$.\\
The classical way of determining the template intensity amounts to averaging eq.\mref{classum} over several pulsar periods, noticing that the term mixing signal and noise vanishes leaving us with the desired quantity of $\langle \vv s^2\rangle$ when subtracting $\sigma_{\mathrm n}$ from the averaged data. To uses Bayes' theorem, we will instead calculate the exact probability distribution to measure $\vv d$ at a certain time, given a certain signal value $\vv s$ at that time:
 \begin{align}
\cali P(\vv d|\vv s) &= \Dint \vv n \cali P(\vv n,\vv d|\vv s) = \Dint \vv n \cali P(\vv d|\vv n,\cdots) \cali P(\vv n)\\
&= \Dint \vv n \delta(\vv d-(\vv n + \vv s)^2) \frac 1 { \sqrt{|2\pi\vv N|}}\Exp{-\hlf \frac{||\vv n||^2}{\sigma_{\mathrm n}^2}}\\
\intertext{We substitute $\wtvv n=(\vv n + \vv s)^2, \vv n = \pm \sqrt {\wtvv n} -\vv s , \mathrm d \vv n = \frac{\mathrm d \wtvv n}{2\sqrt{ \wtvv n}}$ and notice that $\sqrt {|\vv N|} = \sigma_n^{N_\mathrm{tot}}$, where $N_{tot}$ is the total number of samples in all bins.}
&= \iint^\infty_{0} \mathrm D \frac{\wtvv n}{\sqrt{\wtvv n}} \quad \delta(\vv d-\wtvv n) \frac 1 {|\sqrt{2\pi}\sigma_{\mathrm n}^{N_\mathrm{tot}}|}\Exp{-\hlf \frac{||\sqrt {\wtvv n} \pm \vv s||^2}{\sigma_\vv n^2}}\\
&= \frac 1 {\sqrt{|2\pi \vv d|}\sigma_{\mathrm n}^{N_\mathrm{tot}}}\Exp{-\hlf \frac{||\sqrt \vv d \pm\vv s||^2}{\sigma_{\mathrm n}^2}}\label{datacon}\\
\intertext{where we now have to assign a signal model for $s$. We assume}
\vv s&= \Exp {\vv f} \cdot \vv g \sigma_{\mathrm n}
\end{align}
Where we have decided for the signal to be measured in units of noise temperature of the receiver-antenna system.
The $\pm$-operator means that the whole density would be a sum of the densities with the respective sign. This operator vanishes later on, when integrating over $g$ and using $\PP{g} = \PP {-g}$.
$\vv f$ and $\vv g$ are Gaussian random variables with also to be determined covariances $\vv F$ and $\vv G$ which describe the statistical properties of the envelope respectively the radiation process. The exponential form for the envelope was chosen since on one hand it is able to grasp the occurring fluctuations to high radiation densities of single pulses and on the other we may easily assume a scale-free prior for the amplitude, as we will see below.\\
We assume the radiation process to have stationary statistics given by 
$$
 \vv G=\vv G(t-t') \label{statG}
$$
and the part that is generating the observed envelope to be periodic in time. Even though the covariance matrix of $\Exp f$ alone has non-diagonal components, the product's covariance matrix $\langle g(t)\Exp f(t) g'(t) \Exp f'(t) \rangle$ has only diagonal components if we marginalize over g, because of the stationarity of g as assumed in \mref{statG}. Thus we describe the covariance matrix solely by it's Fourier-coefficients at $\omega_k = k\cdot \frac {2\pi} T$. Since the expectation of the signal's covariance matrix for a certain g integrated over a specific pulse certainly does not vanish, the non-diagonal components of the signal's matrix carry information about $g$ which we will discard in this paper. $T$ denotes the real pulsar period, which we will discriminate from the assumed period $\tau$ later on.

\subsubsection{Joint Probability and Information Hamiltonian}
For a moment, let us assume the covariance matrices $\vv F$ and $\vv G$ are known. Then we may write for the joint probability
\[
 P(\underbrace{\vv f,\vv g,\vv s,\tau}_{=:A \cup\{\vv s\}}|\underbrace{\vv F,\vv G}_{=:B}) = \delta(\Exp{\vv f}\vv g\sigma_{\mathrm n}-\vv s) \cali G_{\vv f}(0,\vv F)\cali G_{\vv g}(0,\vv G) \cali P(\tau)
\]
and we define the parameter sets $A$ and $B$ where $\PP \tau$ is the probability density over the assumed period $\tau$. In order to define the Information Hamiltonian $H_B[A,\vv d] := -\log P(A,\vv d|B)$ we notice that $\PP {A,\vv d|B} = \Dint \vv s \cali P(\vv d|\vv s) P(A\cup\{\vv s\}|B)$ and use the measurement model given by \mref{datacon}:
\begin{align}
-\log &P(A,\vv d|B) = H_B[A] = \nonumber\\
\hlf &\left[(\sqrt{\vv d}-\Exp{\vv f}\cdot \vv g \sigma_{\mathrm n})\dgr\frac{1}{\sigma_{\mathrm n}^2}(\sqrt{\vv d}-\Exp{\vv f}\cdot \vv g \sigma_{\mathrm n}) +\right.\nonumber\\
&\left. \vv f\dgr \inv{ \vv F} \vv f +  \vv g\dgr \inv{\vv G}\vv g + \right] -\nonumber\\
&-\log \cali P(\tau) +\frac 1 2 \log(|2\pi \sigma_{\mathrm n}^2|^{N_\mathrm{tot}}|2\pi \vv F||2\pi \vv G|| \vv d|) \label{beforegint}
\end{align}
Since we are interested in the pulsar's statistical template given by the envelope characteristics we marginalize over the radiation process g by integrating it out. This calculation is outlined in Sec. \ref{gint}. We are left with
\begin{align}
 H_B[A] &= \hlf \left[ \sqrt{\vv d}\dgr \frac{1}{\sigma_{\mathrm n}^2}\inv{[1+\widehat{\Exp{\vv f}}\vv G \widehat{\Exp{\vv f}}]}\sqrt \vv d +\vv f\dgr\inv{\vv F} \vv f+\right. \nonumber \\ &\left. +\log(|2\pi\inv{\vv D_{\vv f}}||2\pi \sigma_{\mathrm n}^2|^{N_\mathrm{tot}}|2\pi \vv F||2\pi \vv G|| \vv d |)\right] -\log \cali P(\tau) \label{finalwithoutg}
\end{align}
where the operator $\inv{[1+\widehat{\Exp{\vv f}}\vv G \widehat{\Exp{\vv f}}]}$ together with the noise variance is a variation of the operator leading to the well-known case of the Wiener filter \citep{Wiener:1949}. This similarity arises due to the integration over the Gaussian random field $\vv g$ added upon the signal field $\vv s$. However, the normalisation w.r.t. the data differs and we further distinguish between the periodic part described by $(\vv f,\vv F)$ and the radiation part now showing up only as $\vv G$. The functioning of the operator can be understood by considering its interplay with given data. Maximizing probability equals minimizing the Hamiltonian. Thus high data values will be compensated by large mean values of f. Conversely, for too high values of f the exact value of the data becomes irrelevant, leading to a sub-optimal solution independent of the data. The optimal solution resembles the actual probability distribution the data would be drawn from. As a function of $\vv d$, the joint probability is a special case of a gamma distribution $\propto \Exp{-||\frac{\vv d}{\vv m}||}\vv ||d||^{-.5}$ where $\vv m$ is the mean value of the distribution. In this light the operator resembles the classical formula $\langle \vv d\rangle = \vv m = \sigma_{\mathrm n}^2 + \vv \sigma_s^2$ since $\vv m \overset \wedge = \sigma_{\mathrm n}^2[1+\frac{\vv s\dgr \vv s}{\sigma_{\mathrm n}}]$ for our case.\\
$\vv G$ becomes diagonal in Fourier space since $\vv g$ is assumed to be stationary. As $\Exp{\vv f}$ is a periodic signal, its Fourier space is limited to discrete frequencies that are a multiple of the assumed periodicity. We may absorb the values of $G_\omega$, where $\omega_k = k\cdot \frac {2\pi} \tau$ into $\vv F$. In the following we will discard information about G from single pulses. Since our main interest lies on the statistics of the envelope, we may take this loss.\\
Under these assumptions, and neglecting non-diagonal terms arising when Fourier transforming $f$ (see Sec. \ref{whytdisbad} and \ref{whywedoit}), the $G_\omega=:\sigma^2_{g_k}$ may be absorbed into an effective mean $\vv f$ as is evident in eq. \mref{finalwithoutg}:
 \[
  \Exp{2f_k}\cdot \sigma^2_{g_k} = \Exp{2(f_k+\ol f_k)}
 \]
 where $\ol f_k = \log \sigma_{g_k}$. We redefine $\vv f$ in \mref{finalwithoutg} in that way and yield
 \begin{align}
  H_B&[A\setminus{\vv g}]= \hlf \left[\left\|\frac{\vv d}{\sigma_{\mathrm n}^2}\frac{1}{1+\Exp{2\vv f}}\right\| +(\vv f-\ol{ \vv f})\dgr\inv {\vv F} (\vv f-\ol {\vv f})+ \right.\nonumber\\&\left. +\log(|2\pi\inv{\vv D_{\vv f}}||2\pi \sigma_{\mathrm n}^2|^{N_\mathrm{tot}}|2\pi \vv F||\vv d|)\right] -\log \cali P(\tau)
 \end{align}
 where $\vv G$ is now the unity matrix and we assume $\inv {\vv F}$ to be diagonal with components $\frac 1 {\sigma_{f_k}^2}$. We arrive at our model for a single pulse's amplitudes in Fourier space, assuming all timing parameters to be given:
 \begin{align}
  H_{\ol {\vv f},\sigma_{\vv f}^2}[\vv f,\vv d]= \hlf &\left[\left\|\frac{\vv d}{\sigma_{\mathrm n}^2}\frac{1}{1+\Exp{2\vv f}}\right\| +(\vv f-\ol {\vv f}) \dgr \frac 1 {\vv \sigma_{\vv f}^2} (\vv f-\ol {\vv f}) +\right.\nonumber\\&\left. + \log C(\vv F,\sigma_{\mathrm n},\vv d) \right]
 \end{align}
 This is the negative logarithm of the likelihood function. Bayes' law gives us the posterior
 \begin{equation}
  P(\vv f|\vv d) = \frac{\Exp{-H_{\ol {\vv f},\sigma_{\vv f}^2}[\vv f,\vv d]}}{\Dint \vv f \Exp{-H_{\ol {\vv f},\sigma_{\vv f}^2}[\vv f,\vv d]}} \label{posterior}
 \end{equation}
 when we assume a flat prior on $\ol {\vv f}$ which effects in a scale-free prior on the assumed amplitude, since $\vv s\propto \Exp {\vv f}$ is exponentiated.
\begin{figure*}
 \input{graphics/fftwholee}
 \caption{\label{fig:fftwholee} Fourier transform of a whole epoch of pulsar B0329+54. The x axis has been scaled to the pulsar's signal harmonics. Inset: detail of one spike.}
\end{figure*}
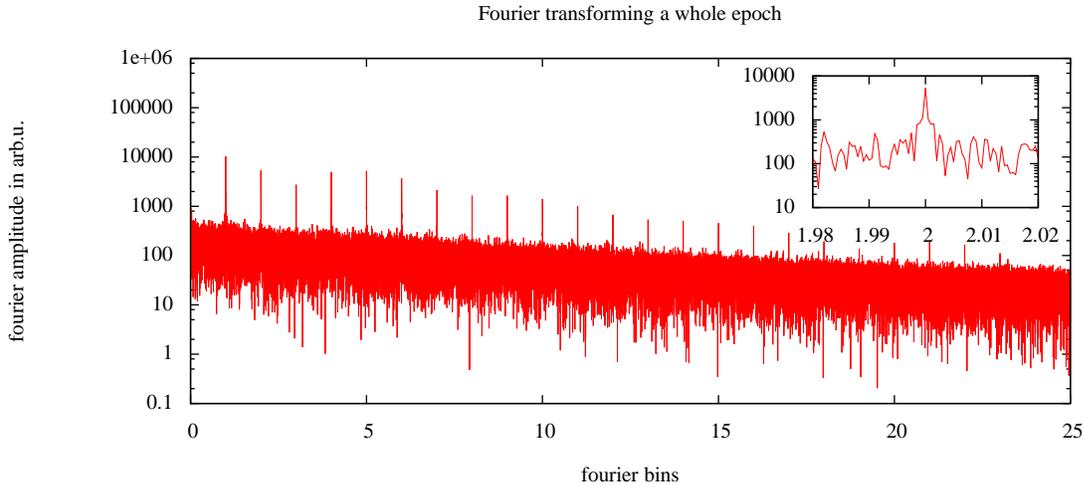

\subsubsection{The receiver equation's pdf}
Assuming $G$ to be a diagonal of ones (meaning vanishing correlations) for this argument, the pdf of the receiver equation of each bin decouples and for a specific bin and different pulses $K$ takes the form of
$$\PP{f|\vec d} \propto \prod_K \Exp{-\frac 1 2 \frac {d_K} {1+f^2}}\sqrt{\frac 1 {(1+f^2)2\pi}}/\sqrt{d_K}$$
which simply states that the variance of $\sqrt{d_K}$ can be explained as sum of the variance of the noise (the one in the denominator of the exponentiated term) plus the variance of $g$ ( equals unity per definition) times the signal strength of the envelope $f$. The square root term of $d_K$ after the exponential is a normalization factor w.r.t. $f$. We deduce that for a constant signal envelope $f$ and vanishing correlations of $g$
\begin{align}
\PP{f|\vec d} &\propto \Exp{-\frac 1 2 \frac {\sum_K{d_K}}{1+f^2}}\sqrt{\frac 1 {(1+f^2)}}\nonumber \\
	      &=\Exp{-\frac 1 2 \frac {\langle d \rangle}{\frac{(1+f^2)}{N_K}})}\sqrt{\frac 1 {{(1+f^2)}^{N_K}}}
\end{align}
where $N_K$ is the number of pulses involved. This probability density function attains it's maximum for $\langle d \rangle \geqq 1$ at $f = \sqrt{\langle d \rangle -1}$ justifying the average data subtracting the noise background as the maximum likely guess. Carrying out a saddle point analysis, we conclude that 
$$\sigma_f = \sqrt{\frac {(1+f^2)^2}  {(2N_K f^2)}}\quad=\begin{cases}
                                                          \sqrt{\frac {1}  {2N_K}}\frac 1 f &\mathrm{ for } f \ll 1\\
							  \sqrt{\frac 1  {2N_K}} f &\mathrm{ for } f \gg 1
                                                         \end{cases}
$$
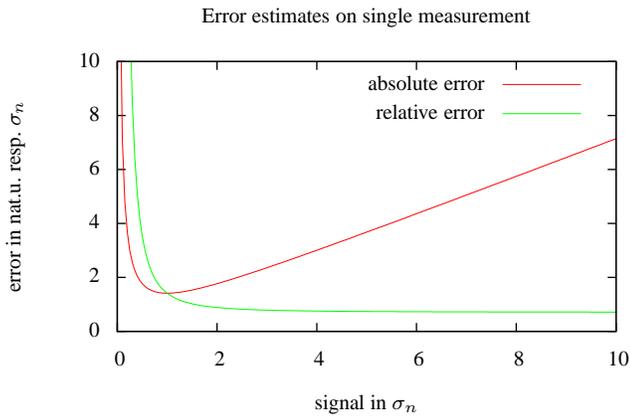
\begin{figure}
 \input{graphics/receiver_error}
 \caption{\label{fig:receiver_error} Comparison of absolute and relative error (as uncertainty over actual signal strength) for a single measurement of a signal $f$}
\end{figure}
For a signal that is constant and not the modulation of a stationary noise process, we would have expected $\sigma_c \propto \sqrt{c / N_K }$ as $c\gg1$ such that the relative error decreases with signal strength. In our case the relative error decreases too, but only to reach $1$ as a limit (see Fig. \ref{fig:receiver_error}). Against our intuition we expect the shape of the profile to be most uncertain at the highest signal values while the areas of lowest uncertainty can be found where $f\approx 1$.\\
We emphasize that this relative uncertainty will not be integrated out by measuring more pulses, as the uncertainty in both high and low signal areas equally fall with the sqare root of $N_K$ leaving the ratio of the remaining error untouched. This fact explains the often expressed observation that stronger pulses do exhibit larger timing errors \citep{2014MNRAS.443.1463S,2011MNRAS.418.1258O}. The reason was found  implicitly by numerous authors (e.g. \citet{1989AJ.....98.1112K,1975ApJ...197..185R}) for the case of self-correlation of the signal. We emphasize that even without self-correlation, we expect the higher pulses to have a higher intrinsic uncertainty.
\\
The non-diagonal terms introduced by $G$ mark the departure from simply averaging the data. Let us review the influence of a non-white stationary process on the data's spectrum.
\subsubsection{The influence of $g$ on the signal's spectrum}\label{whywedoit}
When investigating a Fourier transform of a whole observation's time series containing several hundred pulses (see fig. \ref{fig:fftwholee}) two prominent features dictate the spectrum: Spikes of the pulsar's signal dominate over receiver background noise. If the signal was strictly periodic, we would expect it to be described solely in terms of the Fourier coeffieents of harmonics of the base frequency $1/P$. As the pulsar's signal contains periodicity, we expect spikes to show up in this case too\footnote{these spikes are not an artifact of the folding period for the archive as this folding is discarded in the analysis. Of course the folding period of observations is set as close to the true period $P$ as possible to be able to integrate the data easily by adding up the receiver bins' values (in soft- or hardware)}. However, we are facing a stochastic process $g$ that is modulated via a periodic envelope $f$. As is commonly known, multiplication in time domain translates to convolution in Fourier domain. As $g(\omega)$ is expected to be a red noise process, we expect it to cast the sharp pulsar signal to finite width in Fourier space. Indeed - as the inlay of the figure shows - the signal's power is leaked to the neighbouring bins. The width of the spectral line however is rather narrow, pointing to a fast decline of the spectrum of $g$ in Fourier space. Consequently, the spectrum of $g$ in time domain is spread out widely, pointing to non-vanishing correlations over a long time. These correlations diminish the independence of both adjacent lying bins and the bins of neighbouring pulses.\\
As the multiplication $f(t)\cdot g(t)$ amounts to a convolution in Fourier space, evaluating the posterior distribution in Fourier space exactly was found to be unfeasible. 
Instead, we approximated the convolution by taking only the diagonal terms, losing mathematical rigorosity and precision but keeping the problem solvable in reasonable computational time. The interested reader may refer to Sec. \ref{whytdisbad} of the appendix for a more detailed discussion why solving this in time domain would be favorable, but currently is not feasible.\\
The clear departure from the average comes from the off-diagonal elements of $\sqrt{\vv d} M \sqrt{\vv d\dgr }$ where $M$ is the operator derived. As is evident from the structure of M, it is expected to be off-diagonal only until the correlations introduced by $G(t-t')$ decay. Thus, writing this equation as a matrix equation, symmetry tells us that we can possibly store only a limited combination of $\sqrt{d}(t-\Delta t) \sqrt{d}^*(t+\Delta t)$ values and average them as they are multiplied by the same number when performing the matrix multiplication. The matrix $M$, subject to the exact envelope shape and correlations of the radiation process can be calculated afterwards and former data may be reprocessed with the new $M$ in that way without losing the statistical information from the single pulse level, even though we may integrate parts of the single pulse data to yield smaller data sets. This could possibly be exploited in the future to profit from single pulse statistics without having to store all single pulses.

\subsubsection{Maximum-a-posteriori-Filter}
 We are interested in finding the $\ol {\vv f},\vv \sigma_{\vv f}$ values most compatible with the data and thus set out to maximize the a-posteriori distribution\footnote{Maximum A Posteriori or MAP refers to taking the most likely value of the posterior as best guess} given the data. For a prior specified by $\sigma_{f,p}$ and $\ol{\vv f}_p$ we derive the integral over f of \mref{posterior} w.r.t. $\ol {\vv f}$ and expect the derivative to vanish. This procedure yields an implicit formula for maximizing the posterior distribution of $\ol {\vv f}$.
 \begin{align}
  0 &\overset ! =\Dint \vv f \sum_{K=0}^{K\smax}\sum _{k=0}^{k\smax} \frac{(f_{K,k} - \ol f_k)}{\sigma_k^2} \Exp {-H_{\ol f_{k,p},\sigma_{f_p}^2}[f_{K,k},d_{K,k}]} \nonumber \\
  0 &= \sum _{K=0}^{K\smax}\sum _{k=0}^{k\smax} \dint f_{K,k} \frac{(f_{K,k} - \ol f_k)}{\sigma_k^2} \Exp {-H_{\ol f_{k,p},\sigma_{f_p}^2}[f_{K,k},d_{K,k}]}\nonumber \\
  \intertext{which may be solved for every k independently as}
  0 &= \sum _{K=0}^{K\smax}\langle f_{K,k} - \ol f_k \rangle_{\Exp {-H_{\ol f_{k,p},\sigma_{f_p}^2}[f_{K,k},d_{K,k}]}}\nonumber \\ 
  \ol f_k &= \langle f_{K,k}\rangle_{\Exp {-H_{\ol f_{k,p},\sigma_{f_p}^2}[f_{K,k},d_{K,k}]}}\label{newprior}
 \end{align}
 where we organize our data in pulse numbers indexed $K$ and have summed over single Fourier coefficients $d_k$ of these data of pulse $K$.\\
 The new variance assumed of $\sigma_{\vv f}$ may be calculated equally by evaluating the expectation value of $(f_{k,K}-\ol f_k)^2$, since deriving w.r.t. $\sigma_k$ yields:
 \begin{align}
  0 &\overset ! =\dint f \sum_{K=0}^{K\smax} (\frac{(f_{K,k} - \ol f_k)^2}{\sigma_k^3}-\frac{1}{\sigma_k}) \Exp {-H_{\ol f_{k,p},\sigma_{f_p}^2}[f_{K,k},d_{K,k}]} \nonumber \\
  0 &= \sum _{K=0}^{K\smax}\sum _{k=0}^{k\smax} \dint f_{K,k} (\frac{(f_{K,k} - \ol f_k)^2}{\sigma_k^3}-\frac{1}{\sigma_k}) \Exp {\cdots}\nonumber \\
  \intertext{which may be solved for every k independently as}
  0 &= \sum _{K=0}^{K\smax}\langle((f_{K,k} - \ol f_k)^2 - \sigma_k^2) \rangle_{\Exp {-H_{\ol f_{k,p},\sigma_{f_p}^2}[f_{K,k},d_{K,k}]}}\nonumber \\ 
  \sigma_k &= \sqrt{\langle (f_{K,k} - \ol f_k)^2\rangle_{\Exp {\cdots}}}\label{newpriorsigma}
 \end{align}
Doing this once is assuming a prior of the form of the initial $\ol {\vv f},\vv \sigma_{\vv f}$. A very general choice may be suitable. Then taking the expectancy values of \mref{newprior} as prior input to \mref{posterior} and iterating is known as the expectation-maximization (EM) algorithm proposed by \citet{1977DEMPSTER}. We may in this way in principle get rid of assumptions going into the first prior by iterating since this procedure converges on a prior fully compliant with the data, assuming gaussian distribution of f. However the EM algorithm may only find local maximum which equals a global maximum if the pulsar is really radiating with a log-normal probability distribution. The optimization problem in both $\ol {\vv f}$ and $\vv \sigma_{\vv f}$ may be solved using a suitable iteration method (see e.g. appendix \ref{newtoniter}). We assumed here that the pulsar is sufficiently described by a log-normal distribution and that the dataset converges to the same (global) fix point for all initial pairs of values. 
\subsection{Detuning model}\label{detuning}
The problem of determining ToAs is closely related to a phase shift over the analyzed single pulses. As a wrongly set folding period can cause phase-drifts over the dataset's pulses, we will study the influence of such a detuning on the coefficients of a signal to exclude mistaking a phase-drift over the dataset as a phase shift when determining ToAs. Analysing Fourier transformations of finite (consecutive) pulses allow us to measure a phase-drift corresponding to the detuning of the assumed periodicity $\tau$ and the real periodicity $T= \frac{\tau}{a}$, where $a$ is a correction factor. The equations denoted with a letter before the number are derived in appendix \ref{fourderiv}. For the signal from a single Fourier coefficient with frequency $\omega' = \frac{2\pi n}{T} = \frac{2\pi na}{\tau}$ we will get a Fourier coefficient over the period $[K\tau -\frac \tau 2 :K\tau +\frac \tau 2]$ of pulse $K$:
\begin{equation}
 \wt d_{k,K} = s_n \Exp{2\pi i k (\frac n k a - 1) K} \cdot \Sinc{\pi k (\frac n k a -1 )}\label{phaseandamp}
\end{equation}
where $s$ and $d$ with subscripts $k,K$ denotes the $k^{th}$ complex Fourier coefficient of the $K^{th}$ sample. For $n = k$ eq \mref{phaseandamp} reads
\begin{equation}
\wt d_{k,K} = s_k \Exp{2\pi i k (a - 1) K} \cdot \Sinc{\pi k (a -1 )}
\end{equation}
which means that, since $a \approx 1$, the sinc-factor is approximately 1 for $k\ll \frac 1 {(a-1)}$.Thus the relative phase of the $K^{\mathrm{th}}$ and the $(K+1)^{\mathrm{th}}$ pulse gives access to the model parameter $a$.\\
Let us now derive the Fourier coefficients for the full spectrum of a periodic, purely real signal. for such a signal, $s_n = s_{-n}^*$. Inserting this relation and denoting the phase as $\arg( s_n) =: \Phi_{s_n}$ we get:
\begin{align}
 d_{k,K} &=\sum_{n=1}^\infty a_n \Sinc{\pi(na-k)}\cdot \\ &\cdot \left[\underbrace{\Exp{i(2\pi naK + \Phi_{s_n})}}_I+\underbrace{\frac d {\ol k} \Exp{-i(2\pi naK + \Phi_{s_n})}}_{II}\right]\label{ftcoeff}
 \end{align}
 where $\ol k = \frac{an + k}{2}, d = \ol k-k$ . This formula can be written as $d_{k,K}=\sum_{n=1}^\infty A_{kn} a_n$. The main phase information again is to be found in term I of \mref{ftcoeff} for $n = k$. However this is not the only term relevant in analysing a single coefficient $d_{k,K}$. Assuming $n=k \pm m$ the sinc function slowly converges towards zero as $\frac 1 m$ and determines the factor for coefficient 1. Coefficient two together with the sinc function goes like $\frac 1 {2k\pm m}$ for $m>0$. Thus in the limit of big difference in $k$ and $m$, both terms equally contribute and the phase relation deforms more and more to an elliptic one.\\
 We emphasize that for small detuning, the sinc-term is nearly zero for $k \neq n$. In this case we may calculate only with the diagonal matrix and thus correct the data for the phase $\Delta\Phi(a) = \Exp{i2\pi naK}$. Thus for $a\cdot n_{max} \cdot K_{max}\ll1$ the whole problem of detuning may be described, with a negligible error by the phase shift model in Sec. \ref{psm}.
 \subsection{Phase model}
 Even for a perfectly tuned signal, we still have to face the noise introduced by the receiver and antenna system temperature on the phase. Again we will have to assign a measurement model for the data phase according to the signal. The noise passes the same linear transformation than the true Fourier coefficients and we conclude that
 \begin{equation}
  d_{k,K} = \sum_{n=1}^\infty A_{kn} (s_n + n_{n,K}) \label{datatosigphase} 
 \end{equation}
 where n is a white noise process with variance $\sigma^2_n$. Since $A$ is dependent on the true signal phases, the inverse problem as a whole is quite complex. But if we neglect term II of \mref{ftcoeff} we may invert the remaining matrix $\wt A$ without considering the true signal phases and apply it to a given dataset. 
 \begin{align}
  \underbrace{\sum_{k=1}^\infty \wt A^{-1}_{nk} d_{k,K}}_{=:d'_{k,K}} &= ||s_n||\Exp{i\Phi_{s_n}} +n_{n,K}\\
  \wt A_{kn} &:= \Sinc{\pi(na-k)}\Exp{i2\pi naK}\label{phasecor}
 \end{align}
 This is possible since term II destroys phase information in a very smooth way and - as will be concluded below - high coefficient numbers will have higher phase errors due to the drop in signal amplitude and the relative growth of error. Consequently, higher coefficients do not give as much information about the phase and thus about $a$ anyway.
 Having preprocessed the data with $\wt A^{-1}$ we arrive at the noisy signal and the phase carries information about $\Phi_{s_n}$. Thus, given $a$ and having processed the data vector, we may now fit for the average signal phase and variance by imposing a suitable model. We may estimate the error introduced in the true signal's phase by inspecting eq. \mref{datatosigphase} and Fig. \ref{fig:angularerror}:
 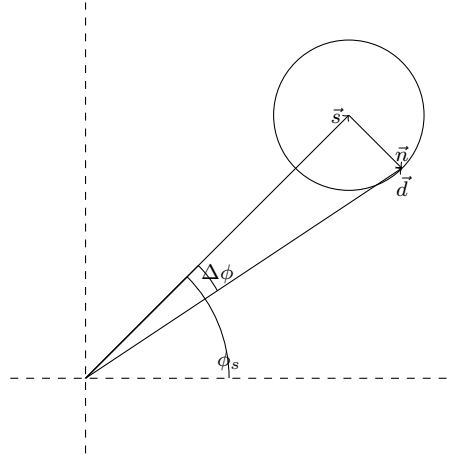
\begin{figure}
 \begin{tikzpicture}[x=0.5cm,y=0.5cm,scale=1.]
 \draw [dashed] (-2,0) --(10,0);
 \draw [dashed] (0,-2,0) --(0,10);
 \draw [->] (0,0) -- (7,7)  node[left] {$\vec s$};
 \draw [->] (7,7)--(8.41,5.59) node[above] {$\vec n$};
 \draw (7,7) circle(2);
 \draw (0,0)--(3,3) arc (45:29:3) node[above] {$\Delta\phi$};
 \draw (0,0)--(2.7,2.7) arc (45:0:3.81) node[above] {$\phi_s$};
 \draw [->] (0,0)--(8.41,5.59) node[below] {$\vec d$};
\end{tikzpicture}
\caption{\label{fig:angularerror} The phase error introduced to a signal by white noise scales as the expectancy value of the arctan of the signal to noise ratio}
 \end{figure}
\begin{align} \tan(\Delta \Phi) \approx \frac{||n||}{||s||} =& \frac{\sigma_{\mathrm n}}{\sigma_{\mathrm n}\Exp{2f}}\\ \intertext{Rather than integrating this precisely one can use the following approximation without significant loss of accuracy}
\vv \sigma_{\Phi_s} :=& \langle \arctan(\Exp{-2\vv f}) \rangle_{\PP{\vv f|\vv d}}\\
\PP{\vv \Phi_{\vv d}|\vv \Phi_{\vv s}} =& \prod_i \sum _{k=-\infty}^\infty \Exp{-\frac{(\Phi_{d'_i}- \Phi_{s_i}+2\pi k)^2}{2\vv \sigma^2_{\Phi_{n}}}}/C
\end{align}
This approximation is valid only for a good signal to noise ratio (S/N), where we also expect the most significant evidence to be. For low S/N it underestimates the error and we must numerically integrate the complete formula to get an error estimate. We emphasize that this is the very step where the confidence of the phase data and thus the weight in later calculations is calculated using the actual amplitude data point measured together with the statistics gathered. $\vv \Phi_{\vv s}$ itself is parametrized by a so-called wrapped Gaussian\footnote{a wrapped Gaussian is literally a gaussian distribution wrapped around the unit circle by identifying all points modulo $2\pi$} and a mean of $\ol {\vv \Phi_{\vv s}}$, and variance $\vv \sigma_{\vv \Phi_{\vv s}}$. we assume a prior of zero average and large variance.
\subsection{Joint probability for phase}
Finally we arrive at the joint probability of the phase model calculating analogue to the amplitude:
\begin{align}
 \PP{\vv \Phi_{\vv d},\vv \Phi_{\vv s}|T,\sigma_{\Phi_n},a} = \Exp{-\frac 1 2 &\left[\frac{||(\vv \Phi_{\vv d'}(\vv \Phi_{\vv d})-\vv \Phi_{\vv s})\mod 2\pi||^2}{\sigma^2_{\Phi_{n}}}+\right.\nonumber\\&+\left.\frac{||(\vv \Phi_{\vv s}-\ol {\vv \Phi_{\vv s}})\mod 2\pi||^2}{\sigma^2_{\vv \Phi_{\vv s}}} \right]}/C\label{phasemappre}
\end{align}
Where $T:=\{\ol{\vv  f},\sigma_{\vv f},\sigma_{\vv \Phi_{\vv s}},\ol {\vv \Phi_{\vv s}}\}$ will be the parameters of the full statistical template reconstructed and $\vv \Phi_{\vv d'} = \vv \Phi_{\vv d'}(\vv \Phi_{\vv d})$ is the phase correction according to \mref{phasecor} or other corrections for the assumed phase shift. Integrating out $\vv \Phi_{\vv s}$ we find:
\begin{align}
 &\PP{\vv \Phi_{\vv d}|T,\sigma_{\Phi_n},a}=\nonumber\\& = \Exp{-\frac 1 2 \frac{ ||(\vv \Phi_{\vv d'}(\vv \Phi_{\vv d})-\vv {\ol \Phi}_{\vv s})\mod 2\pi||^2}{\vv \sigma^2_{\vv \Phi_{n}}+\sigma^2_{\vv \Phi_{\vv s}}}}/C \label{phasemap}
\end{align}

We now again invert this relation using Bayes' theorem and take a MAP-Ansatz on our model parameters $\vv \sigma_{\vv \Phi_{\vv s}},\ol{\vv \Phi_{\vv s}},a$.
\subsection{Phase shift model for a single epoch as reference}\label{psm}
In order to generate ToAs we will have to determine the relative shift in the time of arrival of the pulses compared to another epoch taken as a reference. For the sake of simplicity, we assumed that the pulsar period does not to change with time (which is only a matter of interpreting the relative phase correctly or further imposing the detuning model). The pulsar's signal is assumed to be slipped by a time $\Delta t$ compared to the overall period. This amounts to a change in the phase of the Fourier coefficients given by:
\begin{equation}
d_k = \int _{ -\frac \tau 2}^{\frac \tau 2} \mathrm d t \quad d'_t \cdot \Exp{-i\omega_k (t-\Delta t} = d'_k \cdot  \Exp{i\omega_k \Delta t} \label{phaseshift}
\end{equation}
Thus, having reconstructed the template characteristics $T = \{\ol f_k,\sigma_k,\ol \Phi_k, \sigma_{\Phi_s}\}$ for the first time interval, the subsequent intervals may be analysed by carrying out a parameter study over $\Delta t$. This is done by shifting the measured phases $\ol {\vv \Phi}_{\vv d}\rightarrow \ol {\vv \Phi_{\vv d'}}$ according to \mref{phaseshift} and taking e.g. a MAP-approach over the joint probability of the phase \mref{phasemap}, now varying $\Delta t$, not $a$ which is assumed to be $1$.\\
We now are also able to give the approximation for $a \approx 1$. We derived in this case a phase shift of $\Delta\Phi(a) = \Exp{i2\pi naK}$. Comparing with \mref{phaseshift}, the problem of detuning for small $a$ may be approximated by testing all pulses for a shift of
\begin{equation}
 \Delta t = a K \cdot T
\end{equation}
dependent on the pulsar period number K. All practical cases of detuning fall in the category of the aforementioned reduction to systematic phase shift.
\subsection{Reference independent difference phase model}\label{rim}
\begin{figure*}
\input{graphics/toascheme.tex}
\label{fig:toascheme}
\caption{{\bf Overview of ToA generation}: The statistical imprint of the various epochs ({\bf A}) is used to generate statistical templates ({\bf B}). The data of the epochs ({\bf D}) is non-linearly filtered using the statistical templates to yield a probability distribution ({\bf C}) for the shift of every epoch relative to a reference epoch. For every epoch, the mean phaseshift in topocentric pulsar periods is then added to epochs MJD to calculate the specific ToA ({\bf E}). The error is calculated as the variance of ({\bf C}) for this shift. The reference epoch's error on the ToA is determined as if it would have been timed by every other epoch. }
\end{figure*}
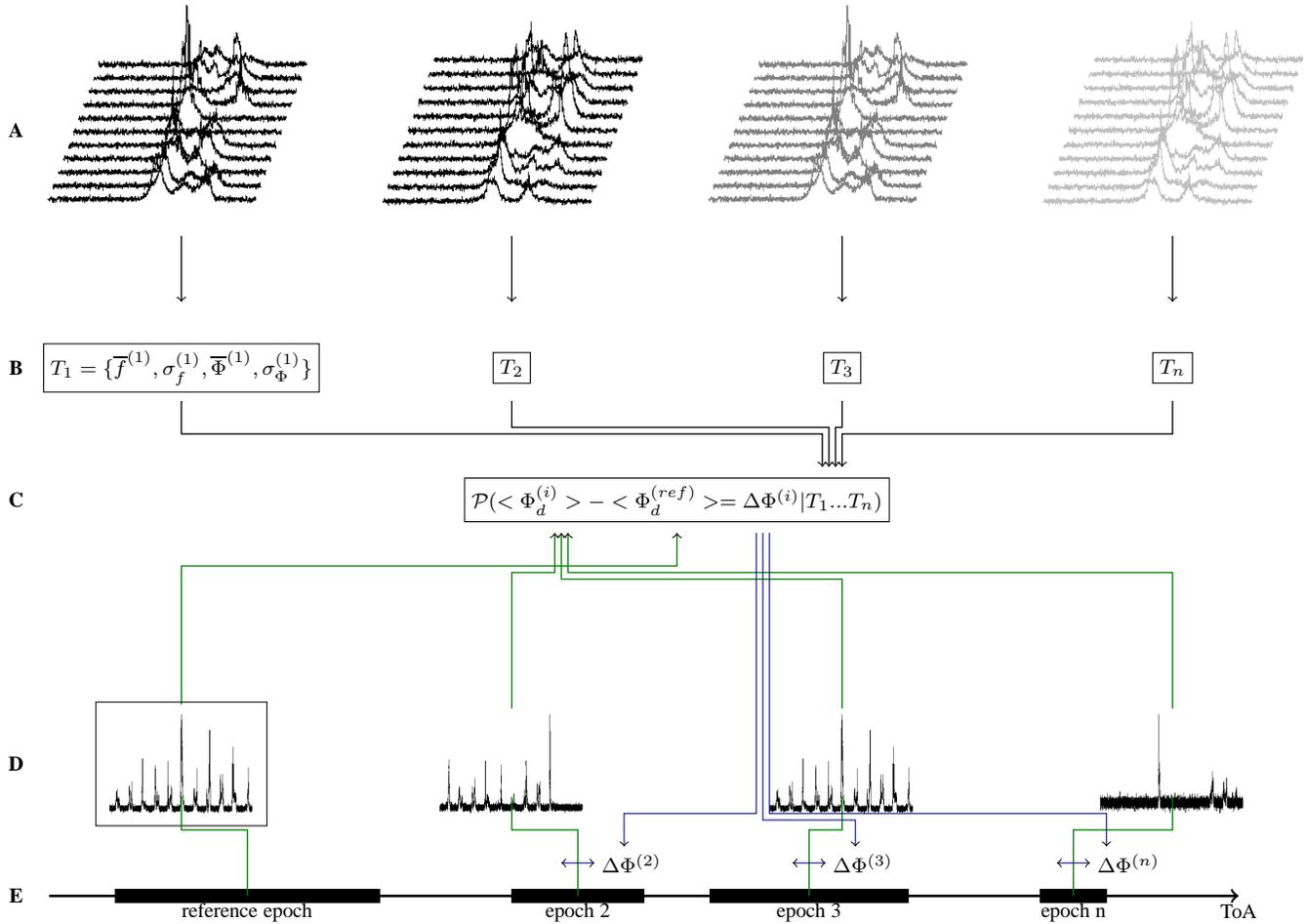

The simplest way to form a reference template is to select a single observation. However this relies on the chosen observation being statistically representative of the sample as a whole and would discard the extra
information available in the whole dataset.\\
This in principle could be mitigated by building a total statistical template over all epochs observed. We discarded this method for two reasons. First of all it assumes that the pulsar template shape and statistical behaviour is the same in every single epoch (which is not the case for e.g. moding pulsars). Furthermore forcing the template to take the same probability density in signal amplitude for distinct epochs is too strict an assumption as for example ISM variations modulate the signal amplitude systematically. Instead, one may formulate a probability for the relative shift of two distinct epochs, given a certain epoch's statistical template is correct, and then demand that the probability distribution has to agree for all templates at once. We depicted this process schematically in Fig. \ref{fig:toascheme}. The calculated probability on a relative phase between two epochs, given all knowledge gathered, is a clear, basic statement coming without the notion of a ``reference epoch''. It can be incorporated in further analyses without introducing a direct bias caused by the phase information of statistical templates used since the exact phase assumed in the template drops out of the calculation. The knowledge gathered just filters the data to be compared in a clever way.\\
Mathematically speaking, we demand that 
\begin{align}
&\PP{\langle\Phi_d^{(i)}\rangle-\langle\Phi_d^{(ref)}\rangle=\Delta\Phi^{(i)}|T_1...T_n} =\nonumber\\&= \prod_j \PP{\langle\Phi_d^{(i)}\rangle_{T_j}-\langle\Phi_d^{(ref)}\rangle_{T_j}=\Delta\Phi^{(i)}|T_j}\label{pdifsch}
\end{align}
The formula for one factor of this product may be deduced from \mref{phasemap} by taking the product of the equation for two epochs' shifts $\Delta \Phi_1 , \Delta \Phi_2 = \Delta \Phi_1 + \Delta \Phi$ and integrating out the difference to the used template, $\Delta \Phi_1$. A factor then reads:
\begin{align}
&\PP{\langle\Phi_d^{(i)}\rangle_{T_j}-\langle\Phi_d^{(\rm{ref})}\rangle_{T_j}=\Delta\Phi^{(i)}|T_j} = \prod_k \frac{1}{\sqrt{2 \pi {\sigma^{(i)}_k}^2}} \cdot \nonumber\\ &\cdot
 \Exp{\frac 1 {2 {\sigma^{(i)}_k}^2 }(\langle\Phi_d,k^{(i)}\rangle_{T_j}-\langle\Phi_d,k^{(\rm{ref})}\rangle_{T_j}-k\Delta\Phi)^2}
\end{align}
Where ${\sigma^{(i)}}^2={\sigma_{s',k}^{(i)}}^2+{\sigma_{s',k}^{(\rm{ref})}}^2$ is the sum of the new phase variances estimated from the epochs' data, given $T_j$ holds. The formula deduced simply states that the difference of the filtered mean phases of the two epochs under consideration is the mean phase difference times the Fourier coefficient number.\\
While the gathered probabilities on phase differences are reference free, in order to output a TOA, we have to define that the absolute phase of a certain epoch amounts to zero. In the classical procedure, the template was assumed to have zero phase. Additionally there exist several conventions on defining the physical point on a profile where the phase is zero, including but not limited to defining the mid of the profile at the ``center of mass'' of the profile or at the highest peak. To that extent we declare a ``reference epoch'' which's ToA is declared to be exactly the timestamp of the rising edge of the first bin of the pulse. As we assume the epoch with the highest S/N-ratio to have the lowest variances in its statistical template as seen from the others, we choose this one to be the reference epoch. While, per definitionem, this TOA's relative phase is exactly zero, the corresponding ToA has to have a variance to be fitted into TEMPO2. Thus we calculate the variance of the reference as it would have been determined by all other epoch's relative shifts as 
\begin{equation}
\frac 1 {{\sigma^{(\rm{ref})}}^2} = \sum_j \frac 1 {{\sigma^{(j)}}^2}
\end{equation}
\\
For the other epochs, the probability distribution of the relative phase shift provides the correct error estimate on the ToA automatically.\\
Another big advantage of parametrizing the ToAs by their difference is that \mref{pdifsch} can be calculated independently for every template and added logarithmic. This can be done since the measured variable, phase difference, is the same for every pair of epochs, no matter which statistical template is assumed for the moment. This gives us the freedom to rasterize the probability distribution as a whole without preferring one template over the others and stay in the Bayesian picture until reducing the probability distribution measured into mean phase differences and variances in the very last step. This becomes essential when also short epochs with weak signal over noise ratio are to be included in the analysis. In this case multiple phases are still plausible e.g. mistaking one peak for the other (see also fig. \ref{fig:sgrob} and Sec. \ref{dps}) since the yielded probability distribution for one template may be too far from the gaussian form to be reduced to a mean and variance. The ToA $t_{\text{ToA}}$ of epoch $i$ with mean phase shift $\Delta\Phi^{(i)}$ and variance ${\sigma^{(i)}}^2$w.r.t. the declared reference then amounts to:
\begin{equation}
 t_{\text{ToA}} = t_{\text{\emph rising edge}} + P_{0,i}\times(\Delta\Phi^{(i)} \pm \sigma^{(i)})
\end{equation}
where $P_{0,i}$ is the folding period of epoch $i$ and the phase difference takes values in the interval $[-.5:.5]$.\\
Developing an interface to Bayesian extensions of TEMPO2, like TempoNest\citep{2014MNRAS.437.3004L}, to communicate the whole probability distributions on the relative phase information to the pulsar timing code would be a further step to enhance the statistics on the ephemeris' parameter sets. While this may be desirable in the future we simply reduce the yielded probability distributions to their mean and variance and calculated ToAs directly comparable to the ones from tools like pat included in the PSRCHIVE toolset and processable by TEMPO2 without modifications.

\section{Application to data}\label{application}
Having derived all relevant equations we may now apply the formulae upon given datasets and test the approach. Starting with simulated datasets we afterwards evaluate a dataset of PSR B1133+16 with 2041 consecutive single pulses consisting of 1024 bins each. Furthermore we determined ToAs from data of PSR J1713+0747 and analysed these ToAs with the TEMPO2 software package comparing with ToAs generated using ``pat'' from PSRCHIVE and a very accurate classical analytical template.

\subsection{Simulated data}
We wrote a simulated data generator serving single pulses of non-fluctuating shape $\ol {\vv f}$ but multiplied with gaussian red noise to account for a highly fluctuating radiation process and analysed the convergence of the amplitude model in Fourier space. The classical average is in this case the statistically optimal procedure to determine the pulse profile since the assumption that there is exactly one profile holds, and a quickly converging law of large numbers applies. Fig.\ref{fig:fittest} depicts the reconstruction from ten, hundred and a thousand pulses comparing classical template integration with Bayesian reconstruction. 
\begin{figure*}
\begin{tabular}{cr}

\begin{minipage}{12.5cm}\input{graphics/p_fit10}\end{minipage}&\begin{minipage}{5cm}After ten pulses, the Bayesian method has equally well reconstructed the signal as the average method, however it provides a handle on the signal fluctuations still possible. The seamingly systematic overestimation of the signal shape by the expected mean amplitueds comes from the log-normal exponential model used and the non vanishing $\sigma_{\vv f}$\end{minipage}\\
 \begin{minipage}{12.5cm}\input{graphics/p_fit100}\end{minipage}&\begin{minipage}{5cm}A hundred pulses already give a profile accurate to over 90\% over the essential Fourier modes of the signal. Notice how the real error on the averaged $\ol {\vv f}$ is limited to the still possible statistical fluctuations $\sigma_{\vv f}$. \end{minipage}\\
 \begin{minipage}{12.5cm}\input{graphics/p_fit1000}\\\end{minipage}&\begin{minipage}{5cm}A thousand pulses give a confident template up to high coefficients. The Bayesian reconstruction estimates the average shape to fluctuate below 20\% of its mean. The absolute error not detected at the very high coefficients may be explained due to low assumed signal and low real signal to noise ratio. The dominant error source seems to be a systematic overestimation due to the log-normal signal model.\end{minipage}\\
\end{tabular}
\caption{\label{fig:fittest} Reconstructions of the original signal shape by both Bayesian and classical methods and their errors after 10, 100 and 1000 pulses respectively. The lower half shows the errors of the estimation to the original signal shape and the Bayesian estimation of the remaining uncertainty $\sigma_{\vv f}$ on the expected log-normal mean $\ol {\vv f}$}.
\end{figure*}
Both methods perform equally well, as was expected. However, the Bayesian reconstruction also gives us an estimate on the profile stability and thus a better understanding of the remaining statistical uncertainty. Reconstructing a constant signal shape is a difficult task for a log-normal Bayesian model since the guess of a non-fluctuating shape is statistically only feasible by collecting lots of data. Thus, even though the test seems simple, it is actually a very strict test for the Bayesian measurement model to pass as the correct solution for infinite observation time is delta peaked. The algorithm is expected to perform much better on real data for which the profile shape measured is much more fluctuating for small integrations since it has a way of incorporating the instabilities of the profile. Since in an observation of finite length the algorithm will never show zero variance on the profile shape, it estimates the profile intensity slightly but systematically too high. This gives rise to a problem if one is set out to measure the intensity of radiation naively by interpreting the statistical template as a way of describing the profile. The template gives a statistical way to determine all compatible intensities or the probability distribution of these, but not the ``true'' intensity directly. \\
The high modes exhibit a large absolute error. This is an effect of low signal to noise ratio at imperfect numerical integration. For low amplitudes, the Hamiltonian for different values of $\ol {\vv f}$ becomes rather flat and thus the outcome just reflects the prior we have been taken. In this case the prior was a log-normal Gaussian model with $2\sigma_{\vv f}$ around the average found. Notice that the $\sigma_{\vv f}$ the algorithm suggests (leaving aside the singularities of the profile) captures the true error (except for singularities and the imperfect numerical integration of high modes) giving a handle on the remaining statistical uncertainty.
\subsection{Determining phase shift}\label{dps}
\begin{figure}
 \input{graphics/s_probs_grob}
 \caption{\label{fig:sgrob}Unnormalized logarithmic probabilities for phase shifts. The comb structure comes from the four peaks present in the template.}
\end{figure}
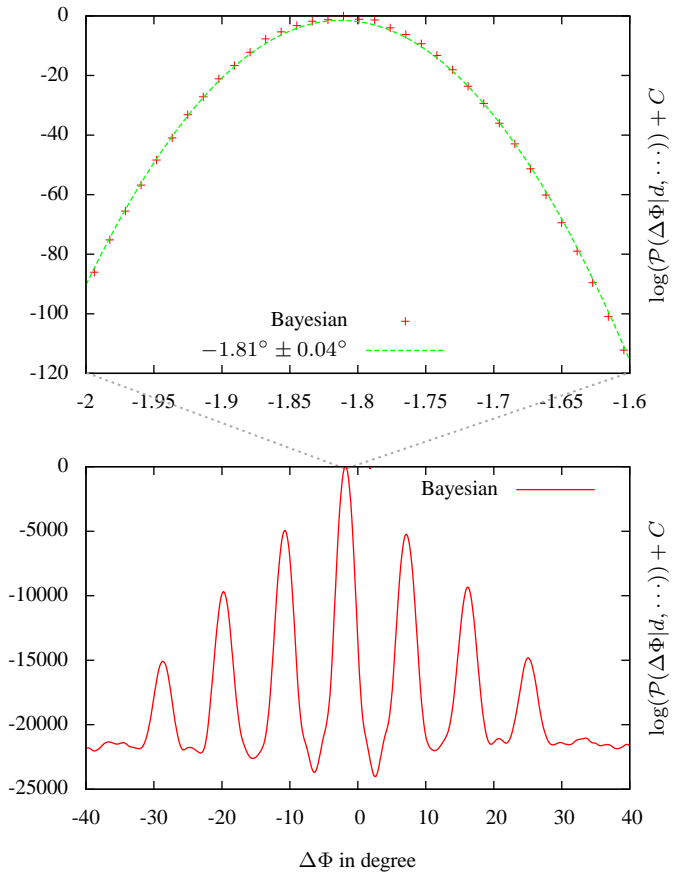

We used a template with four peaks of Gaussian shape, sizes ranging from $1.8^\circ$ to $3.6^\circ$ and generated a dataset with an overall signal over noise ratio of $-17dB\approx \frac 1 {50}$. We compared two independently generated datasets of one hundred pulses with 1024 bins each, shifted by $1.8^\circ$, which amounts to a shift of 5.12 bins. As Fig.\ref{fig:sgrob} shows, the Bayesian reconstruction yields $\Delta\Phi = -1.81^\circ \pm 0.04^\circ$. The other maxima in the probability distribution are of smaller size but reflect the the spacing of the templates peaks and the possibility of mistaking one peak for another. Having a local signal to noise ratio of four for every peak and a resolution of $360^\circ/1024 \approx 0.35 ^\circ$ available, we may estimate the uncertainty of every peaks location to be $1.3^\circ$\footnote{Calculating the exact timing precision achievable is highly nontrivial. We estimated this error as follows: Assume a single peak at a S/N of one to be locateable to about its width, having a signal to noise ratio of four should increase this precision by a factor of $\sqrt4$. Adding the errors on the bin and resolution quadratically, we end up with an average of $1.3^\circ$ per peak.} in every single peak in every single pulse. Having derived the phase drift using $200 \times 4$ peaks, we classically calculate the remaining error due to the receiver noise to $\frac{1.3^\circ}{\sqrt{2\cdot400}}\approx 0.046^\circ$. In the fluctuation free case (the simulated emitted signal is exactly the four peaked shape before the receiver noise is added and the signal is squared) it can be shown that this error is the minimum reachable error. The test of our algorithm reaches this accuracy, too.

\subsection{Real data}
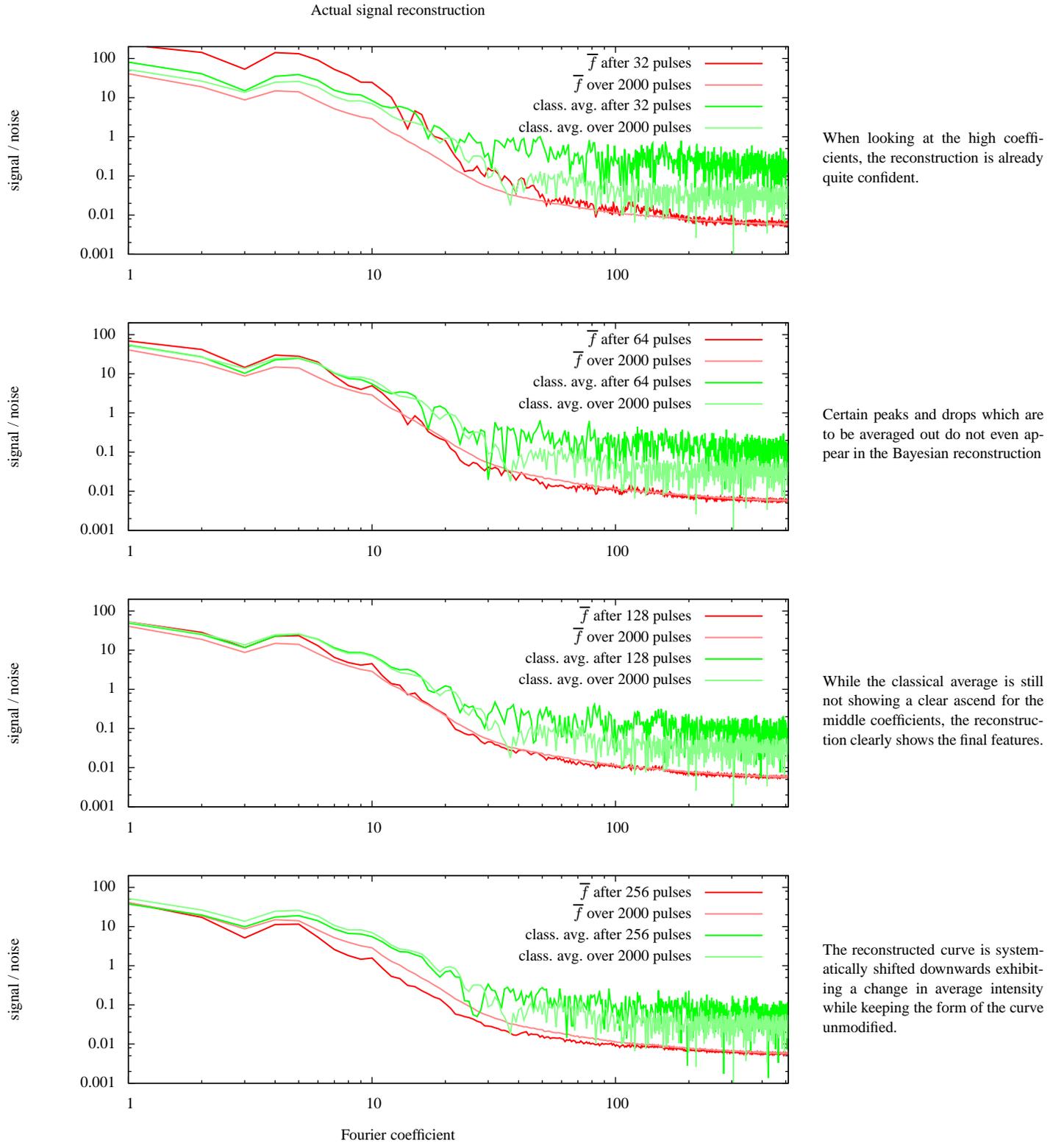
\begin{figure*}
 \begin{tabular}{cr}
 \centering Actual signal reconstruction&\\
 \begin{minipage}{14.5cm}
\input{graphics/methodcomp32}
 \end{minipage}
 & \begin{minipage}{4cm}When looking at the high coefficients, the reconstruction is already quite confident.\end{minipage}\\
 \begin{minipage}{14.5cm}
\input{graphics/methodcomp64}
 \end{minipage}
 & \begin{minipage}{4cm}Certain peaks and drops which are to be averaged out do not even appear in the Bayesian reconstruction\end{minipage}\\
 \begin{minipage}{14.5cm}
\input{graphics/methodcomp128}
 \end{minipage}
 &\begin{minipage}{4cm}While the classical average is still not showing a clear ascend for the middle coefficients, the reconstruction clearly shows the final features.\end{minipage}\\
 \begin{minipage}{14.5cm}
\input{graphics/methodcomp256}
 \end{minipage}
 &\begin{minipage}{4cm}The reconstructed curve is systematically shifted downwards exhibiting a change in average intensity while keeping the form of the curve unmodified.\end{minipage}\\
 \centering Fourier coefficient&
 \end{tabular}
 \caption{Template reconstruction for a subset of pulses}\label{fig:rec}
\end{figure*}

\begin{figure*}
 \input{graphics/reconstruction10}
 \caption{\label{fig:finishedtemplate}Template reconstruction after ten respectively. a thousand pulses. While the classical averaging assumes a certain fluctuating form to be true, the Bayesian template and single pulse model shows which single pulses are typically compatible with the statistics of the data gathered}
\end{figure*}
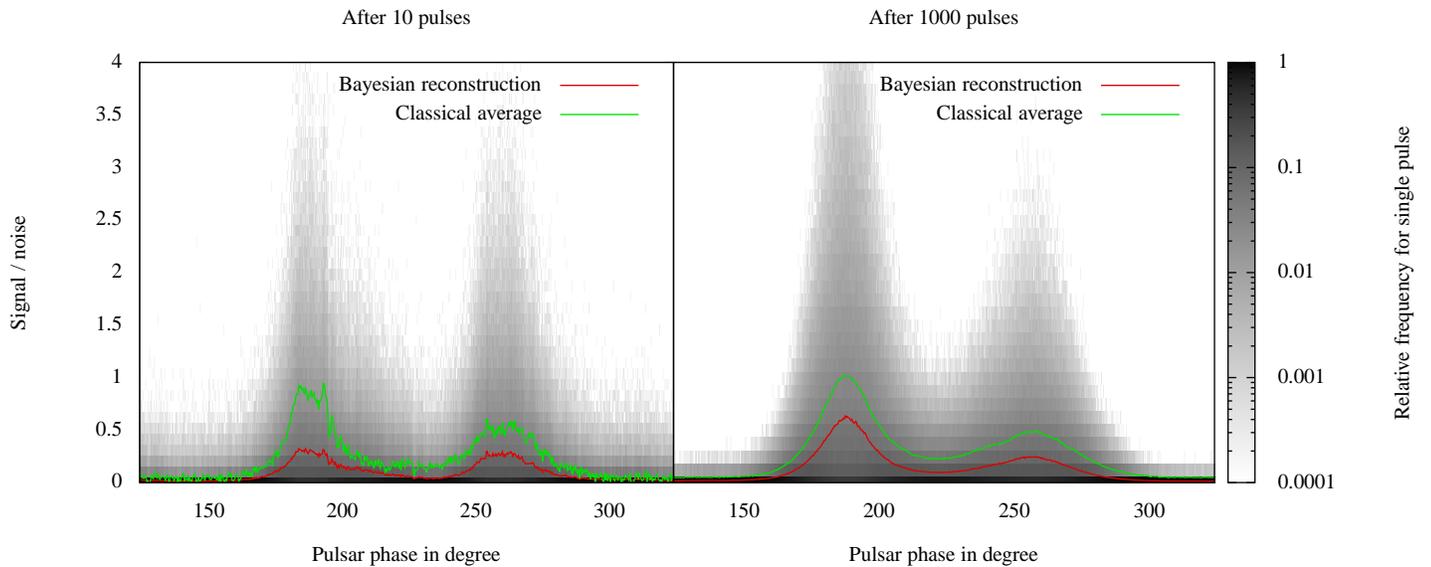

Generating templates from real data yields reconstructions of surprising confidence. Fig. \ref{fig:rec} depicts the amplitude reconstructions from certain numbers of consecutive pulses. In each figure, the long-term convergence gathered from the whole dataset may serve as a reference. We emphasize that for data gathered by a pulsar with fluctuating shape, classical and Bayesian reconstruction are in general not expected to yield the same curve, since fluctuations are handled differently by both models. Both can only be compared to their own long-term limit. Furthermore the statistics of the profile seem to change with time for the dataset examined. We will analyse this short term behaviour in the next section.\\
The Bayesian reconstruction does exhibit lower noise at high fourier coefficients and converges rather quickly to its ultimate form. The main features such as peaks and curvature appear already at 64 pulses while the classical average remains dominated by fluctuations occurring. However systematic offsets in the Bayesian reconstruction appear. These may largely be backtracked to the actual pulsars intensity changes and nulling periods. A numerical fluctuation of the algorithm is rendered unlikely since each point of the construction is calculated independently and thus an equal rise in all coefficients is highly improbable. The classical average also shifts upwards and downwards with varying intensity, but the effect is mostly indistinguishable from noise. For this real life example, the method derived is clearly superior to the 
classical method in areas of low signal to noise ratio, given that it converges very fast to its ultimate form and showing denoised systematic profile changes. It is reproducing the classical outcomes in the lower coefficients and not picking up the noise of the higher coefficients. After 128 pulses a confidence of about ten percent is already reached for all coefficients present. This paves the way to investigate changes in the radiation profiles on these scales. We could further diversify our pulses in intensity classes and perhaps in different modes of radiation given the contrast and fastness of the envelope estimate. Fig. \ref{fig:finishedtemplate} shows the differences of the two reconstruction methods in time domain and the single pulse probability density function as described by the Bayesian statistical template (depicted as grey shade): While the classical average just assumes the pulsar to have a certain profile shape, the Bayesian method is rather based on exclusion and describes profile shapes compatible with the data gathered. Thus, after as few as ten pulses, the Bayesian reconstruction in time domain already resembles the final form of the radiation profile. Notice that it still allows for signals out of the profile, since statistics do not exclude low signals after ten pulses. 

\subsection{Reconstruction stability to fluctuations}
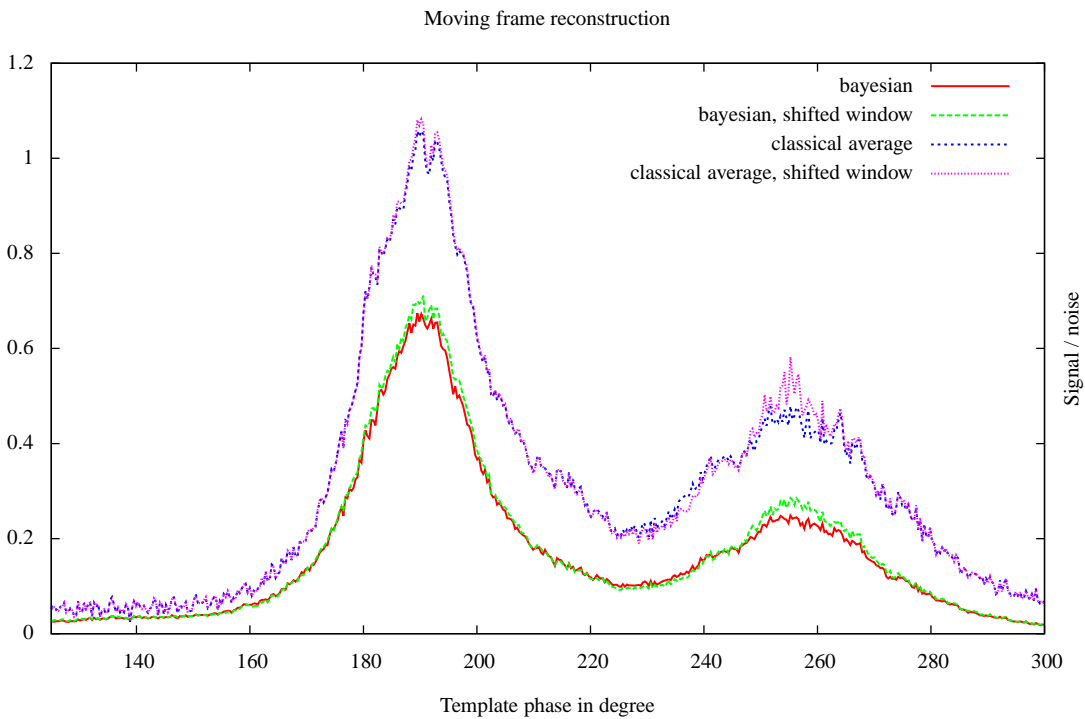
\begin{figure*}
\input{graphics/movingframe}
 \caption{\label{fig:movingframe} Depicted are the Bayesian and classic reconstructions from two sets of 50 pulses shifted by a single pulse. The change of the classical template is dominated by the spiky fluctuations as is evident from the change in the right peak.}
\end{figure*}
The workings and main differences of the two algorithms become evident when analysing the change of the classical profile and statistical template introduced by small changes of the dataset. This is examined by comparing the reconstructions of a window of fifty pulses moved along the dataset. An example of stepping one pulse further is shown in Fig. \ref{fig:movingframe}. Dropping one pulse and inserting another should not change the overall template much. If the template changes, we expect it to change smoothly with time. For the Bayesian reconstruction, this property can be easily observed whilst individual fluctuations dominate the classical average picture. This can be deduced from the right peak of the template. While the algorithm derived gives a rather smooth increase in intensity over the whole peak, the classical average is dominated by spikes of individual fluctuation. One could argue that the classical spikes might just be smoothed out of the form of the fourier transform. This is contradicted by highly localized changes in the Bayesian reconstruction that are statistically significant. The change of the left peak may be seen as an example for a localized statistical change. The classical average does not change much there, but the Bayesian reconstruction shows a significant rise of the left flank of the peak. The algorithm is found to behave in such a way over the whole dataset. Furthermore it incorporates the fluctuation seen in real envelopes. This may increase accuracy of derived parameters like the ToAs since e.g. the confidence of a peak appearing at a certain point should be, but is not, taken into consideration by classical timing procedures.
\subsection{Nulling}\label{nulling}
\begin{figure}
\input{graphics/nulling_simple}
\caption{\label{fig:nulling}Discrimination-less decision: Bayesian analysis gives a probability for a certain pulse candidate to be explained by white noise only. In subsequent data-processing, this probability may be used to weight signals that are only present in non-nulling phase instead of a threshold algorithm producing false positives or negatives.}
\end{figure}
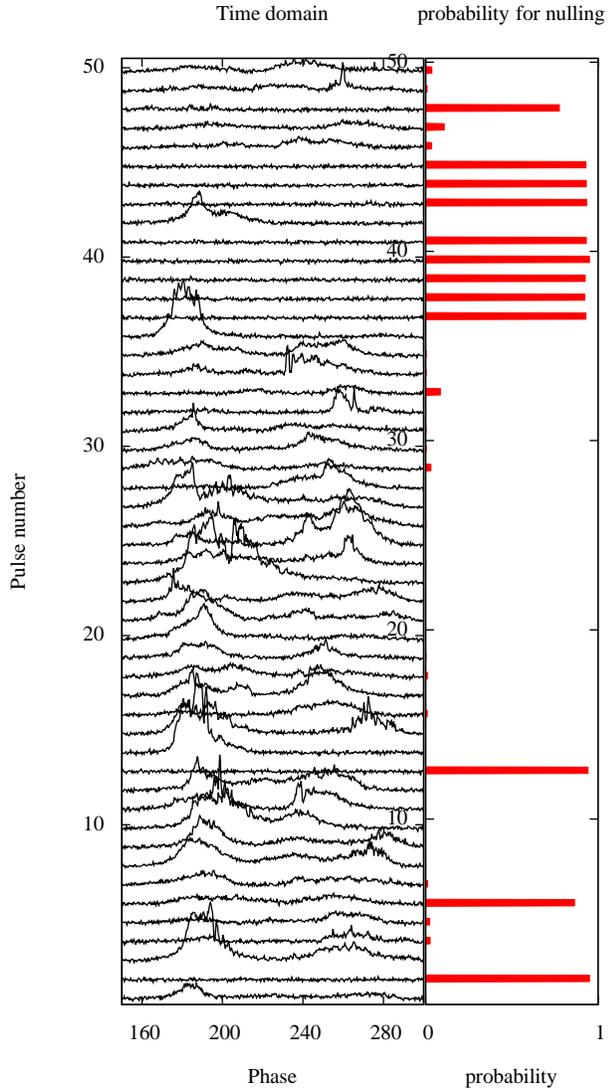

Given a set of proposed models, like different statistical templates, Bayesian statistics can also assign a probability on how likely every single model is to describe the data, given that one of the proposed models is true. An example for the ability to discriminate between different models is nulling detection. Two sets of parameters are analysed within the framework: One set of $\ol {\vv f},\sigma_{\vv f}$is generated from a training set of nulling periods taken from the first 50 pulses, the other one is taken from the overall reconstruction as above. The likelihood of a model $i$ described by a set of parameters $m_i$ is compared as
\begin{equation}
 \PP{m_i|\vv d} = \frac {\Dint \vv s \PP {m_i,\vv s,\vv d}}{\sum_i \Dint \vv s\PP {m_i,\vv s,\vv d} } \label{modingeq}
\end{equation}
which may be derived straight-forwardly from Bayes' theorem. Deciding between the white noise model and the model derived from all thousand pulses, this formula gives the probabilities given in Fig. \ref{fig:nulling} for a certain frame to be a nulling pulsar period.
These probabilities could be used as weights when determining further parameters without biasing the statistics by deciding whether there was a pulse or nulling. Furthermore when looking at the plot, the Bayesian probabilities resemble our state of uncertainty for pulses which are in the 50\% region when a by eye decision should be made. Fortunately, the method is able to quantify this uncertainty for us automatically. Nulling information could in principle further improve the accuracy of detuning and time of arrival analysis.
\subsection{Determining Times Of Arrival}
\subsubsection{ToAs from simulated data}
\begin{figure}
\input{graphics/toa_fixedref}
\caption{\label{fig:fixedtoas}Test of accuracy over epoch size of the phase shift model as described in sec. \ref{psm}. a variable number of pulses is used to generate ToAs and the algorithm is examined for statistical (red) and systematical (green) errors.} 
\end{figure}
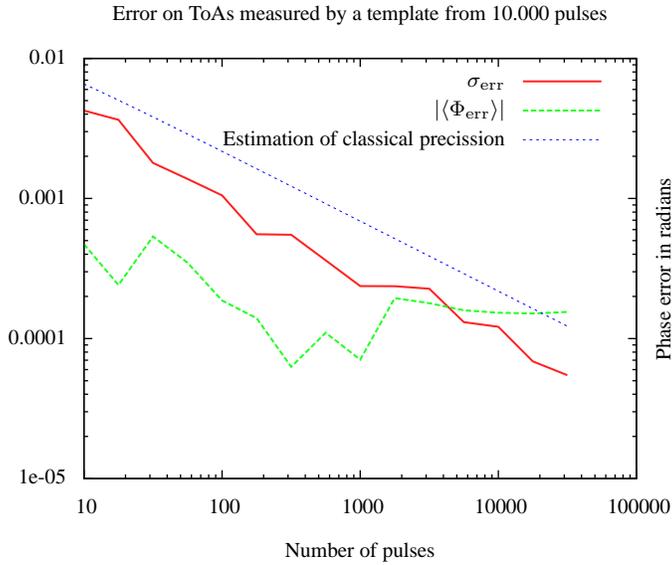
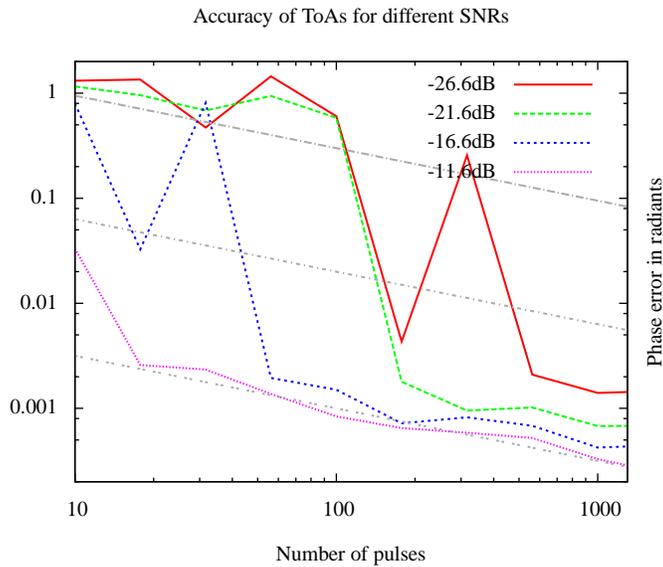
\begin{figure}
\input{graphics/toa_snr}
\caption{\label{fig:fixedtoasnr}Test of accuracy for different S/N-ratios of the phase shift model as described in sec.\ref{psm} the grey dashed lines all follow a $1/\sqrt N$ slope.} 

\end{figure}
Timing simulated data sets is a precise way to test the algorithm for theoretical performance since the true signal to be found out of noisy data are known. We implemented the phase shift model (as described in Sec. \ref{psm}). We generated a statistical template out of 10,000 simulated pulses and then measured the accuracy on timing $N=10,\cdots,50,000$ pulses with that fixed reference, where we used a MAP approach. Statistics was gathered over 20 randomly shifted datasets for each value of $N$ recording the absolute error on the ToA and deriving the mean error and standard deviation to quantify the accuracy of the algorithm. The algorithm performed as depicted in Fig. \ref{fig:fixedtoas}. As expected, the systematic phase error was well bounded by the fluctuations given by the test datasets. The deviation from the true value followed a $1/\sqrt{N}$ law for large $N$ as expected and reached the accuracy that is information theoretical possible. When there were more than 10,000 pulses in the dataset to time, a systematic error appeared. This was expected, since the reference statistical template itself (generated from 10,000 pulses) does not contain more accurate phase information. Since we were measuring an absolute, not a relative shift here, the error contained in the template becomes systematic. The sudden drop of the systematic error between 10 and hundred pulses is a random sampling artifact of the simulation involved. The average over the discrete maximum values taken was in this case accidentally zero. We also evaluate the reached accuracy for different S/N-ratios in Fig. \ref{fig:fixedtoasnr}. It is interesting to notice that the algorithm returns a rather flat probability distribution rendering a lot of phase values likely. This can be deduced from the high errors encountered leading to a plateau for low values of $N$. At a certain point the probability distribution begins to peak and accuracy increases with a jump, locking onto the signal. This may be explained by passing the point where different peaks of the profile shape can be distinguished statistically. This is not a drawback of the method but reflects the fact that, also in the ideal case, the error follows the $1/\sqrt N$ slope only for large $N$ and a unique maximum. After that point the performance follows a stable $1/\sqrt{N}$ curve.\\
\begin{figure}
 \input{graphics/toa_coordfree_20}
\caption{\label{fig:toa_free_sim}The upper panel depicts the remaining uncertainty (in an arbitrary scale) when the row epoch is measured by the column epoch. Notice that there are some combinations of epochs which lack statistical similarities and give greater uncertainties. The lower plot shows the remaining error on the ToAs of the epoch when combining the measurements from above.}
\end{figure}
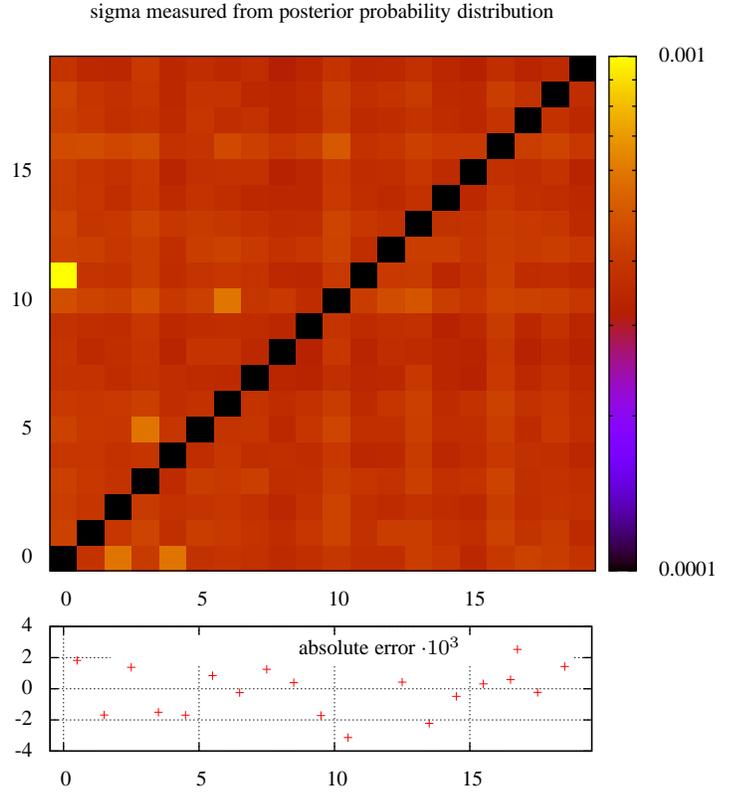
We also simulated the whole process from measuring epochs to determine the ToAs therein. To this end we simulated 20 epochs with 100 pulses each. Then we generated statistical templates from each epoch and used that to time the relative shifts of every other epoch. Then we naively averaged the maximum a-posteriori values of the shifts and their variances using classical formulae for measurement value addition to find the ToAs and compare them with their true values. The dataset contained 2000 single pulses. The results of this simulated measurement is displayed in Fig. \ref{fig:toa_free_sim}. We reach an accuracy of $1.9 \times 10^{-3}$ radians, which is nearly the theoretical maximum for 100 pulses per epoch as can be seen from Fig.\ref{fig:fixedtoas}. When timing with only 10 epochs, the accuracy dropped to about $3\times10^{-2}$ radians. In that case, the theoretically possible accuracy for each epoch was missed. That practically means, that gathering knowledge of later epochs can improve measurements of the past by reprocessing them with the newer statistical templates.
\subsubsection{ToAs from real data}
The method carried out in such a way fails when tested with real data. Reducing the phase shift information of every statistical template to classical mean shifts and their variances before correlating the different templates' measurement flaws the precision of the method. This can be understood easily considering very noisy templates. These may cause different very likely shift values separated by a larger interval of unlikely phase values, e.g. different peaks of the profile may be mistaken for each other. When reducing, the mean and variance of only the most likely peak was determined giving a rather low sigma value on it, since in its neighbourhood, the probability density can be approximated as a gaussian. This sets a too large weight on the statistical template compared with the other templates if the algorithm accidentally locked on the wrong peak. The consequence is a large bias in the direction of the mistaken peak. However, if one does not reduce the data template by template, but examines the probability density on the relative phase shift of two epochs using all statistical templates at once, one circumvents the possibility of this local fallacy and gets an overall correct probability distribution.\\
An algorithm obeying these caveats has been outlined in Sec. \ref{rim}. By measuring the same quantity (relative shift of two epochs) with different statistical templates we may rasterize the probability distribution for that quantity considering all references at once and in parallel. Since pulsar timing expects us to state a single ToA per epoch and the expected error on it, the MAP ansatz taken in the very last step is now justified. The reported errors on the ToAs are found to be in agreement with the RMS reported by TEMPO2 yielding low $\chi^2$ values.\\
Using 10 second observations of PSR J1713+0747 at $1350\rm{MHz}$ with a total of about $19\rm h$ observation time in 36 epochs observed in Effelsberg over about 1.9 years we tested the algorithm. Even though single pulse observations carry much more statistical information that the data at hand, the statistical templates generated show variations. The ToAs generated reach an RMS of $431\rm{ns}$ and a weighted RMS of $163\rm{ns}$ at a fit $\chi^2$ of $35$. The classical method yields $551\rm{ns}$ unweighted or $240\rm{ns}$ weighted RMS with a $\chi^2$ of $455$. We collected the results in Table \ref{tab:timcompare}.\\
\begin{table*}
\caption{\label{tab:timcompare}Comparison of timing residuals of observations taken at $1350\rm{MHz}$  and model accuracy as reported by TEMPO2 for classical and Bayesian ToAs. The smallest values in each category have been highlighted.}
 \begin{tabular}{| c | c | c c c | c c c |}
& &\multicolumn{3}{c}{Classical timing}				& \multicolumn{3}{c}{Bayesian timing}\\
  TEMPO2 fitting mode&rms/Wrms	& $\chi^2$ 	& red. $\chi^2$ & rms/Wrms 	& $\chi^2$ 	& red. $\chi^2$ \\\hline\hline
	
RMS	&$551\rm{ns}$ 	& - 	  	& -		& $\mathbf{431ns}$	&  -		& - 		\\

weighted RMS 	&$240\rm{ns}$ 	& 455 	  	& 23		& $\mathbf{163ns}$	&  $\mathbf{35}$& $\mathbf{2}$ 	\\
\hline
\end{tabular}
\end{table*}
\subsection{Moding}\label{moding}
\begin{figure}
 \input{graphics/scatter}
 \caption{\label{fig:scatter}We overlayed the average profile curves of different modes over a scatter plot of the single pulses coloured according to the probability of belonging to mode 1-3. The dotted shape amounts to an average over all pulses.}
\end{figure}
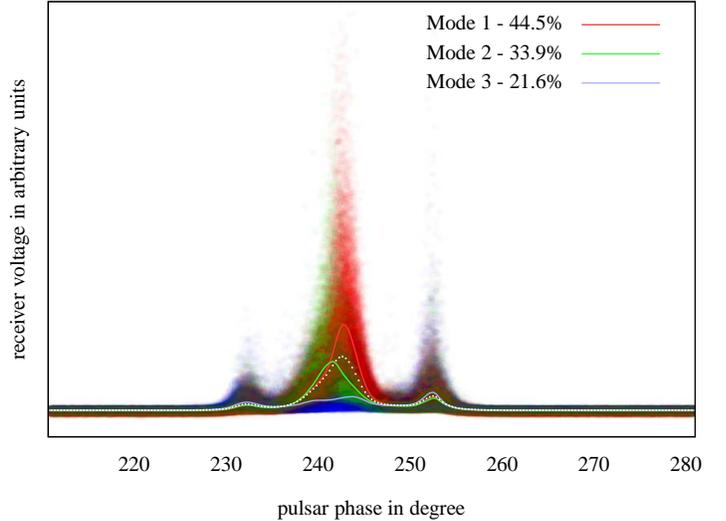
\begin{figure*}
 \input{graphics/histo_5000_10_200}
 \caption{\label{fig:histo} This figure compares the phase detection performance of a single or a moding statistical template on 25 respectively 50 pulse integrations.}
\end{figure*}
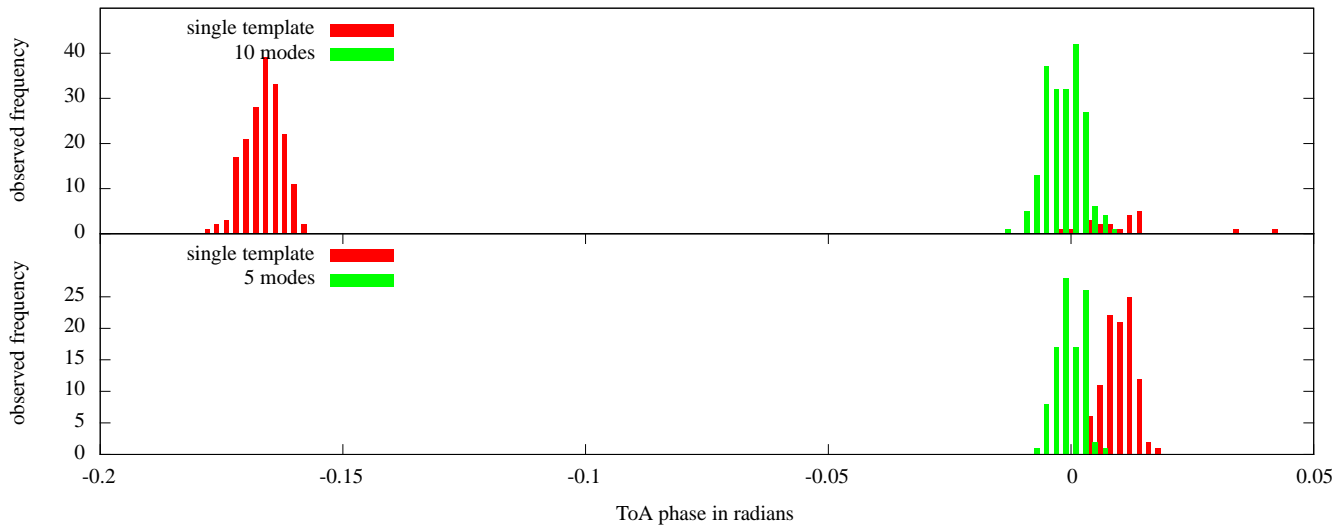

We examined the algorithm's timing behaviour on a moding pulsar. We used a dataset of pulsar B0329+54 at 21 cm consisting of 5000 single pulse observations with 1024 bins each. Using the same formula \mref{modingeq} as for the nulling anlysis, the alorithm decides if a single pulse or a subset of pulses was emitted in a certain pulsar mode described by statistical template $m_i$ (consisting of the same set of parameters like $T_i$ ). We auto-generated a moding analysis by generating a statistical template $T$ over all data. Then we analysed the probability of each given single pulse to appear, given the template $T$ describes the radiation correctly. Basically this means evaluating the joint probability for the phase \mref{phasemappre} and amplitude (by integrating \mref{finalwithoutg} over s).  We ordered this list of probabilities descending and initially divided it into $n_{modes}$ sections of equal length, where we assumed a certain $n_{modes}$ to be the correct number.\\
In principle, the number of statistically distinct modes is also subject to uncertainty. One may also calculate the probability, that a certain $n_{modes}$ holds. This could in principle further improve the results derived in the following but is of secondary interest for a first analysis of the behaviour of the algorithm.\\
Given this initial assignment of the single pulses to modes, we started an iteration procedure. In each step, a new set of statistical templates according to the weights of the previous step are calculated: $\{T^{(0)}_1,...,T^{(0)}_{n_{modes}}\}$. These were then used as models in eq. \mref{modingeq} to assign the probability for each individual pulse to belong to a certain mode. These probabilities become the new weights for that pulse. We experimented with this iterating procedure and found about five steps to be sufficient that the difference from step to step is negligible.\\
The algorithm allows an unambiguous assignment of single pulses to a certain statistical template except for a very small number of pulses. The statistical nature of single pulses makes it difficult to grasp what classification is happening. Thus we decided to make a scatter plot (Fig. \ref{fig:scatter}) of 5000 single pulses assigned to three modes. The single pulses were blurred to simulate a density kernel and plotted with an opacity of 1\% in the colour of the probability to belong to a certain mode. For orientation, we overlayed the average pulse shapes each mode would generate if it was the sole mode observed along with an average over all pulses, drawn in dotted white.\\
The term moding usually refers to the appearance of few distinguishable shapes of the integrated pulse profile. As a single pulse profile is very different from its neighbouring pulses, it was very common to assign different modes to consecutive pulses. Thus the question arises, whether to still call this behaviour moding or not. On one hand, we could try to further develop the algorithm to make mode switching on such small timescales very unlikely. Then the algorithm would pick up ``modes'' in a more classical sense. On the other hand, the integrated profiles of these ``sub''-modes are quite distinct and certainly their ratio has an impact on the average profile shape over a few minutes integrating and thus also on timing (as we will examine below). Understanding the relationship between these sub-modes and the astonishingly stable moding behaviour over a larger integration time could lead to a deeper understanding of the conditions in the pulsar magnetosphere.\\
These submodes also seem to be correlated over the whole profile. For example, mode three in the figure is the only one having an earlier rise of intensity at 230 degrees and additionally a very low slope in the middle of the profile.\\
Taking these different correlations into account can also archive a more accurate timing and make the timing stable against different integration times and noise. Fig. \ref{fig:histo} shows two examples of how a moding template on single pulse level can improve detecting the phaseshift of a few tens to hundreds of pulses. Statistical templates were generated from 5000 pulses and later used to time simulated epochs consisting of consecutive subsets of 25 in the upper and 50 single pulses in the lower panel, taken from the same 5000 pulses. The most likely value for the shift of each set  was binned to generate histograms of the observed frequency. Knowing that there was no subpulse drift over the 5000 pulses as a whole, we expect the phase to be measured as zero for every subset. This is indeed the case, if we assume the pulsar to radiate in different modes (depicted in green). When run over subsets of 25 pulses, the non-moding template (depicted in red) fails to detect the correct shift by repeatedly mistaking one peak for the other. Unfortunately the few outliers which would have identified the right peak are scattered around zero with a systematic bias. We suspect this bias to be caused by the single statistical template trying to cover two or more very distinct modes of radiation. Doubling the number of pulses in a subset fixes the problem of not matching the right peaks. However, there is still a bias and the results have a larger variance than the moding template. The moding template detects the correct pulse phase with satisfying accuracy while showing a lower variance. This can be understood having a look at the generated modes. The maximum value of the second peak for every single mode is further left for pulses which are less intense. In the average picture over all pulses this correlation cannot be accounted for. When looking at subsets of pulses however, the lower probability of reaching a high intensity mode shifts the peak to the left. Consequently, the reference template is detected to be shifted to the right. The ansatz using one statistical template reduces this inaccuracy to about half a degree, which is still a low value in the light that e.g. the peaks of the first and second mode, if we assume three modes (see Fig.\ref{fig:scatter}), are separated by $1.5$ degrees and the average taken from 50 pulses is still very noisy. Using a statistical moding template, the variance in variability exceeds the inaccuracy introduced by integrating a smaller subset than the one the template was generated from.\\
We conclude that assuming even a low number of modes to be present can significantly improve the timing results both in variance and systematic error.

\section{Conclusions}\label{conclusion}
We developed and evaluated a log-normal Bayesian model for single pulse analysis of pulsars. The algorithm described is able to reproduce the results of classical averaging procedures when considering non-fluctuating templates up to the approximations taken. The method is surprisingly versatile in situations of weak signal or a fluctuating template shape, since it is shown to be more robust against fluctuations. Thus it may enable one to examine fast changes in the radiation profile or distinguish different modes of radiation, such as the nulling analysis, an example for model discrimination, has shown. \\
Secondary parameters' reconstruction may be implemented easily by calculating their imprint on the data and then evaluating the joint probability using the model derived, as was demonstrated with phase drift on simulated epochs. The parameter studies derived in such a way profit from additional statistical information incorporated in the template automatically since only the statistically significant part of the data will have an imprint on the probability distribution of the parameters under examination.\\
The benefit of rather understanding which parameter values are compatible with the data gathered than trying to reconstruct one possible signal out of the dataset might be worth the effort implementing a Bayesian reconstruction for the parameter of your choice. This may give statistical access to hitherto poorly examined parameters and topics about the radiation process and magnetosphere by providing strong and reasonable reconstructions for noisy and fluctuating data where a classical estimator leaves us with a definite but perhaps insignificant answer.\\
A first test to generate ToAs from real data showed that the algorithm developed can reduce $\chi^2$ values while reaching the accuracy of classical methods even though we did not use single pulses but few second integrations and approximated some calculations. It gave correct ToAs even though the statistical templates used were generated from the same data that was to time. The classical way of template generation is known to affect the results of timing if the same data is used for template generation \citep{2005MNRAS.362.1267H}.\\
If one intends to use the method on routine basis as an alternative to the classical one, the data volume to keep for logging and generating reproducible output increases drastically. One would have to find a way of reducing the data without loosing the possibility to reprocess it or re-validate it.\\Data compression and performance might be improved at the same time for Bayesian methods, as \citet{2013MNRAS.429...55V} already points out. Computational feasibility sets a limit to the potential use of Bayesian analysis, but there exist ready to use methods for speeding up Bayesian calculations, such as \citet{2012arXiv1210.3489T} or the Metropolis-algorithm \citep{1953JChPh..21.1087M}. The formulae presented may also be sped up by analysing them in perturbative manner. \citet{2009PhRvD..80j5005E} and references therein develop an information field theory, taking the information Hamiltonian $H = -\log \PP \cdots $ as a starting point. This allows for approximating the numerical integrations involved analytically and breaking their influence down to computationally simpler matrix arithmetic.\\
Furthermore consolidating different data channels, like radio data at different frequencies, is easily possible in a Bayesian framework.\\
Once the computational complexity is dealt with, the Bayesian method might provide a way to reduce observation time for low signal to noise pulsars and give a handle on the statistical uncertainties involved in template generation. It gives access to a broad range of secondary applications such as sophisticated methods of interference detection or testing pulsars long-term behaviour for template changes or short- and long-term moding behaviour affecting ToA-analyses.\\
Hence the basic analysis carried out has shown that Bayesian reconstruction of pulsar templates is not only feasible, but also forms a more complete picture on the pulsars template shape. The additional data turns out to be very valuable for subsequent analysis of secondary parameters mitigating the effects of insignificant fluctuations. Benefits known to arise by inspection of the stochastic behaviour in the analysis will be automatically propagated by implementing a measurement using the proposed Bayesian template. Having these manifold possibilities in mind we suspect the method to bear fruit worth picking and suggest further examination.

\section{Acknowledgements}
The authors would like to thank Thorsten Enßlin for valuable input on the measurement model. We thank Axel Jessner to provide a dataset of B1133+16 conveniently preprocessed to access in the code and valuable discussion. We thank Kejia Lee for the extensive comments shaping and sharping the argumentative line of the paper. We thank Niels Oppermann for extended discussion on the mathematical difficulties associated with the evaluation of the posterior pdf.
\bibliographystyle{mn2e}
\bibliography{info-measure}

\appendix
\onecolumn

\section{Calculations}
\subsection{Integration over g}\label{gint}                
 Starting from eq. \mref{beforegint}:
\begin{align}
 H_B[A] &= \hlf \left[(\sqrt{\vv d}-\widehat{\Exp{\vv f}}\cdot \vv g \sigma_{\mathrm n})\dgr\frac{1}{\sigma_{\mathrm n}^2}(\sqrt{\vv d}-\widehat{\Exp{\vv f}}\cdot \vv g \sigma_{\mathrm n}) + \vv f\dgr \inv {\vv F} \vv f +  \vv g\dgr \inv {\vv G} \vv g + \right] \underbrace{-\log \cali P(\tau) +\frac 1 2 \log(|2\pi \sigma_{\mathrm n}^2|^{N_\mathrm{tot}}|2\pi \vv F||2\pi \vv G||\vv d|)}_{=: \blacktriangle - .5\cdot \vv f\dgr \inv {\vv F} \vv f}\nonumber \\
&= \hlf \left[ \sqrt{\vv d}\dgr \frac{1}{\sigma_{\mathrm n}^2}\sqrt \vv d -2 \frac{(\sqrt \vv d \widehat{\Exp{\vv f}})\dgr}{\sigma_{\mathrm n}} \vv g + \vv g\dgr \widehat{\Exp{\vv f}}\dgr \widehat{\Exp{\vv f}}  \vv g + \vv g\dgr \inv {\vv G} \vv g\right]+ {\blacktriangle}=\nonumber\\
 &= \hlf \left[ \sqrt{\vv d}\dgr \frac{1}{\sigma_{\mathrm n}^2}\sqrt \vv d +\vv g\dgr \underbrace{\left(\inv {\vv G} + \widehat{\Exp{\vv f}}\dgr \widehat{\Exp{\vv f}} \right)}_{=:\vv D_{\vv f}} \vv g -2\frac{(\sqrt \vv d \widehat{\Exp{\vv f}})\dgr}{\sigma_{\mathrm n}} \vv g \right]+ {\blacktriangle}=\nonumber\\
 &= \hlf \left[ \sqrt{\vv d}\dgr \frac{1}{\sigma_{\mathrm n}^2}\sqrt \vv d +(\vv g-...)\dgr D_{\vv f}(\vv g-...)-(\frac{\sqrt \vv d \widehat{\Exp{\vv f}}}{\sigma_{\mathrm n}})\dgr \inv {\vv D_{\vv f}}(\frac{\sqrt \vv d\widehat{\Exp{\vv f}}}{\sigma_{\mathrm n}}) \right]+ {\blacktriangle} \nonumber 
 \end{align}
 Integrating over g leads to:
 \begin{align}
 H_B[A\setminus{\vv g}]&= \hlf \left[ \frac{\sqrt{\vv d}}{\sigma_{\mathrm n}^2}\dgr \left(1- \widehat{\Exp{\vv f}}\dgr \inv{\vv D_{\vv f}} \widehat{\Exp{\vv f}}\right)\frac{\sqrt{\vv d}}{\sigma_{\mathrm n}} +\log |2\pi\inv{\vv D_{\vv f}}|\right]+ {\blacktriangle}=\nonumber\\
  &=\hlf \left[ \sqrt{\vv d}\dgr \frac{1}{\sigma_{\mathrm n}^2}\left(1- \inv {\wtvv D_{\vv f}}\right)\sqrt \vv d +\log |2\pi\inv{\vv D_{\vv f}}|\right]+ {\blacktriangle}=\nonumber\\
  &= \hlf \left[ \sqrt{\vv d}\dgr \frac{1}{\sigma_{\mathrm n}^2}\left(1- \inv {\wtvv D_{\vv f}}\right)\sqrt \vv d +\vv f\dgr\inv {\vv F} \vv f+\log(|2\pi\inv{\vv D_{\vv f}}||2\pi \sigma_{\mathrm n}^2|^{N_\mathrm{tot}}|2\pi \vv F||2\pi \vv G||\vv d|)\right] -\log \cali P(\tau)\nonumber\\
\end{align}
where $\wtvv D_{\vv f} = \widehat{\Exp{-\vv f}}\inv {\vv G}\widehat{\Exp{-\vv f}}+ 1$ is always greater than 1 leading to a broadening of the likelihood in case of bad signal noise ratio. We may reformulate the matrix in the data-dependent term as
\begin{align}
 1- \inv {\wtvv D_{\vv f}} &= [\inv {\wtvv D_{\vv f}}][\wtvv D_f-1] = \inv {[\widehat{\Exp{-\vv f}}\inv {\vv G}\widehat{\Exp{-\vv f}}+ 1]}\widehat{\Exp{-\vv f}}\inv {\vv G}\widehat{\Exp{-\vv f}} = \inv{[1+\widehat{\Exp{\vv f}}\vv G \widehat{\Exp{\vv f}}]}\label{inver}\\
 \intertext{ which leads us to the final form}
 H_B[A] &= \hlf \left[ \sqrt{\vv d}\dgr \frac{1}{\sigma_{\mathrm n}^2}\inv{[1+\widehat{\Exp{\vv f}}\vv G \widehat{\Exp{\vv f}}]}\sqrt \vv d +\vv f\dgr\inv {\vv F} \vv f+\log(|2\pi\inv{\vv D_{\vv f}}||2\pi \sigma_{\mathrm n}^2|^{N_\mathrm{tot}}|2\pi \vv F||2\pi \vv G||\vv d|)\right] -\log \cali P(\tau) \tag{\ref{finalwithoutg}}
\end{align}

\subsection{Discrete Fourier coefficients of a detuned signal}\label{fourderiv}
For the signal consisting of a single Fourier coefficient with frequency $\omega' = \frac{2\pi n}{T} = \frac{2\pi na}{\tau}$ the measured coefficient over a period $[K\tau -\frac \tau 2 :K\tau +\frac \tau 2]$ of pulse $K$ is readily calculated as 
\begin{align}
 \wt d_{k,K} &= \int _{K\tau -\frac \tau 2}^{K\tau +\frac \tau 2} \mathrm d t \quad s_n\Exp{i (\omega'_m - \omega_k) t} = \nonumber \\
	 &= s_n \left[\frac{\Exp{2\pi i \frac {(na -k)}{\tau} t}}{2\pi i(na -k)}\right]_{K\tau -\frac \tau 2}^{K\tau +\frac \tau 2}= \nonumber \\
	 &= s_n \left[\frac{\Exp{2\pi i k \frac {(\frac n k a -1)}{\tau} t}}{2\pi k i(\frac n k a -1)}\right]_{K\tau -\frac \tau 2}^{K\tau +\frac \tau 2}= \nonumber \\
	 &= s_n \Exp{2\pi i k (\frac n k a - 1) K}\frac{\sin \pi k (\frac n k a-1 )}{\pi k (\frac n k a -1)}\nonumber \\ 
	 &= s_n \Exp{2\pi i k (\frac n k a - 1) K} \cdot \Sinc{\pi k (\frac n k a -1 )}\tag{\ref{phaseandamp}}
\intertext{For the full spectrum of a real signal, where $s_n = s_{-n}^*$ we get}
 d_{k,K} &= \int _{K\tau -\frac \tau 2}^{K\tau +\frac \tau 2} \mathrm d t \quad \sum _{n=-\infty}^\infty s_n\Exp{i (\omega'_m - \omega_k) t} = \nonumber \\
 	 &= \left[\sum _{n=-\infty}^\infty s_n \frac{\Exp{2\pi i \frac {(na -k)}{\tau} t}}{2\pi i(na -k)}\right]_{K\tau -\frac \tau 2}^{K\tau +\frac \tau 2}= \nonumber \\
 	 &= \sum _{n=1}^\infty \left[ s_n \frac{\Exp{2\pi i \frac {(na -k)}{\tau} t}}{2\pi i(na -k)}+s_n^* \frac{\Exp{2\pi i \frac {(-na -k)}{\tau} t}}{2\pi i(-na -k)}\right]_{K\tau -\frac \tau 2}^{K\tau +\frac \tau 2}= \nonumber \\
	 &= \sum _{n=1}^\infty a_n\left[\frac{(k+na)(\cos \Phi_{s_n} + i\sin \Phi_{s_n})\Exp{2\pi i \frac {(na -k)}{\tau} t} + (k-na)(\cos \Phi_{s_n} - i\sin \Phi_{s_n})\Exp{2\pi i \frac {(-na -k)}{\tau} t}} {2\pi i (k^2 -(na)^2)}\right]_{K\tau -\frac \tau 2}^{K\tau +\frac \tau 2} =\nonumber \\
	 &= \sum _{n=1}^\infty a_n\left[\Exp{2\pi i \frac {-k}{\tau} t}\left(\cos \Phi_{s_n} \frac{-ik\cos(\frac{2\pi na}{\tau}t)+na\sin(\frac{2\pi na}{\tau}t)}{\pi(k^2-(na)^2)}+\sin \Phi_{s_n} \frac{ik\sin(\frac{2\pi na}{\tau}t)+na\cos(\frac{2\pi na}{\tau}t)}{\pi(k^2-(na)^2)}\right)\right]_{K\tau -\frac \tau 2}^{K\tau +\frac \tau 2}=\nonumber \\
	 &=\pm \sum _{n=1}^\infty a_n \left[2\sin(\pi na)\frac{\cos \Phi_{s_n} (ik\sin(2\pi na K)+na\cos(2\pi na K))+\sin \Phi_{s_n} (-na\sin(2\pi na K)+ik\cos(2\pi na K))}{\pi(k^2-(na)^2)}\right]=\nonumber \\
	 &=\pm \sum _{n=1}^\infty a_n \left[2\sin(\pi na)\frac{i\sin (2\pi na K)(\cos \Phi_{s_n} k + i\sin \Phi_{s_n} na) + \cos (2\pi na K)(\cos \Phi_{s_n} na + i\sin \Phi_{s_n} k)}{\pi(k^2-(na)^2)}\right]=\nonumber\\
	 &=\sum_{n=1}^\infty a_n\left[2\Sinc{\pi(na-k)}\frac{i\sin (2\pi na K)(\cos \Phi_{s_n} k + i\sin \Phi_{s_n} na) + \cos (2\pi na K)(\cos \Phi_{s_n} na + i\sin \Phi_{s_n} k)}{k+na}\right]=\nonumber \\
	 \intertext{where we now introduce $\ol k = \frac{an + k}{2}, d = \ol k-k$ and simplify}
	 &=\sum_{n=1}^\infty a_n\left[\Sinc{\pi(na-k)}\cdot \right. \\
	 &\cdot \left. \frac{i\sin (2\pi na K)(\cos \Phi_{s_n} (\ol k-d) + i\sin \Phi_{s_n} (\ol k+d)) + \cos (2\pi na K)(\cos \Phi_{s_n} (\ol k+d) + i\sin \Phi_{s_n} (\ol k-d))}{\ol k}\right]=\nonumber \\
	 &=\sum_{n=1}^\infty a_n \left[\Sinc{\pi(na-k)} \cdot \nonumber \right.\\
	   &\qquad \cdot \left.\frac{\ol k(i\sin(2\pi naK)+\cos(2\pi naK))(\cos \Phi_{s_n} +i \sin \Phi_{s_n})+d(-i\sin(2\pi naK)+\cos(2\pi naK))(\cos \Phi_{s_n} -i \sin \Phi_{s_n})}{\ol k}\right]=\nonumber\\
	 &=\sum_{n=1}^\infty a_n \Sinc{\pi(na-k)}\left[\underbrace{\Exp{i(2\pi naK + \Phi_{s_n})}}_I+\underbrace{\frac d {\ol k} \Exp{-i(2\pi naK + \Phi_{s_n})}}_{II}\right]=:\sum_{n=1}^\infty A_{kn} a_n \tag{\ref{ftcoeff}}
 \end{align}
where we assumed $b_0$ to be zero and used $s_n =: a_n (i \sin \Phi_{s_n} + \cos \Phi_{s_n}), a_n \mathrm{real}$.
\section{Some considerations on the evaluation of the receiver posterior pdf}\label{whytdisbad}
Having rigorously marginalized out the stationary process $g$, we seek to maximize the posterior probability density function with respect to the template parameters. When projecting this equation into Fourier space, $G$ is a diagonal matrix and the multiplication with $\hat f$ transforms to a convolution that breaks this diagonality, leading to a coupling of the Fourier coefficients. While in the diagonal case the pdf completely factorizes in separate Fourier coefficients, which are independently and quickly integrated, there is only one $N_{bin}$-dimensional integral in the exact case. For typical $N_{bin}$s of $1024$ the computational demand is simply too high. Thus we decided for imposing diagonality in Fourier space, which discards the non-diagonal terms of f leading to a solvable integral. However, this discards phase information as now $f$ has also the properties of a stationary process. Consequently we have derived an amplitude-only model. An Amplitude model alone does not contain valuable information about TOAs since the amplitude is invariant to time shifts.  Therefore, we impose a model on the phase of the signal based on wrapped gaussians and being evaluated phenomenologically. This model uses the amplitude information to estimate the phase error of an observation at hand. The phenomenological way was chosen as the exact calculations for quadrature of the stochastic signal mix the phases of different Fourier channels in a highly non-trivial fashion again leading to an unwanted coupling of phases present only at the single pulse level.\\
This ansatz tries to find a compromise between numerical feasibility and accuracy. For the case of real measurements containing additional noise and fluctuations, we expected to grasp the uncertainties left with acceptable precision and lead to acceptable $\chi^2$-values at the end. However losing the mathematical precision, we expect not to reach the accuracy of the classical timing codes when it comes to evaluating exact, nonfluctuating test data sets. Indeed our round mean square errors turned out to be $25\%$ worse in these cases while the reached $\chi^2$ was comparable accurate.\\
The question surely arises why we did not evaluate the pdf in time domain. In time domain, the main operator acting on the data is the inverse of a diagonal matrix ($\hat f$) plus a Toeplitz-matrix $G$ \footnote{a matrix whose entries depend only on the distance to the diagonal} that can be further reduced to a circular Toeplitz-Matrix by the symmetry in $\hat f$. There exist fast ways (up to $\mathcal O (N\log N)$) of calculating the operators scalar product with the data \citep{doi:10.1137/080720280} however this does not reduce the dimensionality of the integration when it comes to varying the template parameters of interest. Unfortunately, expanding the operator linearly (to develop a mean field theory for the signal field, which should lead to a very good estimate and first order corrections for the correlations present in the problem) does not help in this case, as the derivatives refer to specific entries of the inverse matrix. Calculating the inverse matrix itself scales $\mathcal O (N^3)$ or best  $\mathcal O (N^2)$ in our case, and as we need $N$ entries for the calculation of the mean field gradient, the computational demand is again unacceptable. A class of transformations on f that leave the determinant of the operator invariant, combined with a pdf on the invariants of the transformation, could make exact evaluation of the derived equations possible. In lack of such a tool we decided to approximate the problem in Fourier space.
\section{Iterative solver for fields}\label{newtoniter}
Finding the maximum a-posteriori (MAP) probability set equals finding a global maximum for $\ol {\vv f}$ and $\sigma_{\vv f}$ both at the same time. Equations \mref{newprior} and \mref{newpriorsigma} explicate optimization w.r.t. one parameter only in the context of correct prior knowledge. Furthermore, we want to find the optimal solution w.r.t. both parameters. We may resolve these issues by assuming that there is only one maximum (which might not be the case as there might be several modes of radiation in the system etc.) and rely on a suitable iteration scheme. We implemented a method of steepest ascend. The Hamiltonian may be seen as a potential to be broken up in parts dependent on $\ol f_k$ and $\sigma_k$ for every Fourier coefficient. For the sake of shortness let us call this part $\Phi(\ol{\vv f}(t), \vv \sigma(t))$. We now try to find the path of the steepest ascend, parametrized by t: $\gamma_t = (\ol {\vv f}, \vv \sigma)$. Infinitesimally, isolines may be found using $\mathrm d \Phi(\gamma_t) = 0$. Working this out for our Hamiltonian leads to
\begin{align}
 0 &= \langle \frac{(\vv f-\ol {\vv f})^2}{\sigma^3}- \frac{1}{\vv \sigma}\rangle \frac{\partial \vv \sigma}{\partial t} + \langle \frac{\vv f-\ol {\vv f}}{\vv \sigma^2}\rangle  \frac{\partial \ol {\vv f}}{\partial t}\\
\intertext{Consequently, the direction of the steepest ascend/descent follows the differential equation}0&=-\langle \frac{\vv f-\ol{\vv f}}{\sigma^2}\rangle  \frac{\partial \vv \sigma}{\partial t} + \left( \frac{\langle(\vv f-\ol {\vv f})^2\rangle}{\vv \sigma^3}- \frac{1}{\sigma}\right) \frac{\partial \ol {\vv f}}{\partial t}\label{infmet}
\end{align}
where orthogonality is easily shown. Since for most cases occurring setting $\ol {\vv f}$ according to \mref{newprior} is an acceptable method while a reasonable update procedure of $\vv \sigma$ prior to optimizing it was missing, we update $\vv \sigma$ when updating $\ol {\vv f}$ by the method, but not vice versa. Integrating \mref{infmet} is approximated by
\begin{align}
 \Delta \sigma &= \frac{\partial \vv \sigma}{\partial \ol {\vv f}} \Delta \ol {\vv f} = \left(\frac{\langle(\vv f-\ol {\vv f} )^2\rangle}{\sigma} - \vv \sigma\right) \frac{\Delta \ol {\vv f}}{\langle \vv f-\ol {\vv f} \rangle}\\
 \Rightarrow \sigma_{new} &= \frac{\langle(\vv f-\ol {\vv f}_{new} )^2\rangle + \langle(\vv f-\ol {\vv f}_{old} )^2\rangle }{2\vv \sigma_{old}}
\end{align}
and determining $\ol {\vv f}_{new}$ as in \mref{newprior}. As is evident by derivation, this method is generic for a prior of the form of a Gaussian, however could be refined by also updating the $\ol {\vv f}$ value by similar considerations.
\end{document}

%% file: hacks.tex
\newcommand{\Exp}[1]{\mathrm{exp}[#1]}
\newcommand{\Sinc}[1]{\mathrm{sinc}[#1]}
\newcommand{\mref}[1]{(\ref{#1})}
\newcommand{\dint}{\int \mathrm d}
\newcommand{\Dint}{\iint \mathrm D}

\newcommand{\smax}{_\text{max}}

\newcommand{\cali}[1]{{\cal #1}}
\newcommand{\inv}[1]{{#1^{-1}}}
\newcommand{\dgr}{^\dagger}
\newcommand{\PP}[1]{\cali P({#1})}

\newcommand{\ol}{\overline}
\newcommand{\hlf}{\frac{1}{2}}
\newcommand{\wt}{\widetilde}
\newcommand{\vv}[1]{\mathbf{#1}}
\newcommand{\wtvv}[1]{\widetilde{\mathbf{#1}}}

\definecolor{MyDarkBlue}{rgb}{0,0,0.5}
\definecolor{MyLightBlue}{rgb}{.9,.9,1}
\definecolor{MyDarkRed}{rgb}{0.5,0,0}
\definecolor{MyDarkGreen}{rgb}{0,0.5,0}

%% file: graphics/profile_exp.tex
\begingroup
  \makeatletter
  \providecommand\color[2][]{%
    \GenericError{(gnuplot) \space\space\space\@spaces}{%
      Package color not loaded in conjunction with
      terminal option `colourtext'%
    }{See the gnuplot documentation for explanation.%
    }{Either use 'blacktext' in gnuplot or load the package
      color.sty in LaTeX.}%
    \renewcommand\color[2][]{}%
  }%
  \providecommand\includegraphics[2][]{%
    \GenericError{(gnuplot) \space\space\space\@spaces}{%
      Package graphicx or graphics not loaded%
    }{See the gnuplot documentation for explanation.%
    }{The gnuplot epslatex terminal needs graphicx.sty or graphics.sty.}%
    \renewcommand\includegraphics[2][]{}%
  }%
  \providecommand\rotatebox[2]{#2}%
  \@ifundefined{ifGPcolor}{%
    \newif\ifGPcolor
    \GPcolortrue
  }{}%
  \@ifundefined{ifGPblacktext}{%
    \newif\ifGPblacktext
    \GPblacktexttrue
  }{}%
  \let\gplgaddtomacro\g@addto@macro
  \gdef\gplbacktext{}%
  \gdef\gplfronttext{}%
  \makeatother
  \ifGPblacktext
    \def\colorrgb#1{}%
    \def\colorgray#1{}%
  \else
    \ifGPcolor
      \def\colorrgb#1{\color[rgb]{#1}}%
      \def\colorgray#1{\color[gray]{#1}}%
      \expandafter\def\csname LTw\endcsname{\color{white}}%
      \expandafter\def\csname LTb\endcsname{\color{black}}%
      \expandafter\def\csname LTa\endcsname{\color{black}}%
      \expandafter\def\csname LT0\endcsname{\color[rgb]{1,0,0}}%
      \expandafter\def\csname LT1\endcsname{\color[rgb]{0,1,0}}%
      \expandafter\def\csname LT2\endcsname{\color[rgb]{0,0,1}}%
      \expandafter\def\csname LT3\endcsname{\color[rgb]{1,0,1}}%
      \expandafter\def\csname LT4\endcsname{\color[rgb]{0,1,1}}%
      \expandafter\def\csname LT5\endcsname{\color[rgb]{1,1,0}}%
      \expandafter\def\csname LT6\endcsname{\color[rgb]{0,0,0}}%
      \expandafter\def\csname LT7\endcsname{\color[rgb]{1,0.3,0}}%
      \expandafter\def\csname LT8\endcsname{\color[rgb]{0.5,0.5,0.5}}%
    \else
      \def\colorrgb#1{\color{black}}%
      \def\colorgray#1{\color[gray]{#1}}%
      \expandafter\def\csname LTw\endcsname{\color{white}}%
      \expandafter\def\csname LTb\endcsname{\color{black}}%
      \expandafter\def\csname LTa\endcsname{\color{black}}%
      \expandafter\def\csname LT0\endcsname{\color{black}}%
      \expandafter\def\csname LT1\endcsname{\color{black}}%
      \expandafter\def\csname LT2\endcsname{\color{black}}%
      \expandafter\def\csname LT3\endcsname{\color{black}}%
      \expandafter\def\csname LT4\endcsname{\color{black}}%
      \expandafter\def\csname LT5\endcsname{\color{black}}%
      \expandafter\def\csname LT6\endcsname{\color{black}}%
      \expandafter\def\csname LT7\endcsname{\color{black}}%
      \expandafter\def\csname LT8\endcsname{\color{black}}%
    \fi
  \fi
  \setlength{\unitlength}{0.0500bp}%
  \begin{picture}(5102.00,3400.00)%
    \gplgaddtomacro\gplbacktext{%
      \csname LTb\endcsname%
      \put(176,1556){\rotatebox{-270}{\makebox(0,0){\strut{}intensity in arb.u.}}}%
      \put(2550,154){\makebox(0,0){\strut{}pulsar phase}}%
      \put(2550,3069){\makebox(0,0){\strut{}Model for pulse profile}}%
    }%
    \gplgaddtomacro\gplfronttext{%
      \csname LTb\endcsname%
      \put(3718,2566){\makebox(0,0)[r]{\strut{}fluctuating/uncertain envelope}}%
      \csname LTb\endcsname%
      \put(3718,2346){\makebox(0,0)[r]{\strut{}average profile}}%
      \csname LTb\endcsname%
      \put(3718,2126){\makebox(0,0)[r]{\strut{}single pulse data}}%
    }%
    \gplbacktext
    \put(0,0){\includegraphics{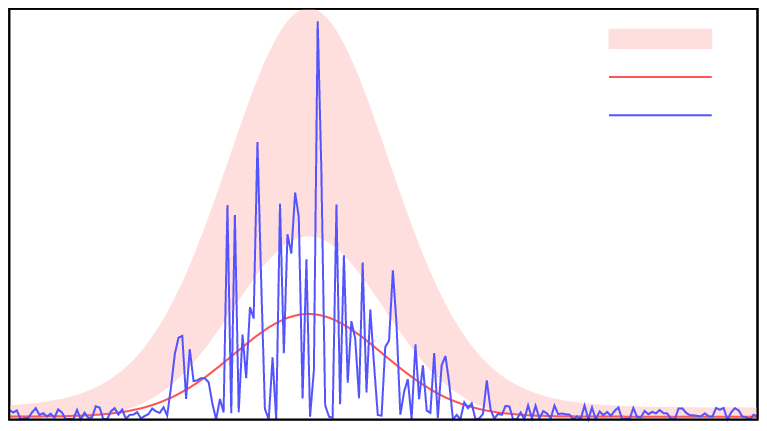}}%
    \gplfronttext
  \end{picture}%
\endgroup

%% file: graphics/fftwholee.tex
\begingroup
  \makeatletter
  \providecommand\color[2][]{%
    \GenericError{(gnuplot) \space\space\space\@spaces}{%
      Package color not loaded in conjunction with
      terminal option `colourtext'%
    }{See the gnuplot documentation for explanation.%
    }{Either use 'blacktext' in gnuplot or load the package
      color.sty in LaTeX.}%
    \renewcommand\color[2][]{}%
  }%
  \providecommand\includegraphics[2][]{%
    \GenericError{(gnuplot) \space\space\space\@spaces}{%
      Package graphicx or graphics not loaded%
    }{See the gnuplot documentation for explanation.%
    }{The gnuplot epslatex terminal needs graphicx.sty or graphics.sty.}%
    \renewcommand\includegraphics[2][]{}%
  }%
  \providecommand\rotatebox[2]{#2}%
  \@ifundefined{ifGPcolor}{%
    \newif\ifGPcolor
    \GPcolortrue
  }{}%
  \@ifundefined{ifGPblacktext}{%
    \newif\ifGPblacktext
    \GPblacktexttrue
  }{}%
  \let\gplgaddtomacro\g@addto@macro
  \gdef\gplbacktext{}%
  \gdef\gplfronttext{}%
  \makeatother
  \ifGPblacktext
    \def\colorrgb#1{}%
    \def\colorgray#1{}%
  \else
    \ifGPcolor
      \def\colorrgb#1{\color[rgb]{#1}}%
      \def\colorgray#1{\color[gray]{#1}}%
      \expandafter\def\csname LTw\endcsname{\color{white}}%
      \expandafter\def\csname LTb\endcsname{\color{black}}%
      \expandafter\def\csname LTa\endcsname{\color{black}}%
      \expandafter\def\csname LT0\endcsname{\color[rgb]{1,0,0}}%
      \expandafter\def\csname LT1\endcsname{\color[rgb]{0,1,0}}%
      \expandafter\def\csname LT2\endcsname{\color[rgb]{0,0,1}}%
      \expandafter\def\csname LT3\endcsname{\color[rgb]{1,0,1}}%
      \expandafter\def\csname LT4\endcsname{\color[rgb]{0,1,1}}%
      \expandafter\def\csname LT5\endcsname{\color[rgb]{1,1,0}}%
      \expandafter\def\csname LT6\endcsname{\color[rgb]{0,0,0}}%
      \expandafter\def\csname LT7\endcsname{\color[rgb]{1,0.3,0}}%
      \expandafter\def\csname LT8\endcsname{\color[rgb]{0.5,0.5,0.5}}%
    \else
      \def\colorrgb#1{\color{black}}%
      \def\colorgray#1{\color[gray]{#1}}%
      \expandafter\def\csname LTw\endcsname{\color{white}}%
      \expandafter\def\csname LTb\endcsname{\color{black}}%
      \expandafter\def\csname LTa\endcsname{\color{black}}%
      \expandafter\def\csname LT0\endcsname{\color{black}}%
      \expandafter\def\csname LT1\endcsname{\color{black}}%
      \expandafter\def\csname LT2\endcsname{\color{black}}%
      \expandafter\def\csname LT3\endcsname{\color{black}}%
      \expandafter\def\csname LT4\endcsname{\color{black}}%
      \expandafter\def\csname LT5\endcsname{\color{black}}%
      \expandafter\def\csname LT6\endcsname{\color{black}}%
      \expandafter\def\csname LT7\endcsname{\color{black}}%
      \expandafter\def\csname LT8\endcsname{\color{black}}%
    \fi
  \fi
  \setlength{\unitlength}{0.0500bp}%
  \begin{picture}(8502.00,3968.00)%
    \gplgaddtomacro\gplbacktext{%
      \csname LTb\endcsname%
      \put(1342,704){\makebox(0,0)[r]{\strut{} 0.1}}%
      \put(1342,1076){\makebox(0,0)[r]{\strut{} 1}}%
      \put(1342,1448){\makebox(0,0)[r]{\strut{} 10}}%
      \put(1342,1820){\makebox(0,0)[r]{\strut{} 100}}%
      \put(1342,2191){\makebox(0,0)[r]{\strut{} 1000}}%
      \put(1342,2563){\makebox(0,0)[r]{\strut{} 10000}}%
      \put(1342,2935){\makebox(0,0)[r]{\strut{} 100000}}%
      \put(1342,3307){\makebox(0,0)[r]{\strut{} 1e+06}}%
      \put(1474,484){\makebox(0,0){\strut{} 0}}%
      \put(2800,484){\makebox(0,0){\strut{} 5}}%
      \put(4126,484){\makebox(0,0){\strut{} 10}}%
      \put(5453,484){\makebox(0,0){\strut{} 15}}%
      \put(6779,484){\makebox(0,0){\strut{} 20}}%
      \put(8105,484){\makebox(0,0){\strut{} 25}}%
      \put(176,2005){\rotatebox{-270}{\makebox(0,0){\strut{}fourier amplitude in arb.u.}}}%
      \put(4789,154){\makebox(0,0){\strut{}fourier bins}}%
      \put(4789,3637){\makebox(0,0){\strut{}Fourier transforming a whole epoch}}%
    }%
    \gplgaddtomacro\gplfronttext{%
    }%
    \gplgaddtomacro\gplbacktext{%
      \csname LTb\endcsname%
      \put(6031,2182){\makebox(0,0)[r]{\strut{} 10}}%
      \put(6031,2513){\makebox(0,0)[r]{\strut{} 100}}%
      \put(6031,2843){\makebox(0,0)[r]{\strut{} 1000}}%
      \put(6031,3174){\makebox(0,0)[r]{\strut{} 10000}}%
      \put(6163,1962){\makebox(0,0){\strut{} 1.98}}%
      \put(6588,1962){\makebox(0,0){\strut{} 1.99}}%
      \put(7013,1962){\makebox(0,0){\strut{} 2}}%
      \put(7438,1962){\makebox(0,0){\strut{} 2.01}}%
      \put(7863,1962){\makebox(0,0){\strut{} 2.02}}%
    }%
    \gplgaddtomacro\gplfronttext{%
    }%
    \gplbacktext
    \put(0,0){\includegraphics{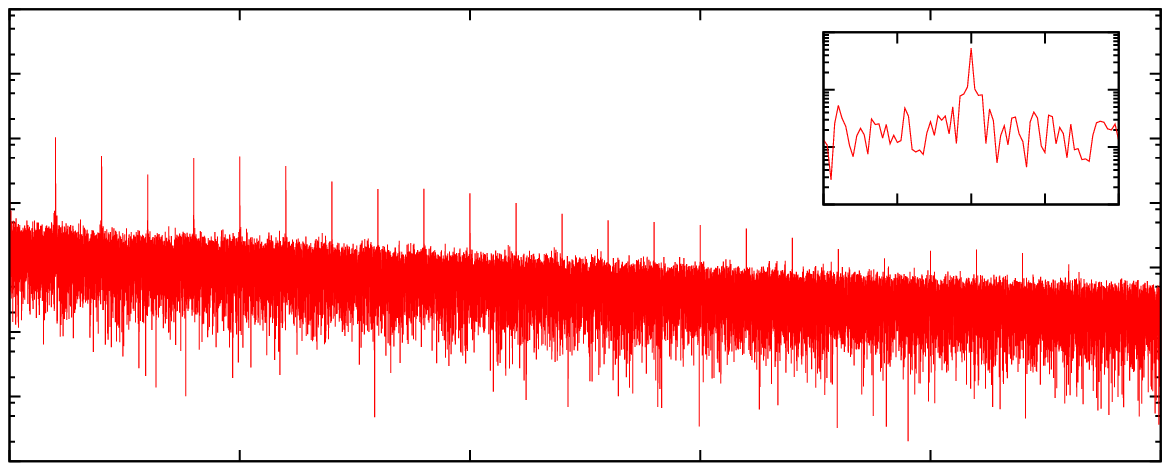}}%
    \gplfronttext
  \end{picture}%
\endgroup

%% file: graphics/receiver_error.tex
\begingroup
  \makeatletter
  \providecommand\color[2][]{%
    \GenericError{(gnuplot) \space\space\space\@spaces}{%
      Package color not loaded in conjunction with
      terminal option `colourtext'%
    }{See the gnuplot documentation for explanation.%
    }{Either use 'blacktext' in gnuplot or load the package
      color.sty in LaTeX.}%
    \renewcommand\color[2][]{}%
  }%
  \providecommand\includegraphics[2][]{%
    \GenericError{(gnuplot) \space\space\space\@spaces}{%
      Package graphicx or graphics not loaded%
    }{See the gnuplot documentation for explanation.%
    }{The gnuplot epslatex terminal needs graphicx.sty or graphics.sty.}%
    \renewcommand\includegraphics[2][]{}%
  }%
  \providecommand\rotatebox[2]{#2}%
  \@ifundefined{ifGPcolor}{%
    \newif\ifGPcolor
    \GPcolortrue
  }{}%
  \@ifundefined{ifGPblacktext}{%
    \newif\ifGPblacktext
    \GPblacktexttrue
  }{}%
  \let\gplgaddtomacro\g@addto@macro
  \gdef\gplbacktext{}%
  \gdef\gplfronttext{}%
  \makeatother
  \ifGPblacktext
    \def\colorrgb#1{}%
    \def\colorgray#1{}%
  \else
    \ifGPcolor
      \def\colorrgb#1{\color[rgb]{#1}}%
      \def\colorgray#1{\color[gray]{#1}}%
      \expandafter\def\csname LTw\endcsname{\color{white}}%
      \expandafter\def\csname LTb\endcsname{\color{black}}%
      \expandafter\def\csname LTa\endcsname{\color{black}}%
      \expandafter\def\csname LT0\endcsname{\color[rgb]{1,0,0}}%
      \expandafter\def\csname LT1\endcsname{\color[rgb]{0,1,0}}%
      \expandafter\def\csname LT2\endcsname{\color[rgb]{0,0,1}}%
      \expandafter\def\csname LT3\endcsname{\color[rgb]{1,0,1}}%
      \expandafter\def\csname LT4\endcsname{\color[rgb]{0,1,1}}%
      \expandafter\def\csname LT5\endcsname{\color[rgb]{1,1,0}}%
      \expandafter\def\csname LT6\endcsname{\color[rgb]{0,0,0}}%
      \expandafter\def\csname LT7\endcsname{\color[rgb]{1,0.3,0}}%
      \expandafter\def\csname LT8\endcsname{\color[rgb]{0.5,0.5,0.5}}%
    \else
      \def\colorrgb#1{\color{black}}%
      \def\colorgray#1{\color[gray]{#1}}%
      \expandafter\def\csname LTw\endcsname{\color{white}}%
      \expandafter\def\csname LTb\endcsname{\color{black}}%
      \expandafter\def\csname LTa\endcsname{\color{black}}%
      \expandafter\def\csname LT0\endcsname{\color{black}}%
      \expandafter\def\csname LT1\endcsname{\color{black}}%
      \expandafter\def\csname LT2\endcsname{\color{black}}%
      \expandafter\def\csname LT3\endcsname{\color{black}}%
      \expandafter\def\csname LT4\endcsname{\color{black}}%
      \expandafter\def\csname LT5\endcsname{\color{black}}%
      \expandafter\def\csname LT6\endcsname{\color{black}}%
      \expandafter\def\csname LT7\endcsname{\color{black}}%
      \expandafter\def\csname LT8\endcsname{\color{black}}%
    \fi
  \fi
  \setlength{\unitlength}{0.0500bp}%
  \begin{picture}(5102.00,3400.00)%
    \gplgaddtomacro\gplbacktext{%
      \csname LTb\endcsname%
      \put(814,704){\makebox(0,0)[r]{\strut{} 0}}%
      \put(814,1111){\makebox(0,0)[r]{\strut{} 2}}%
      \put(814,1518){\makebox(0,0)[r]{\strut{} 4}}%
      \put(814,1925){\makebox(0,0)[r]{\strut{} 6}}%
      \put(814,2332){\makebox(0,0)[r]{\strut{} 8}}%
      \put(814,2739){\makebox(0,0)[r]{\strut{} 10}}%
      \put(946,484){\makebox(0,0){\strut{} 0}}%
      \put(1698,484){\makebox(0,0){\strut{} 2}}%
      \put(2450,484){\makebox(0,0){\strut{} 4}}%
      \put(3201,484){\makebox(0,0){\strut{} 6}}%
      \put(3953,484){\makebox(0,0){\strut{} 8}}%
      \put(4705,484){\makebox(0,0){\strut{} 10}}%
      \put(176,1721){\rotatebox{-270}{\makebox(0,0){\strut{}error in nat.u. resp. $\sigma_n$}}}%
      \put(2825,154){\makebox(0,0){\strut{}signal in $\sigma_n$}}%
      \put(2825,3069){\makebox(0,0){\strut{}Error estimates on single measurement}}%
    }%
    \gplgaddtomacro\gplfronttext{%
      \csname LTb\endcsname%
      \put(3718,2566){\makebox(0,0)[r]{\strut{}absolute error}}%
      \csname LTb\endcsname%
      \put(3718,2346){\makebox(0,0)[r]{\strut{}relative error}}%
    }%
    \gplbacktext
    \put(0,0){\includegraphics{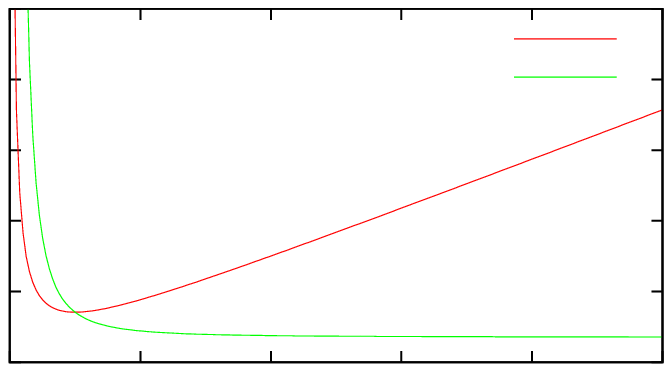}}%
    \gplfronttext
  \end{picture}%
\endgroup

%% file: graphics/toascheme.tex
\begin{tikzpicture}[scale=.9]
\node at (-2.5,9.6) {\bf A};
\node at (-2.5,6) {\bf B};
\node at (-2.5,4) {\bf C};
\node at (-2.5,0) {\bf D};
\node at (-2.5,-2) {\bf E};

\node  at (0,10) {\includegraphics[height=3cm]{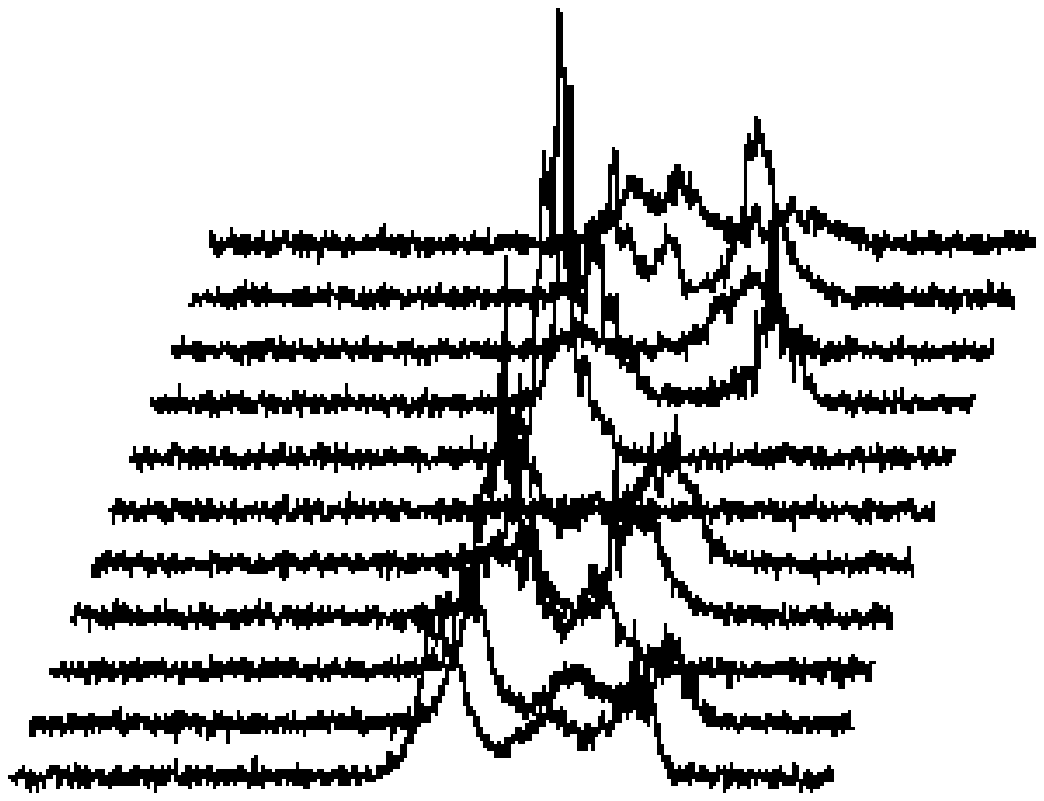}};
\node  at (5,9.85) {\includegraphics[height=2.8cm]{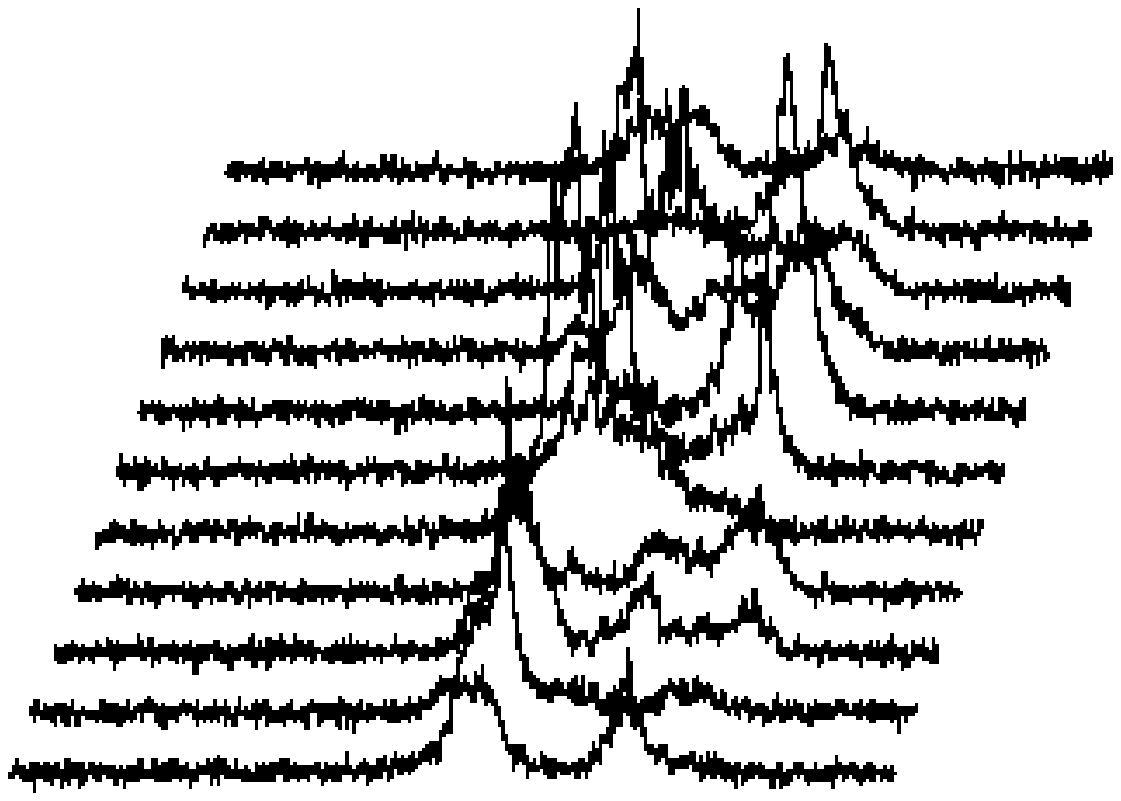}};
\node  at (10,10) [opacity=.5] {\includegraphics[height=3cm]{graphs/randompulses}};
\node  at (15,9.85) [opacity=.25] {\includegraphics[height=2.8cm]{graphs/randompulses2}};

\draw [->] [line width=.5] (0,8) -- (0,7);
\draw [->] [line width=.5] (5,8) -- (5,7);
\draw [->] [line width=.5] (10,8) -- (10,7);
\draw [->] [line width=.5] (15,8) -- (15,7);

\node at (0,6) [rectangle,draw] {$T_1 = {\{\ol f^{(1)},\sigma_f^{(1)},\ol \Phi^{(1)}, \sigma_\Phi^{(1)}}\}$};
\node at (5,6) [rectangle,draw] {$T_2$};
\node at (10,6) [rectangle,draw] {$T_3$};
\node at (15,6) [rectangle,draw] {$T_n$};

\draw [->] [line width=.5] (0,5.5) -- (0,5) -- (9.7,5)-- (9.7,4.5);
\draw [->] [line width=.5] (5,5.5) -- (5,5.1) -- (9.8,5.1)-- (9.8,4.5);
\draw [->] [line width=.5] (10,5.5) -- (10,5.1) -- (9.9,5.1)-- (9.9,4.5);
\draw [->] [line width=.5] (15,5.5) -- (15,5) -- (10,5)-- (10,4.5);

\node at (7.5,4) [rectangle,draw] {$\PP{<\Phi_d^{(i)}>-<\Phi_d^{(ref)}>=\Delta\Phi^{(i)}|T_1...T_n}$};

\definecolor{strokecol}{rgb}{0.0,0.5,0.0};
\pgfsetstrokecolor{strokecol}

\draw [->] [line width=.5] (0,.9) -- (0,3) -- (7.5,3)-- (7.5,3.5);
\draw [->] [line width=.5] (5,.75) -- (5,2.9) -- (5.65,2.9)-- (5.65,3.5);
\draw [->] [line width=.5] (10,.75) -- (10,2.8) -- (5.75,2.8)-- (5.75,3.5);
\draw [->] [line width=.5] (15,.75) -- (15,2.9) -- (5.85,2.9)-- (5.85,3.5);

\definecolor{strokecol}{rgb}{0.0,0.0,0.0};
\pgfsetstrokecolor{strokecol}

\node at (0,0)  [rectangle,draw] {\includegraphics[height=1.5cm]{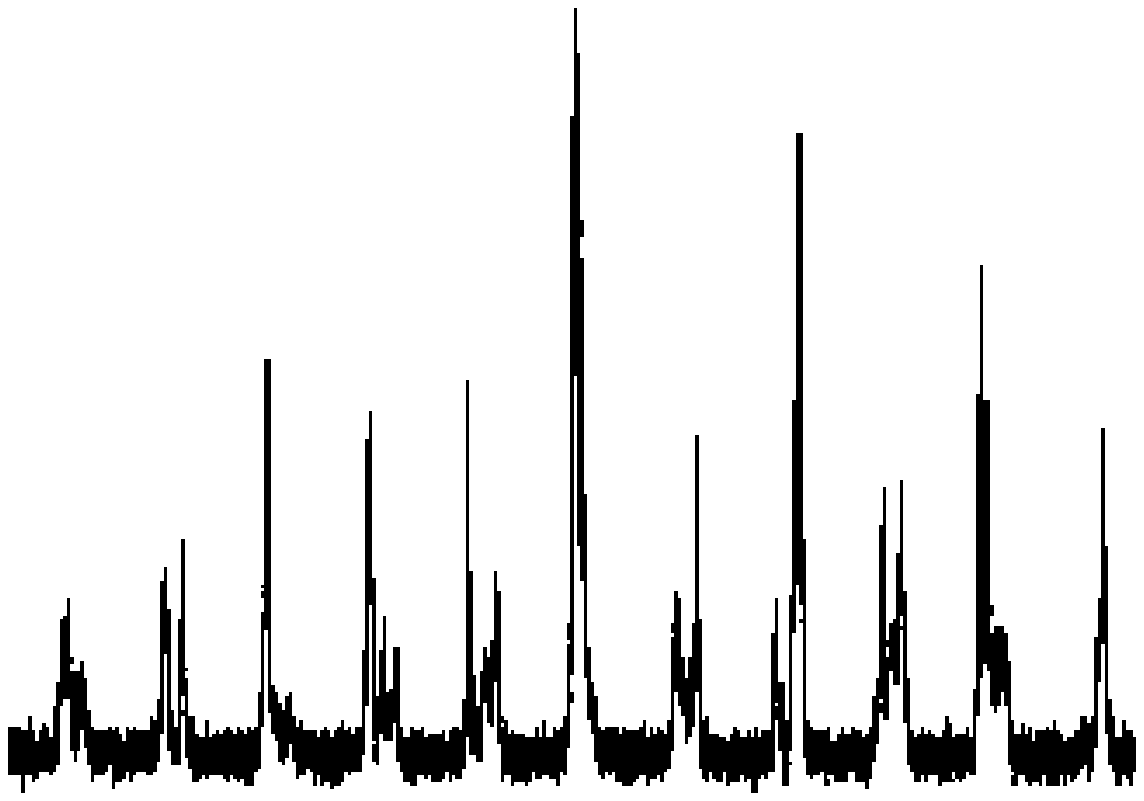}};
\node at (5,0) {\includegraphics[height=1.5cm]{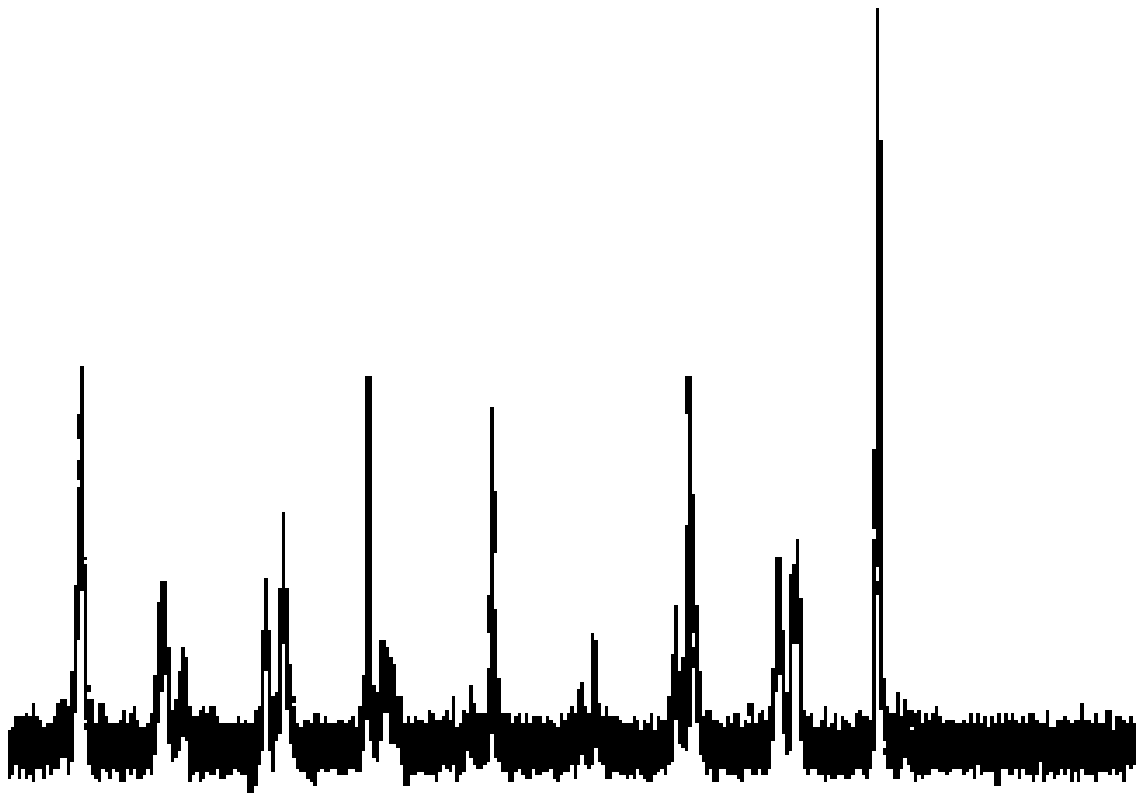}};
\node at (10,0) {\includegraphics[height=1.5cm]{graphs/spiketrain}};
\node at (15,0) {\includegraphics[height=1.5cm]{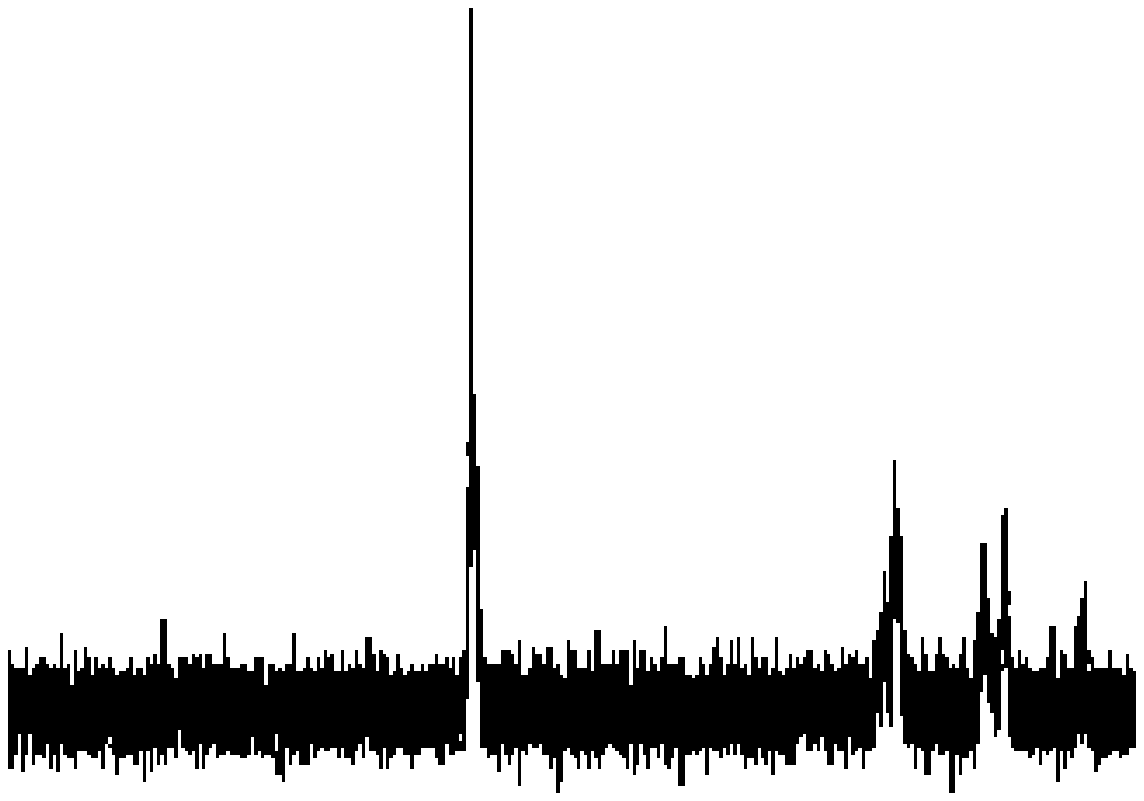}};

\draw [->] [line width=1] (-2,-2) -- (16,-2) node [below] {ToA};
\draw [fill=black] (-1,-1.9) rectangle node [below] {reference epoch} (3,-2.1);
\draw [fill=black] (5,-1.9) rectangle node [below] {epoch 2} (7,-2.1);
\draw [fill=black] (8,-1.9) rectangle node [below] {epoch 3} (11,-2.1);
\draw [fill=black] (13,-1.9)  rectangle node [below] {epoch n}  (14,-2.1);

\definecolor{strokecol}{rgb}{0.0,0.5,0.0};
\pgfsetstrokecolor{strokecol}

\draw  [line width=.5] (1,-2) -- (1,-1) -- (0,-1)-- (0,-.5);
\draw  [line width=.5] (6,-2) -- (6,-1) -- (5,-1)-- (5,-.5);
\draw  [line width=.5] (9.5,-2) -- (9.5,-1) -- (10,-1)-- (10,-.5);
\draw  [line width=.5] (13.5,-2) -- (13.5,-1) -- (15,-1)-- (15,-.5);

\definecolor{strokecol}{rgb}{0.0,0.0,.5};
\pgfsetstrokecolor{strokecol}

\draw  [->] (8.7,3.5) -- (8.7,-.75) -- (6.7,-.75)--(6.7,-1.25);
\draw  [->] (8.8,3.5) -- (8.8,-.85) -- (10.2,-.85)--(10.2,-1.25);
\draw  [->] (8.9,3.5) -- (8.9,-.75) -- (14,-.75)--(14,-1.25);

\draw  [<->] (5.75,-1.5) -- (6.25,-1.5) node [right] {$\Delta\Phi^{(2)}$};
\draw  [<->] (9.25,-1.5) -- (9.75,-1.5) node [right] {$\Delta\Phi^{(3)}$};
\draw  [<->] (13.25,-1.5) -- (13.75,-1.5) node [right] {$\Delta\Phi^{(n)}$};

\end{tikzpicture}

%% file: graphics/p_fit10.tex
\begingroup
  \makeatletter
  \providecommand\color[2][]{%
    \GenericError{(gnuplot) \space\space\space\@spaces}{%
      Package color not loaded in conjunction with
      terminal option `colourtext'%
    }{See the gnuplot documentation for explanation.%
    }{Either use 'blacktext' in gnuplot or load the package
      color.sty in LaTeX.}%
    \renewcommand\color[2][]{}%
  }%
  \providecommand\includegraphics[2][]{%
    \GenericError{(gnuplot) \space\space\space\@spaces}{%
      Package graphicx or graphics not loaded%
    }{See the gnuplot documentation for explanation.%
    }{The gnuplot epslatex terminal needs graphicx.sty or graphics.sty.}%
    \renewcommand\includegraphics[2][]{}%
  }%
  \providecommand\rotatebox[2]{#2}%
  \@ifundefined{ifGPcolor}{%
    \newif\ifGPcolor
    \GPcolortrue
  }{}%
  \@ifundefined{ifGPblacktext}{%
    \newif\ifGPblacktext
    \GPblacktexttrue
  }{}%
  \let\gplgaddtomacro\g@addto@macro
  \gdef\gplbacktext{}%
  \gdef\gplfronttext{}%
  \makeatother
  \ifGPblacktext
    \def\colorrgb#1{}%
    \def\colorgray#1{}%
  \else
    \ifGPcolor
      \def\colorrgb#1{\color[rgb]{#1}}%
      \def\colorgray#1{\color[gray]{#1}}%
      \expandafter\def\csname LTw\endcsname{\color{white}}%
      \expandafter\def\csname LTb\endcsname{\color{black}}%
      \expandafter\def\csname LTa\endcsname{\color{black}}%
      \expandafter\def\csname LT0\endcsname{\color[rgb]{1,0,0}}%
      \expandafter\def\csname LT1\endcsname{\color[rgb]{0,1,0}}%
      \expandafter\def\csname LT2\endcsname{\color[rgb]{0,0,1}}%
      \expandafter\def\csname LT3\endcsname{\color[rgb]{1,0,1}}%
      \expandafter\def\csname LT4\endcsname{\color[rgb]{0,1,1}}%
      \expandafter\def\csname LT5\endcsname{\color[rgb]{1,1,0}}%
      \expandafter\def\csname LT6\endcsname{\color[rgb]{0,0,0}}%
      \expandafter\def\csname LT7\endcsname{\color[rgb]{1,0.3,0}}%
      \expandafter\def\csname LT8\endcsname{\color[rgb]{0.5,0.5,0.5}}%
    \else
      \def\colorrgb#1{\color{black}}%
      \def\colorgray#1{\color[gray]{#1}}%
      \expandafter\def\csname LTw\endcsname{\color{white}}%
      \expandafter\def\csname LTb\endcsname{\color{black}}%
      \expandafter\def\csname LTa\endcsname{\color{black}}%
      \expandafter\def\csname LT0\endcsname{\color{black}}%
      \expandafter\def\csname LT1\endcsname{\color{black}}%
      \expandafter\def\csname LT2\endcsname{\color{black}}%
      \expandafter\def\csname LT3\endcsname{\color{black}}%
      \expandafter\def\csname LT4\endcsname{\color{black}}%
      \expandafter\def\csname LT5\endcsname{\color{black}}%
      \expandafter\def\csname LT6\endcsname{\color{black}}%
      \expandafter\def\csname LT7\endcsname{\color{black}}%
      \expandafter\def\csname LT8\endcsname{\color{black}}%
    \fi
  \fi
  \setlength{\unitlength}{0.0500bp}%
  \begin{picture}(6802.00,4250.00)%
    \gplgaddtomacro\gplbacktext{%
      \csname LTb\endcsname%
      \put(0,1528){\makebox(0,0)[r]{\strut{}-1}}%
      \put(0,2035){\makebox(0,0)[r]{\strut{}-0.5}}%
      \put(0,2542){\makebox(0,0)[r]{\strut{} 0}}%
      \put(0,3049){\makebox(0,0)[r]{\strut{} 0.5}}%
      \put(0,3556){\makebox(0,0)[r]{\strut{} 1}}%
      \put(6888,2542){\rotatebox{-270}{\makebox(0,0){\strut{}$log(S/N)$}}}%
    }%
    \gplgaddtomacro\gplfronttext{%
      \csname LTb\endcsname%
      \put(5814,3699){\makebox(0,0)[r]{\strut{}Bayesian estimated average}}%
      \csname LTb\endcsname%
      \put(5814,3479){\makebox(0,0)[r]{\strut{}Classical average}}%
      \csname LTb\endcsname%
      \put(5814,3259){\makebox(0,0)[r]{\strut{}Original signal shape}}%
    }%
    \gplgaddtomacro\gplbacktext{%
      \csname LTb\endcsname%
      \put(0,440){\makebox(0,0)[r]{\strut{} 0.001}}%
      \put(0,718){\makebox(0,0)[r]{\strut{} 0.01}}%
      \put(0,997){\makebox(0,0)[r]{\strut{} 0.1}}%
      \put(0,1275){\makebox(0,0)[r]{\strut{} 1}}%
      \put(132,220){\makebox(0,0){\strut{} 1}}%
      \put(2545,220){\makebox(0,0){\strut{} 10}}%
      \put(4958,220){\makebox(0,0){\strut{} 100}}%
      \put(6888,857){\rotatebox{-270}{\makebox(0,0){\strut{}absolute error}}}%
      \put(3400,-110){\makebox(0,0){\strut{}Fourier coefficient}}%
    }%
    \gplgaddtomacro\gplfronttext{%
      \csname LTb\endcsname%
      \put(5814,550){\makebox(0,0)[r]{\strut{}$\sigma_f$}}%
    }%
    \gplbacktext
    \put(0,0){\includegraphics{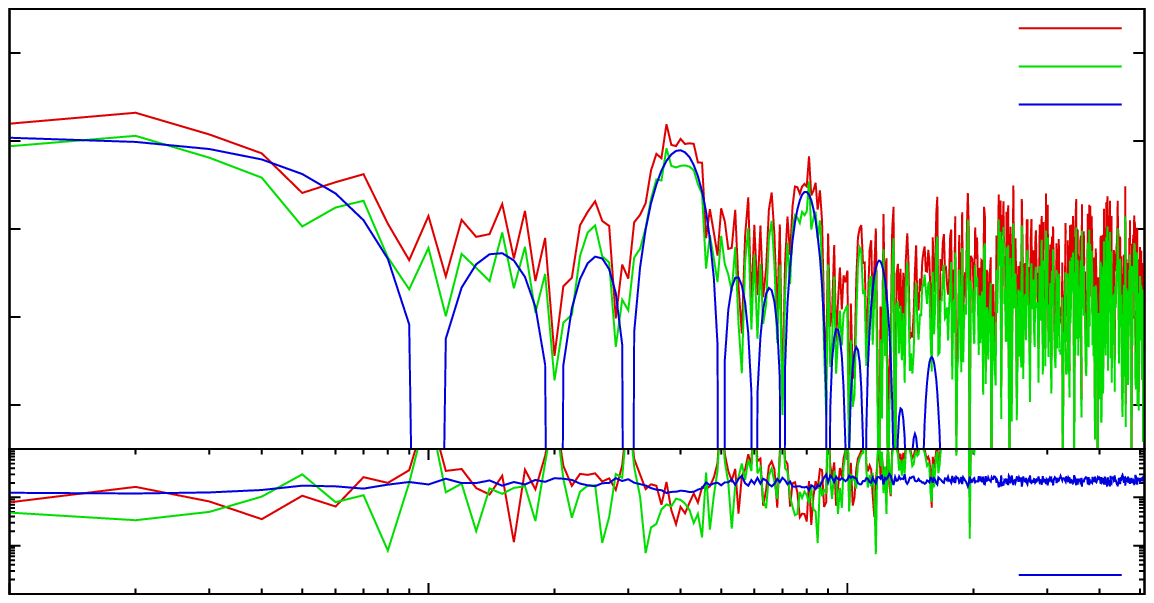}}%
    \gplfronttext
  \end{picture}%
\endgroup

%% file: graphics/p_fit100.tex
\begingroup
  \makeatletter
  \providecommand\color[2][]{%
    \GenericError{(gnuplot) \space\space\space\@spaces}{%
      Package color not loaded in conjunction with
      terminal option `colourtext'%
    }{See the gnuplot documentation for explanation.%
    }{Either use 'blacktext' in gnuplot or load the package
      color.sty in LaTeX.}%
    \renewcommand\color[2][]{}%
  }%
  \providecommand\includegraphics[2][]{%
    \GenericError{(gnuplot) \space\space\space\@spaces}{%
      Package graphicx or graphics not loaded%
    }{See the gnuplot documentation for explanation.%
    }{The gnuplot epslatex terminal needs graphicx.sty or graphics.sty.}%
    \renewcommand\includegraphics[2][]{}%
  }%
  \providecommand\rotatebox[2]{#2}%
  \@ifundefined{ifGPcolor}{%
    \newif\ifGPcolor
    \GPcolortrue
  }{}%
  \@ifundefined{ifGPblacktext}{%
    \newif\ifGPblacktext
    \GPblacktexttrue
  }{}%
  \let\gplgaddtomacro\g@addto@macro
  \gdef\gplbacktext{}%
  \gdef\gplfronttext{}%
  \makeatother
  \ifGPblacktext
    \def\colorrgb#1{}%
    \def\colorgray#1{}%
  \else
    \ifGPcolor
      \def\colorrgb#1{\color[rgb]{#1}}%
      \def\colorgray#1{\color[gray]{#1}}%
      \expandafter\def\csname LTw\endcsname{\color{white}}%
      \expandafter\def\csname LTb\endcsname{\color{black}}%
      \expandafter\def\csname LTa\endcsname{\color{black}}%
      \expandafter\def\csname LT0\endcsname{\color[rgb]{1,0,0}}%
      \expandafter\def\csname LT1\endcsname{\color[rgb]{0,1,0}}%
      \expandafter\def\csname LT2\endcsname{\color[rgb]{0,0,1}}%
      \expandafter\def\csname LT3\endcsname{\color[rgb]{1,0,1}}%
      \expandafter\def\csname LT4\endcsname{\color[rgb]{0,1,1}}%
      \expandafter\def\csname LT5\endcsname{\color[rgb]{1,1,0}}%
      \expandafter\def\csname LT6\endcsname{\color[rgb]{0,0,0}}%
      \expandafter\def\csname LT7\endcsname{\color[rgb]{1,0.3,0}}%
      \expandafter\def\csname LT8\endcsname{\color[rgb]{0.5,0.5,0.5}}%
    \else
      \def\colorrgb#1{\color{black}}%
      \def\colorgray#1{\color[gray]{#1}}%
      \expandafter\def\csname LTw\endcsname{\color{white}}%
      \expandafter\def\csname LTb\endcsname{\color{black}}%
      \expandafter\def\csname LTa\endcsname{\color{black}}%
      \expandafter\def\csname LT0\endcsname{\color{black}}%
      \expandafter\def\csname LT1\endcsname{\color{black}}%
      \expandafter\def\csname LT2\endcsname{\color{black}}%
      \expandafter\def\csname LT3\endcsname{\color{black}}%
      \expandafter\def\csname LT4\endcsname{\color{black}}%
      \expandafter\def\csname LT5\endcsname{\color{black}}%
      \expandafter\def\csname LT6\endcsname{\color{black}}%
      \expandafter\def\csname LT7\endcsname{\color{black}}%
      \expandafter\def\csname LT8\endcsname{\color{black}}%
    \fi
  \fi
  \setlength{\unitlength}{0.0500bp}%
  \begin{picture}(6802.00,4250.00)%
    \gplgaddtomacro\gplbacktext{%
      \csname LTb\endcsname%
      \put(0,1528){\makebox(0,0)[r]{\strut{}-1}}%
      \put(0,2035){\makebox(0,0)[r]{\strut{}-0.5}}%
      \put(0,2542){\makebox(0,0)[r]{\strut{} 0}}%
      \put(0,3049){\makebox(0,0)[r]{\strut{} 0.5}}%
      \put(0,3556){\makebox(0,0)[r]{\strut{} 1}}%
      \put(6888,2542){\rotatebox{-270}{\makebox(0,0){\strut{}$log(S/N)$}}}%
    }%
    \gplgaddtomacro\gplfronttext{%
      \csname LTb\endcsname%
      \put(5814,3699){\makebox(0,0)[r]{\strut{}Bayesian estimated average}}%
      \csname LTb\endcsname%
      \put(5814,3479){\makebox(0,0)[r]{\strut{}Classical average}}%
      \csname LTb\endcsname%
      \put(5814,3259){\makebox(0,0)[r]{\strut{}Original signal shape}}%
    }%
    \gplgaddtomacro\gplbacktext{%
      \csname LTb\endcsname%
      \put(0,440){\makebox(0,0)[r]{\strut{} 0.001}}%
      \put(0,718){\makebox(0,0)[r]{\strut{} 0.01}}%
      \put(0,997){\makebox(0,0)[r]{\strut{} 0.1}}%
      \put(0,1275){\makebox(0,0)[r]{\strut{} 1}}%
      \put(132,220){\makebox(0,0){\strut{} 1}}%
      \put(2545,220){\makebox(0,0){\strut{} 10}}%
      \put(4958,220){\makebox(0,0){\strut{} 100}}%
      \put(6888,857){\rotatebox{-270}{\makebox(0,0){\strut{}absolute error}}}%
      \put(3400,-110){\makebox(0,0){\strut{}Fourier coefficient}}%
    }%
    \gplgaddtomacro\gplfronttext{%
      \csname LTb\endcsname%
      \put(5814,550){\makebox(0,0)[r]{\strut{}$\sigma_f$}}%
    }%
    \gplbacktext
    \put(0,0){\includegraphics{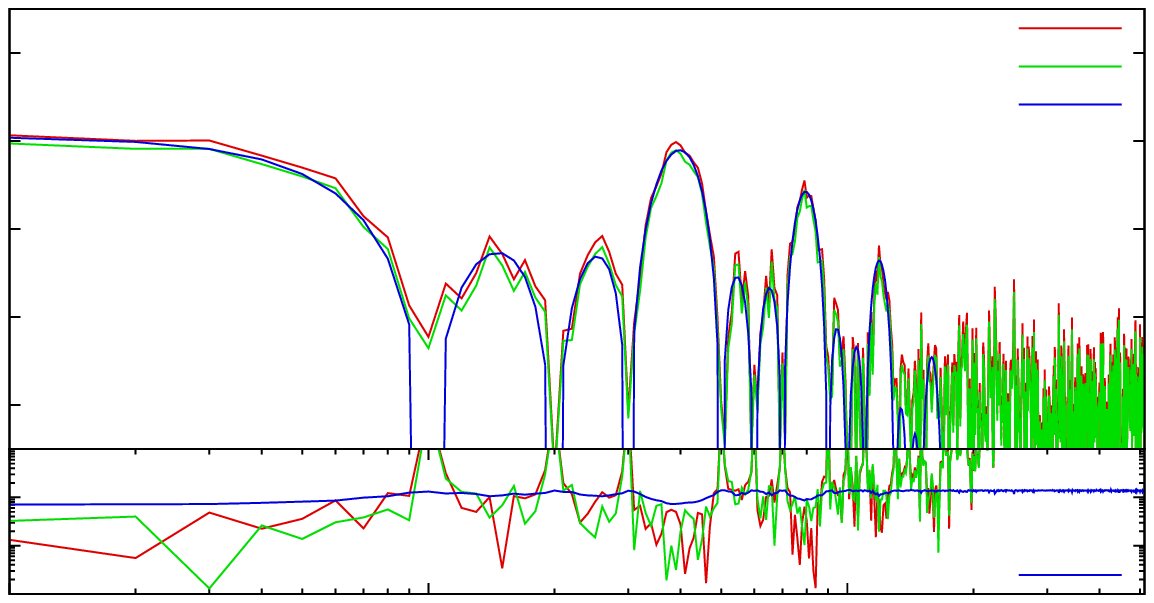}}%
    \gplfronttext
  \end{picture}%
\endgroup

%% file: graphics/p_fit1000.tex
\begingroup
  \makeatletter
  \providecommand\color[2][]{%
    \GenericError{(gnuplot) \space\space\space\@spaces}{%
      Package color not loaded in conjunction with
      terminal option `colourtext'%
    }{See the gnuplot documentation for explanation.%
    }{Either use 'blacktext' in gnuplot or load the package
      color.sty in LaTeX.}%
    \renewcommand\color[2][]{}%
  }%
  \providecommand\includegraphics[2][]{%
    \GenericError{(gnuplot) \space\space\space\@spaces}{%
      Package graphicx or graphics not loaded%
    }{See the gnuplot documentation for explanation.%
    }{The gnuplot epslatex terminal needs graphicx.sty or graphics.sty.}%
    \renewcommand\includegraphics[2][]{}%
  }%
  \providecommand\rotatebox[2]{#2}%
  \@ifundefined{ifGPcolor}{%
    \newif\ifGPcolor
    \GPcolortrue
  }{}%
  \@ifundefined{ifGPblacktext}{%
    \newif\ifGPblacktext
    \GPblacktexttrue
  }{}%
  \let\gplgaddtomacro\g@addto@macro
  \gdef\gplbacktext{}%
  \gdef\gplfronttext{}%
  \makeatother
  \ifGPblacktext
    \def\colorrgb#1{}%
    \def\colorgray#1{}%
  \else
    \ifGPcolor
      \def\colorrgb#1{\color[rgb]{#1}}%
      \def\colorgray#1{\color[gray]{#1}}%
      \expandafter\def\csname LTw\endcsname{\color{white}}%
      \expandafter\def\csname LTb\endcsname{\color{black}}%
      \expandafter\def\csname LTa\endcsname{\color{black}}%
      \expandafter\def\csname LT0\endcsname{\color[rgb]{1,0,0}}%
      \expandafter\def\csname LT1\endcsname{\color[rgb]{0,1,0}}%
      \expandafter\def\csname LT2\endcsname{\color[rgb]{0,0,1}}%
      \expandafter\def\csname LT3\endcsname{\color[rgb]{1,0,1}}%
      \expandafter\def\csname LT4\endcsname{\color[rgb]{0,1,1}}%
      \expandafter\def\csname LT5\endcsname{\color[rgb]{1,1,0}}%
      \expandafter\def\csname LT6\endcsname{\color[rgb]{0,0,0}}%
      \expandafter\def\csname LT7\endcsname{\color[rgb]{1,0.3,0}}%
      \expandafter\def\csname LT8\endcsname{\color[rgb]{0.5,0.5,0.5}}%
    \else
      \def\colorrgb#1{\color{black}}%
      \def\colorgray#1{\color[gray]{#1}}%
      \expandafter\def\csname LTw\endcsname{\color{white}}%
      \expandafter\def\csname LTb\endcsname{\color{black}}%
      \expandafter\def\csname LTa\endcsname{\color{black}}%
      \expandafter\def\csname LT0\endcsname{\color{black}}%
      \expandafter\def\csname LT1\endcsname{\color{black}}%
      \expandafter\def\csname LT2\endcsname{\color{black}}%
      \expandafter\def\csname LT3\endcsname{\color{black}}%
      \expandafter\def\csname LT4\endcsname{\color{black}}%
      \expandafter\def\csname LT5\endcsname{\color{black}}%
      \expandafter\def\csname LT6\endcsname{\color{black}}%
      \expandafter\def\csname LT7\endcsname{\color{black}}%
      \expandafter\def\csname LT8\endcsname{\color{black}}%
    \fi
  \fi
  \setlength{\unitlength}{0.0500bp}%
  \begin{picture}(6802.00,4250.00)%
    \gplgaddtomacro\gplbacktext{%
      \csname LTb\endcsname%
      \put(0,1528){\makebox(0,0)[r]{\strut{}-1}}%
      \put(0,2035){\makebox(0,0)[r]{\strut{}-0.5}}%
      \put(0,2542){\makebox(0,0)[r]{\strut{} 0}}%
      \put(0,3049){\makebox(0,0)[r]{\strut{} 0.5}}%
      \put(0,3556){\makebox(0,0)[r]{\strut{} 1}}%
      \put(6888,2542){\rotatebox{-270}{\makebox(0,0){\strut{}$log(S/N)$}}}%
    }%
    \gplgaddtomacro\gplfronttext{%
      \csname LTb\endcsname%
      \put(5814,3699){\makebox(0,0)[r]{\strut{}Bayesian estimated average}}%
      \csname LTb\endcsname%
      \put(5814,3479){\makebox(0,0)[r]{\strut{}Classical average}}%
      \csname LTb\endcsname%
      \put(5814,3259){\makebox(0,0)[r]{\strut{}Original signal shape}}%
    }%
    \gplgaddtomacro\gplbacktext{%
      \csname LTb\endcsname%
      \put(0,440){\makebox(0,0)[r]{\strut{} 0.001}}%
      \put(0,718){\makebox(0,0)[r]{\strut{} 0.01}}%
      \put(0,997){\makebox(0,0)[r]{\strut{} 0.1}}%
      \put(0,1275){\makebox(0,0)[r]{\strut{} 1}}%
      \put(132,220){\makebox(0,0){\strut{} 1}}%
      \put(2545,220){\makebox(0,0){\strut{} 10}}%
      \put(4958,220){\makebox(0,0){\strut{} 100}}%
      \put(6888,857){\rotatebox{-270}{\makebox(0,0){\strut{}absolute error}}}%
      \put(3400,-110){\makebox(0,0){\strut{}Fourier coefficient}}%
    }%
    \gplgaddtomacro\gplfronttext{%
      \csname LTb\endcsname%
      \put(5814,550){\makebox(0,0)[r]{\strut{}$\sigma_f$}}%
    }%
    \gplbacktext
    \put(0,0){\includegraphics{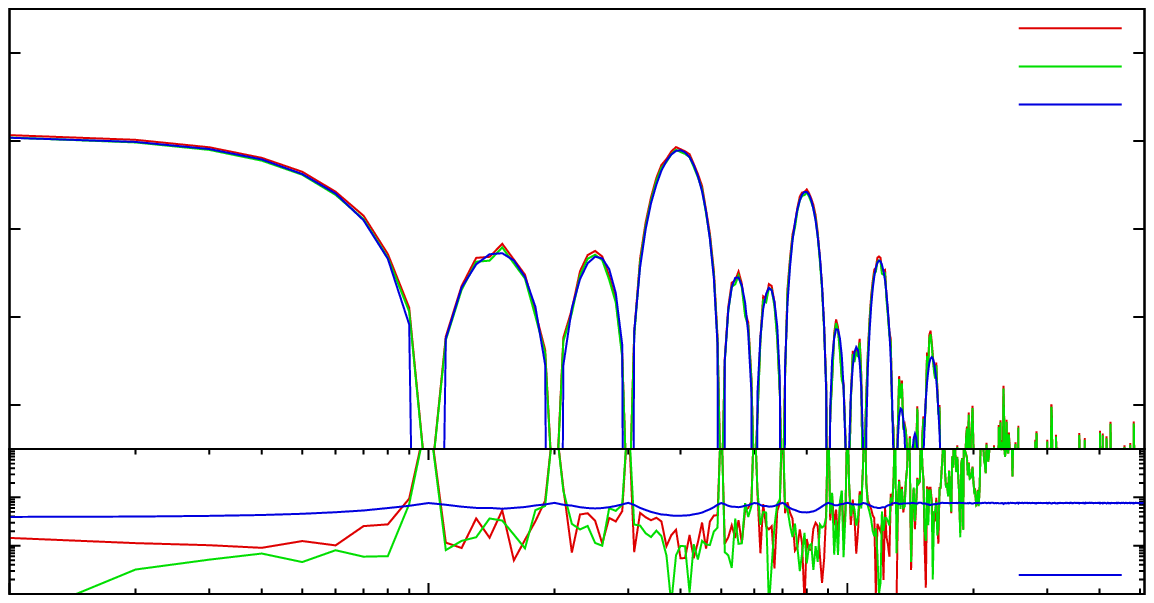}}%
    \gplfronttext
  \end{picture}%
\endgroup

%% file: graphics/s_probs_grob.tex
\begingroup
  \makeatletter
  \providecommand\color[2][]{%
    \GenericError{(gnuplot) \space\space\space\@spaces}{%
      Package color not loaded in conjunction with
      terminal option `colourtext'%
    }{See the gnuplot documentation for explanation.%
    }{Either use 'blacktext' in gnuplot or load the package
      color.sty in LaTeX.}%
    \renewcommand\color[2][]{}%
  }%
  \providecommand\includegraphics[2][]{%
    \GenericError{(gnuplot) \space\space\space\@spaces}{%
      Package graphicx or graphics not loaded%
    }{See the gnuplot documentation for explanation.%
    }{The gnuplot epslatex terminal needs graphicx.sty or graphics.sty.}%
    \renewcommand\includegraphics[2][]{}%
  }%
  \providecommand\rotatebox[2]{#2}%
  \@ifundefined{ifGPcolor}{%
    \newif\ifGPcolor
    \GPcolortrue
  }{}%
  \@ifundefined{ifGPblacktext}{%
    \newif\ifGPblacktext
    \GPblacktexttrue
  }{}%
  \let\gplgaddtomacro\g@addto@macro
  \gdef\gplbacktext{}%
  \gdef\gplfronttext{}%
  \makeatother
  \ifGPblacktext
    \def\colorrgb#1{}%
    \def\colorgray#1{}%
  \else
    \ifGPcolor
      \def\colorrgb#1{\color[rgb]{#1}}%
      \def\colorgray#1{\color[gray]{#1}}%
      \expandafter\def\csname LTw\endcsname{\color{white}}%
      \expandafter\def\csname LTb\endcsname{\color{black}}%
      \expandafter\def\csname LTa\endcsname{\color{black}}%
      \expandafter\def\csname LT0\endcsname{\color[rgb]{1,0,0}}%
      \expandafter\def\csname LT1\endcsname{\color[rgb]{0,1,0}}%
      \expandafter\def\csname LT2\endcsname{\color[rgb]{0,0,1}}%
      \expandafter\def\csname LT3\endcsname{\color[rgb]{1,0,1}}%
      \expandafter\def\csname LT4\endcsname{\color[rgb]{0,1,1}}%
      \expandafter\def\csname LT5\endcsname{\color[rgb]{1,1,0}}%
      \expandafter\def\csname LT6\endcsname{\color[rgb]{0,0,0}}%
      \expandafter\def\csname LT7\endcsname{\color[rgb]{1,0.3,0}}%
      \expandafter\def\csname LT8\endcsname{\color[rgb]{0.5,0.5,0.5}}%
    \else
      \def\colorrgb#1{\color{black}}%
      \def\colorgray#1{\color[gray]{#1}}%
      \expandafter\def\csname LTw\endcsname{\color{white}}%
      \expandafter\def\csname LTb\endcsname{\color{black}}%
      \expandafter\def\csname LTa\endcsname{\color{black}}%
      \expandafter\def\csname LT0\endcsname{\color{black}}%
      \expandafter\def\csname LT1\endcsname{\color{black}}%
      \expandafter\def\csname LT2\endcsname{\color{black}}%
      \expandafter\def\csname LT3\endcsname{\color{black}}%
      \expandafter\def\csname LT4\endcsname{\color{black}}%
      \expandafter\def\csname LT5\endcsname{\color{black}}%
      \expandafter\def\csname LT6\endcsname{\color{black}}%
      \expandafter\def\csname LT7\endcsname{\color{black}}%
      \expandafter\def\csname LT8\endcsname{\color{black}}%
    \fi
  \fi
  \setlength{\unitlength}{0.0500bp}%
  \begin{picture}(4534.00,6802.00)%
    \gplgaddtomacro\gplbacktext{%
      \csname LTb\endcsname%
      \put(171,3841){\makebox(0,0)[r]{\strut{}-120}}%
      \put(171,4290){\makebox(0,0)[r]{\strut{}-100}}%
      \put(171,4740){\makebox(0,0)[r]{\strut{}-80}}%
      \put(171,5189){\makebox(0,0)[r]{\strut{}-60}}%
      \put(171,5638){\makebox(0,0)[r]{\strut{}-40}}%
      \put(171,6088){\makebox(0,0)[r]{\strut{}-20}}%
      \put(171,6537){\makebox(0,0)[r]{\strut{} 0}}%
      \put(303,3621){\makebox(0,0){\strut{}-2}}%
      \put(815,3621){\makebox(0,0){\strut{}-1.95}}%
      \put(1328,3621){\makebox(0,0){\strut{}-1.9}}%
      \put(1840,3621){\makebox(0,0){\strut{}-1.85}}%
      \put(2352,3621){\makebox(0,0){\strut{}-1.8}}%
      \put(2864,3621){\makebox(0,0){\strut{}-1.75}}%
      \put(3377,3621){\makebox(0,0){\strut{}-1.7}}%
      \put(3889,3621){\makebox(0,0){\strut{}-1.65}}%
      \put(4401,3621){\makebox(0,0){\strut{}-1.6}}%
      \put(4620,5189){\rotatebox{-270}{\makebox(0,0){\strut{}$\log(\PP{\Delta\Phi|d,\cdots}) +C$}}}%
    }%
    \gplgaddtomacro\gplfronttext{%
      \csname LTb\endcsname%
      \put(2283,4234){\makebox(0,0)[r]{\strut{}Bayesian}}%
      \csname LTb\endcsname%
      \put(2283,4014){\makebox(0,0)[r]{\strut{}$-1.81^\circ \pm 0.04^\circ$}}%
    }%
    \gplgaddtomacro\gplbacktext{%
      \csname LTb\endcsname%
      \put(171,704){\makebox(0,0)[r]{\strut{}-25000}}%
      \put(171,1191){\makebox(0,0)[r]{\strut{}-20000}}%
      \put(171,1677){\makebox(0,0)[r]{\strut{}-15000}}%
      \put(171,2164){\makebox(0,0)[r]{\strut{}-10000}}%
      \put(171,2650){\makebox(0,0)[r]{\strut{}-5000}}%
      \put(171,3137){\makebox(0,0)[r]{\strut{} 0}}%
      \put(303,484){\makebox(0,0){\strut{}-40}}%
      \put(815,484){\makebox(0,0){\strut{}-30}}%
      \put(1328,484){\makebox(0,0){\strut{}-20}}%
      \put(1840,484){\makebox(0,0){\strut{}-10}}%
      \put(2352,484){\makebox(0,0){\strut{} 0}}%
      \put(2864,484){\makebox(0,0){\strut{} 10}}%
      \put(3377,484){\makebox(0,0){\strut{} 20}}%
      \put(3889,484){\makebox(0,0){\strut{} 30}}%
      \put(4401,484){\makebox(0,0){\strut{} 40}}%
      \put(4620,1920){\rotatebox{-270}{\makebox(0,0){\strut{}$\log(\PP{\Delta\Phi|d,\cdots}) +C$}}}%
      \put(2352,154){\makebox(0,0){\strut{}$\Delta\Phi$ in degree}}%
    }%
    \gplgaddtomacro\gplfronttext{%
      \csname LTb\endcsname%
      \put(3414,2964){\makebox(0,0)[r]{\strut{}Bayesian}}%
    }%
    \gplbacktext
    \put(0,0){\includegraphics{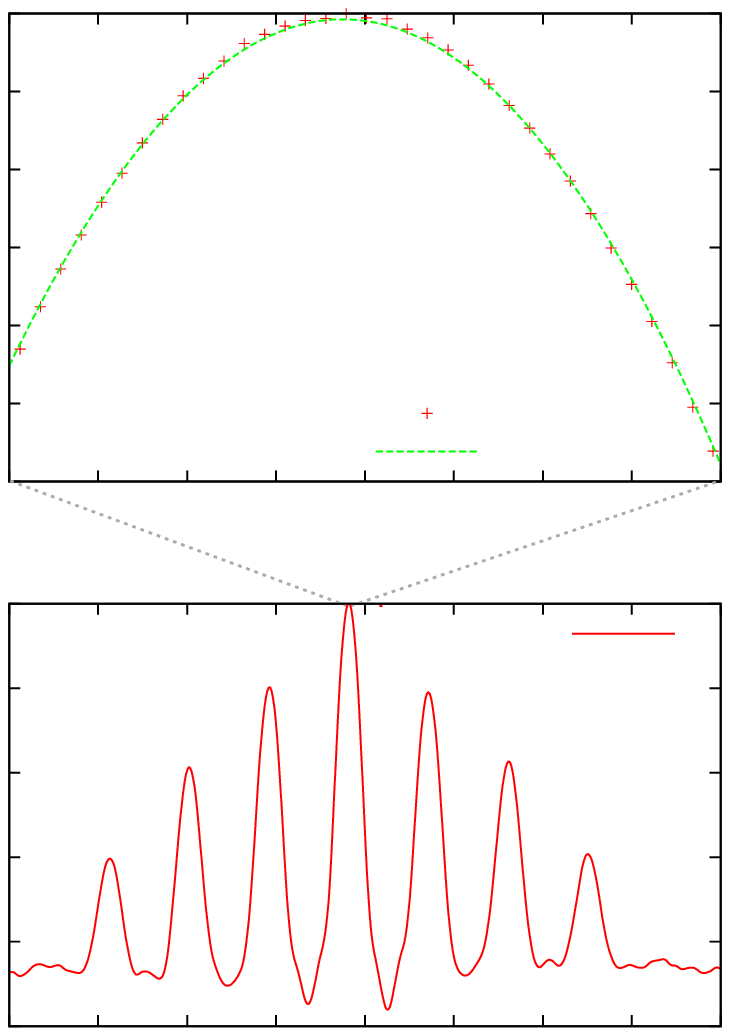}}%
    \gplfronttext
  \end{picture}%
\endgroup

%% file: graphics/methodcomp32.tex
\begingroup
  \makeatletter
  \providecommand\color[2][]{%
    \GenericError{(gnuplot) \space\space\space\@spaces}{%
      Package color not loaded in conjunction with
      terminal option `colourtext'%
    }{See the gnuplot documentation for explanation.%
    }{Either use 'blacktext' in gnuplot or load the package
      color.sty in LaTeX.}%
    \renewcommand\color[2][]{}%
  }%
  \providecommand\includegraphics[2][]{%
    \GenericError{(gnuplot) \space\space\space\@spaces}{%
      Package graphicx or graphics not loaded%
    }{See the gnuplot documentation for explanation.%
    }{The gnuplot epslatex terminal needs graphicx.sty or graphics.sty.}%
    \renewcommand\includegraphics[2][]{}%
  }%
  \providecommand\rotatebox[2]{#2}%
  \@ifundefined{ifGPcolor}{%
    \newif\ifGPcolor
    \GPcolortrue
  }{}%
  \@ifundefined{ifGPblacktext}{%
    \newif\ifGPblacktext
    \GPblacktexttrue
  }{}%
  \let\gplgaddtomacro\g@addto@macro
  \gdef\gplbacktext{}%
  \gdef\gplfronttext{}%
  \makeatother
  \ifGPblacktext
    \def\colorrgb#1{}%
    \def\colorgray#1{}%
  \else
    \ifGPcolor
      \def\colorrgb#1{\color[rgb]{#1}}%
      \def\colorgray#1{\color[gray]{#1}}%
      \expandafter\def\csname LTw\endcsname{\color{white}}%
      \expandafter\def\csname LTb\endcsname{\color{black}}%
      \expandafter\def\csname LTa\endcsname{\color{black}}%
      \expandafter\def\csname LT0\endcsname{\color[rgb]{1,0,0}}%
      \expandafter\def\csname LT1\endcsname{\color[rgb]{0,1,0}}%
      \expandafter\def\csname LT2\endcsname{\color[rgb]{0,0,1}}%
      \expandafter\def\csname LT3\endcsname{\color[rgb]{1,0,1}}%
      \expandafter\def\csname LT4\endcsname{\color[rgb]{0,1,1}}%
      \expandafter\def\csname LT5\endcsname{\color[rgb]{1,1,0}}%
      \expandafter\def\csname LT6\endcsname{\color[rgb]{0,0,0}}%
      \expandafter\def\csname LT7\endcsname{\color[rgb]{1,0.3,0}}%
      \expandafter\def\csname LT8\endcsname{\color[rgb]{0.5,0.5,0.5}}%
    \else
      \def\colorrgb#1{\color{black}}%
      \def\colorgray#1{\color[gray]{#1}}%
      \expandafter\def\csname LTw\endcsname{\color{white}}%
      \expandafter\def\csname LTb\endcsname{\color{black}}%
      \expandafter\def\csname LTa\endcsname{\color{black}}%
      \expandafter\def\csname LT0\endcsname{\color{black}}%
      \expandafter\def\csname LT1\endcsname{\color{black}}%
      \expandafter\def\csname LT2\endcsname{\color{black}}%
      \expandafter\def\csname LT3\endcsname{\color{black}}%
      \expandafter\def\csname LT4\endcsname{\color{black}}%
      \expandafter\def\csname LT5\endcsname{\color{black}}%
      \expandafter\def\csname LT6\endcsname{\color{black}}%
      \expandafter\def\csname LT7\endcsname{\color{black}}%
      \expandafter\def\csname LT8\endcsname{\color{black}}%
    \fi
  \fi
  \setlength{\unitlength}{0.0500bp}%
  \begin{picture}(8502.00,2834.00)%
    \gplgaddtomacro\gplbacktext{%
      \csname LTb\endcsname%
      \put(1210,440){\makebox(0,0)[r]{\strut{} 0.001}}%
      \put(1210,842){\makebox(0,0)[r]{\strut{} 0.01}}%
      \put(1210,1243){\makebox(0,0)[r]{\strut{} 0.1}}%
      \put(1210,1645){\makebox(0,0)[r]{\strut{} 1}}%
      \put(1210,2046){\makebox(0,0)[r]{\strut{} 10}}%
      \put(1210,2448){\makebox(0,0)[r]{\strut{} 100}}%
      \put(1342,220){\makebox(0,0){\strut{} 1}}%
      \put(3838,220){\makebox(0,0){\strut{} 10}}%
      \put(6334,220){\makebox(0,0){\strut{} 100}}%
      \put(176,1504){\rotatebox{-270}{\makebox(0,0){\strut{}signal / noise}}}%
      \put(4723,-66){\makebox(0,0){\strut{}}}%
      \put(4723,2459){\makebox(0,0){\strut{}}}%
    }%
    \gplgaddtomacro\gplfronttext{%
      \csname LTb\endcsname%
      \put(7118,2396){\makebox(0,0)[r]{\strut{}$\ol f$ after 32 pulses}}%
      \csname LTb\endcsname%
      \put(7118,2176){\makebox(0,0)[r]{\strut{}$\ol f$ over 2000 pulses}}%
      \csname LTb\endcsname%
      \put(7118,1956){\makebox(0,0)[r]{\strut{}class. avg. after 32 pulses}}%
      \csname LTb\endcsname%
      \put(7118,1736){\makebox(0,0)[r]{\strut{}class. avg. over 2000 pulses}}%
    }%
    \gplbacktext
    \put(0,0){\includegraphics{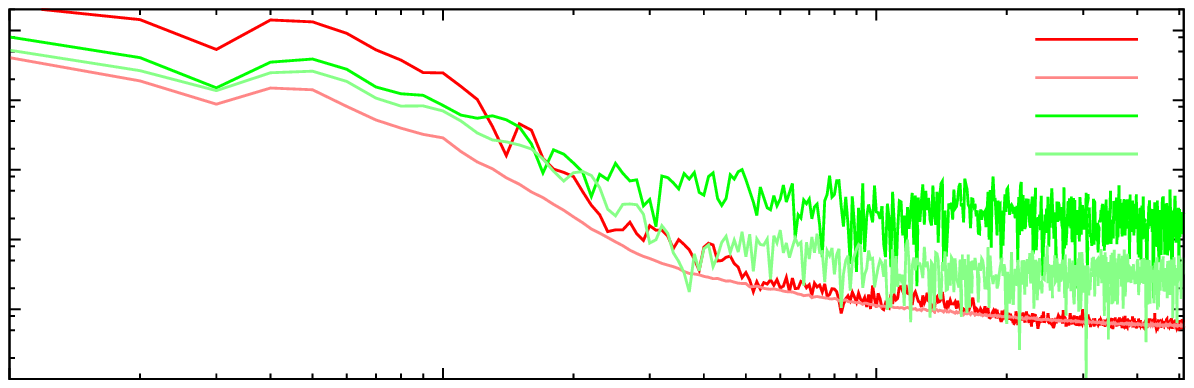}}%
    \gplfronttext
  \end{picture}%
\endgroup

%% file: graphics/methodcomp64.tex
\begingroup
  \makeatletter
  \providecommand\color[2][]{%
    \GenericError{(gnuplot) \space\space\space\@spaces}{%
      Package color not loaded in conjunction with
      terminal option `colourtext'%
    }{See the gnuplot documentation for explanation.%
    }{Either use 'blacktext' in gnuplot or load the package
      color.sty in LaTeX.}%
    \renewcommand\color[2][]{}%
  }%
  \providecommand\includegraphics[2][]{%
    \GenericError{(gnuplot) \space\space\space\@spaces}{%
      Package graphicx or graphics not loaded%
    }{See the gnuplot documentation for explanation.%
    }{The gnuplot epslatex terminal needs graphicx.sty or graphics.sty.}%
    \renewcommand\includegraphics[2][]{}%
  }%
  \providecommand\rotatebox[2]{#2}%
  \@ifundefined{ifGPcolor}{%
    \newif\ifGPcolor
    \GPcolortrue
  }{}%
  \@ifundefined{ifGPblacktext}{%
    \newif\ifGPblacktext
    \GPblacktexttrue
  }{}%
  \let\gplgaddtomacro\g@addto@macro
  \gdef\gplbacktext{}%
  \gdef\gplfronttext{}%
  \makeatother
  \ifGPblacktext
    \def\colorrgb#1{}%
    \def\colorgray#1{}%
  \else
    \ifGPcolor
      \def\colorrgb#1{\color[rgb]{#1}}%
      \def\colorgray#1{\color[gray]{#1}}%
      \expandafter\def\csname LTw\endcsname{\color{white}}%
      \expandafter\def\csname LTb\endcsname{\color{black}}%
      \expandafter\def\csname LTa\endcsname{\color{black}}%
      \expandafter\def\csname LT0\endcsname{\color[rgb]{1,0,0}}%
      \expandafter\def\csname LT1\endcsname{\color[rgb]{0,1,0}}%
      \expandafter\def\csname LT2\endcsname{\color[rgb]{0,0,1}}%
      \expandafter\def\csname LT3\endcsname{\color[rgb]{1,0,1}}%
      \expandafter\def\csname LT4\endcsname{\color[rgb]{0,1,1}}%
      \expandafter\def\csname LT5\endcsname{\color[rgb]{1,1,0}}%
      \expandafter\def\csname LT6\endcsname{\color[rgb]{0,0,0}}%
      \expandafter\def\csname LT7\endcsname{\color[rgb]{1,0.3,0}}%
      \expandafter\def\csname LT8\endcsname{\color[rgb]{0.5,0.5,0.5}}%
    \else
      \def\colorrgb#1{\color{black}}%
      \def\colorgray#1{\color[gray]{#1}}%
      \expandafter\def\csname LTw\endcsname{\color{white}}%
      \expandafter\def\csname LTb\endcsname{\color{black}}%
      \expandafter\def\csname LTa\endcsname{\color{black}}%
      \expandafter\def\csname LT0\endcsname{\color{black}}%
      \expandafter\def\csname LT1\endcsname{\color{black}}%
      \expandafter\def\csname LT2\endcsname{\color{black}}%
      \expandafter\def\csname LT3\endcsname{\color{black}}%
      \expandafter\def\csname LT4\endcsname{\color{black}}%
      \expandafter\def\csname LT5\endcsname{\color{black}}%
      \expandafter\def\csname LT6\endcsname{\color{black}}%
      \expandafter\def\csname LT7\endcsname{\color{black}}%
      \expandafter\def\csname LT8\endcsname{\color{black}}%
    \fi
  \fi
  \setlength{\unitlength}{0.0500bp}%
  \begin{picture}(8502.00,2834.00)%
    \gplgaddtomacro\gplbacktext{%
      \csname LTb\endcsname%
      \put(1210,440){\makebox(0,0)[r]{\strut{} 0.001}}%
      \put(1210,842){\makebox(0,0)[r]{\strut{} 0.01}}%
      \put(1210,1243){\makebox(0,0)[r]{\strut{} 0.1}}%
      \put(1210,1645){\makebox(0,0)[r]{\strut{} 1}}%
      \put(1210,2046){\makebox(0,0)[r]{\strut{} 10}}%
      \put(1210,2448){\makebox(0,0)[r]{\strut{} 100}}%
      \put(1342,220){\makebox(0,0){\strut{} 1}}%
      \put(3838,220){\makebox(0,0){\strut{} 10}}%
      \put(6334,220){\makebox(0,0){\strut{} 100}}%
      \put(176,1504){\rotatebox{-270}{\makebox(0,0){\strut{}signal / noise}}}%
      \put(4723,-66){\makebox(0,0){\strut{}}}%
      \put(4723,2459){\makebox(0,0){\strut{}}}%
    }%
    \gplgaddtomacro\gplfronttext{%
      \csname LTb\endcsname%
      \put(7118,2396){\makebox(0,0)[r]{\strut{}$\ol f$ after 64 pulses}}%
      \csname LTb\endcsname%
      \put(7118,2176){\makebox(0,0)[r]{\strut{}$\ol f$ over 2000 pulses}}%
      \csname LTb\endcsname%
      \put(7118,1956){\makebox(0,0)[r]{\strut{}class. avg. after 64 pulses}}%
      \csname LTb\endcsname%
      \put(7118,1736){\makebox(0,0)[r]{\strut{}class. avg. over 2000 pulses}}%
    }%
    \gplbacktext
    \put(0,0){\includegraphics{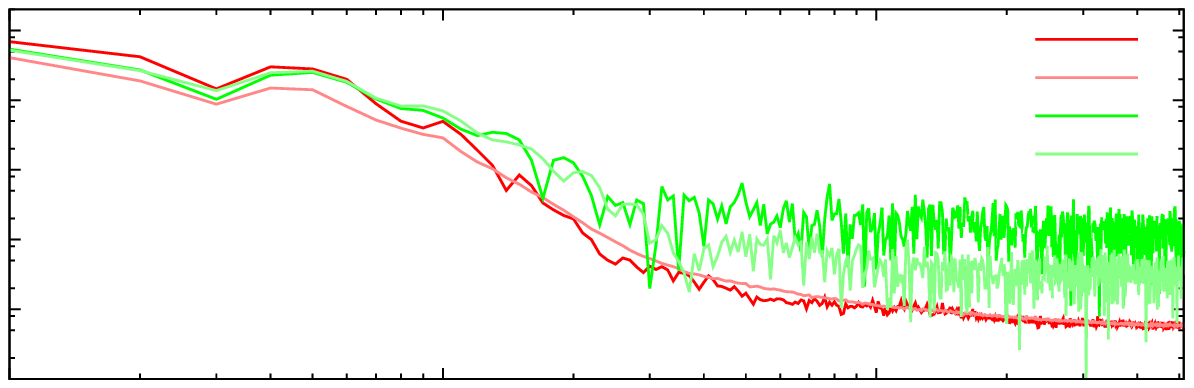}}%
    \gplfronttext
  \end{picture}%
\endgroup

%% file: graphics/methodcomp128.tex
\begingroup
  \makeatletter
  \providecommand\color[2][]{%
    \GenericError{(gnuplot) \space\space\space\@spaces}{%
      Package color not loaded in conjunction with
      terminal option `colourtext'%
    }{See the gnuplot documentation for explanation.%
    }{Either use 'blacktext' in gnuplot or load the package
      color.sty in LaTeX.}%
    \renewcommand\color[2][]{}%
  }%
  \providecommand\includegraphics[2][]{%
    \GenericError{(gnuplot) \space\space\space\@spaces}{%
      Package graphicx or graphics not loaded%
    }{See the gnuplot documentation for explanation.%
    }{The gnuplot epslatex terminal needs graphicx.sty or graphics.sty.}%
    \renewcommand\includegraphics[2][]{}%
  }%
  \providecommand\rotatebox[2]{#2}%
  \@ifundefined{ifGPcolor}{%
    \newif\ifGPcolor
    \GPcolortrue
  }{}%
  \@ifundefined{ifGPblacktext}{%
    \newif\ifGPblacktext
    \GPblacktexttrue
  }{}%
  \let\gplgaddtomacro\g@addto@macro
  \gdef\gplbacktext{}%
  \gdef\gplfronttext{}%
  \makeatother
  \ifGPblacktext
    \def\colorrgb#1{}%
    \def\colorgray#1{}%
  \else
    \ifGPcolor
      \def\colorrgb#1{\color[rgb]{#1}}%
      \def\colorgray#1{\color[gray]{#1}}%
      \expandafter\def\csname LTw\endcsname{\color{white}}%
      \expandafter\def\csname LTb\endcsname{\color{black}}%
      \expandafter\def\csname LTa\endcsname{\color{black}}%
      \expandafter\def\csname LT0\endcsname{\color[rgb]{1,0,0}}%
      \expandafter\def\csname LT1\endcsname{\color[rgb]{0,1,0}}%
      \expandafter\def\csname LT2\endcsname{\color[rgb]{0,0,1}}%
      \expandafter\def\csname LT3\endcsname{\color[rgb]{1,0,1}}%
      \expandafter\def\csname LT4\endcsname{\color[rgb]{0,1,1}}%
      \expandafter\def\csname LT5\endcsname{\color[rgb]{1,1,0}}%
      \expandafter\def\csname LT6\endcsname{\color[rgb]{0,0,0}}%
      \expandafter\def\csname LT7\endcsname{\color[rgb]{1,0.3,0}}%
      \expandafter\def\csname LT8\endcsname{\color[rgb]{0.5,0.5,0.5}}%
    \else
      \def\colorrgb#1{\color{black}}%
      \def\colorgray#1{\color[gray]{#1}}%
      \expandafter\def\csname LTw\endcsname{\color{white}}%
      \expandafter\def\csname LTb\endcsname{\color{black}}%
      \expandafter\def\csname LTa\endcsname{\color{black}}%
      \expandafter\def\csname LT0\endcsname{\color{black}}%
      \expandafter\def\csname LT1\endcsname{\color{black}}%
      \expandafter\def\csname LT2\endcsname{\color{black}}%
      \expandafter\def\csname LT3\endcsname{\color{black}}%
      \expandafter\def\csname LT4\endcsname{\color{black}}%
      \expandafter\def\csname LT5\endcsname{\color{black}}%
      \expandafter\def\csname LT6\endcsname{\color{black}}%
      \expandafter\def\csname LT7\endcsname{\color{black}}%
      \expandafter\def\csname LT8\endcsname{\color{black}}%
    \fi
  \fi
  \setlength{\unitlength}{0.0500bp}%
  \begin{picture}(8502.00,2834.00)%
    \gplgaddtomacro\gplbacktext{%
      \csname LTb\endcsname%
      \put(1210,440){\makebox(0,0)[r]{\strut{} 0.001}}%
      \put(1210,842){\makebox(0,0)[r]{\strut{} 0.01}}%
      \put(1210,1243){\makebox(0,0)[r]{\strut{} 0.1}}%
      \put(1210,1645){\makebox(0,0)[r]{\strut{} 1}}%
      \put(1210,2046){\makebox(0,0)[r]{\strut{} 10}}%
      \put(1210,2448){\makebox(0,0)[r]{\strut{} 100}}%
      \put(1342,220){\makebox(0,0){\strut{} 1}}%
      \put(3838,220){\makebox(0,0){\strut{} 10}}%
      \put(6334,220){\makebox(0,0){\strut{} 100}}%
      \put(176,1504){\rotatebox{-270}{\makebox(0,0){\strut{}signal / noise}}}%
      \put(4723,-66){\makebox(0,0){\strut{}}}%
      \put(4723,2459){\makebox(0,0){\strut{}}}%
    }%
    \gplgaddtomacro\gplfronttext{%
      \csname LTb\endcsname%
      \put(7118,2396){\makebox(0,0)[r]{\strut{}$\ol f$ after 128 pulses}}%
      \csname LTb\endcsname%
      \put(7118,2176){\makebox(0,0)[r]{\strut{}$\ol f$ over 2000 pulses}}%
      \csname LTb\endcsname%
      \put(7118,1956){\makebox(0,0)[r]{\strut{}class. avg. after 128 pulses}}%
      \csname LTb\endcsname%
      \put(7118,1736){\makebox(0,0)[r]{\strut{}class. avg. over 2000 pulses}}%
    }%
    \gplbacktext
    \put(0,0){\includegraphics{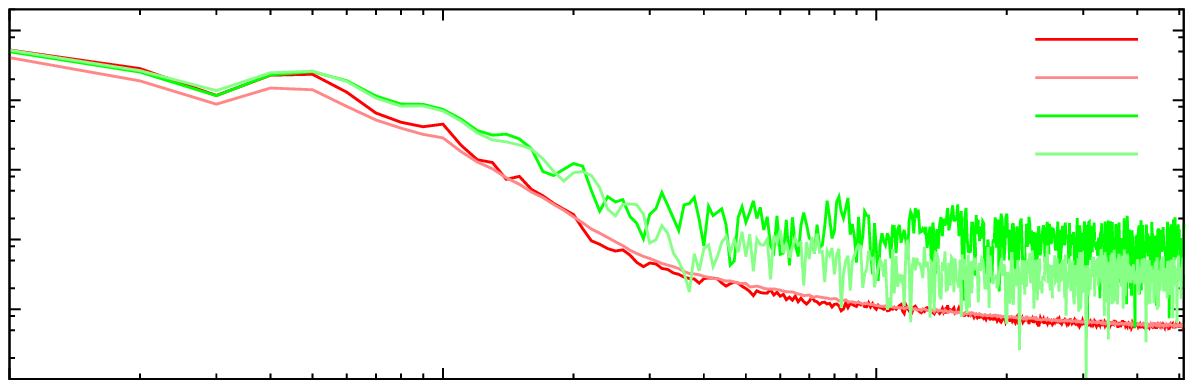}}%
    \gplfronttext
  \end{picture}%
\endgroup

%% file: graphics/methodcomp256.tex
\begingroup
  \makeatletter
  \providecommand\color[2][]{%
    \GenericError{(gnuplot) \space\space\space\@spaces}{%
      Package color not loaded in conjunction with
      terminal option `colourtext'%
    }{See the gnuplot documentation for explanation.%
    }{Either use 'blacktext' in gnuplot or load the package
      color.sty in LaTeX.}%
    \renewcommand\color[2][]{}%
  }%
  \providecommand\includegraphics[2][]{%
    \GenericError{(gnuplot) \space\space\space\@spaces}{%
      Package graphicx or graphics not loaded%
    }{See the gnuplot documentation for explanation.%
    }{The gnuplot epslatex terminal needs graphicx.sty or graphics.sty.}%
    \renewcommand\includegraphics[2][]{}%
  }%
  \providecommand\rotatebox[2]{#2}%
  \@ifundefined{ifGPcolor}{%
    \newif\ifGPcolor
    \GPcolortrue
  }{}%
  \@ifundefined{ifGPblacktext}{%
    \newif\ifGPblacktext
    \GPblacktexttrue
  }{}%
  \let\gplgaddtomacro\g@addto@macro
  \gdef\gplbacktext{}%
  \gdef\gplfronttext{}%
  \makeatother
  \ifGPblacktext
    \def\colorrgb#1{}%
    \def\colorgray#1{}%
  \else
    \ifGPcolor
      \def\colorrgb#1{\color[rgb]{#1}}%
      \def\colorgray#1{\color[gray]{#1}}%
      \expandafter\def\csname LTw\endcsname{\color{white}}%
      \expandafter\def\csname LTb\endcsname{\color{black}}%
      \expandafter\def\csname LTa\endcsname{\color{black}}%
      \expandafter\def\csname LT0\endcsname{\color[rgb]{1,0,0}}%
      \expandafter\def\csname LT1\endcsname{\color[rgb]{0,1,0}}%
      \expandafter\def\csname LT2\endcsname{\color[rgb]{0,0,1}}%
      \expandafter\def\csname LT3\endcsname{\color[rgb]{1,0,1}}%
      \expandafter\def\csname LT4\endcsname{\color[rgb]{0,1,1}}%
      \expandafter\def\csname LT5\endcsname{\color[rgb]{1,1,0}}%
      \expandafter\def\csname LT6\endcsname{\color[rgb]{0,0,0}}%
      \expandafter\def\csname LT7\endcsname{\color[rgb]{1,0.3,0}}%
      \expandafter\def\csname LT8\endcsname{\color[rgb]{0.5,0.5,0.5}}%
    \else
      \def\colorrgb#1{\color{black}}%
      \def\colorgray#1{\color[gray]{#1}}%
      \expandafter\def\csname LTw\endcsname{\color{white}}%
      \expandafter\def\csname LTb\endcsname{\color{black}}%
      \expandafter\def\csname LTa\endcsname{\color{black}}%
      \expandafter\def\csname LT0\endcsname{\color{black}}%
      \expandafter\def\csname LT1\endcsname{\color{black}}%
      \expandafter\def\csname LT2\endcsname{\color{black}}%
      \expandafter\def\csname LT3\endcsname{\color{black}}%
      \expandafter\def\csname LT4\endcsname{\color{black}}%
      \expandafter\def\csname LT5\endcsname{\color{black}}%
      \expandafter\def\csname LT6\endcsname{\color{black}}%
      \expandafter\def\csname LT7\endcsname{\color{black}}%
      \expandafter\def\csname LT8\endcsname{\color{black}}%
    \fi
  \fi
  \setlength{\unitlength}{0.0500bp}%
  \begin{picture}(8502.00,2834.00)%
    \gplgaddtomacro\gplbacktext{%
      \csname LTb\endcsname%
      \put(1210,440){\makebox(0,0)[r]{\strut{} 0.001}}%
      \put(1210,842){\makebox(0,0)[r]{\strut{} 0.01}}%
      \put(1210,1243){\makebox(0,0)[r]{\strut{} 0.1}}%
      \put(1210,1645){\makebox(0,0)[r]{\strut{} 1}}%
      \put(1210,2046){\makebox(0,0)[r]{\strut{} 10}}%
      \put(1210,2448){\makebox(0,0)[r]{\strut{} 100}}%
      \put(1342,220){\makebox(0,0){\strut{} 1}}%
      \put(3838,220){\makebox(0,0){\strut{} 10}}%
      \put(6334,220){\makebox(0,0){\strut{} 100}}%
      \put(176,1504){\rotatebox{-270}{\makebox(0,0){\strut{}signal / noise}}}%
      \put(4723,-66){\makebox(0,0){\strut{}}}%
      \put(4723,2459){\makebox(0,0){\strut{}}}%
    }%
    \gplgaddtomacro\gplfronttext{%
      \csname LTb\endcsname%
      \put(7118,2396){\makebox(0,0)[r]{\strut{}$\ol f$ after 256 pulses}}%
      \csname LTb\endcsname%
      \put(7118,2176){\makebox(0,0)[r]{\strut{}$\ol f$ over 2000 pulses}}%
      \csname LTb\endcsname%
      \put(7118,1956){\makebox(0,0)[r]{\strut{}class. avg. after 256 pulses}}%
      \csname LTb\endcsname%
      \put(7118,1736){\makebox(0,0)[r]{\strut{}class. avg. over 2000 pulses}}%
    }%
    \gplbacktext
    \put(0,0){\includegraphics{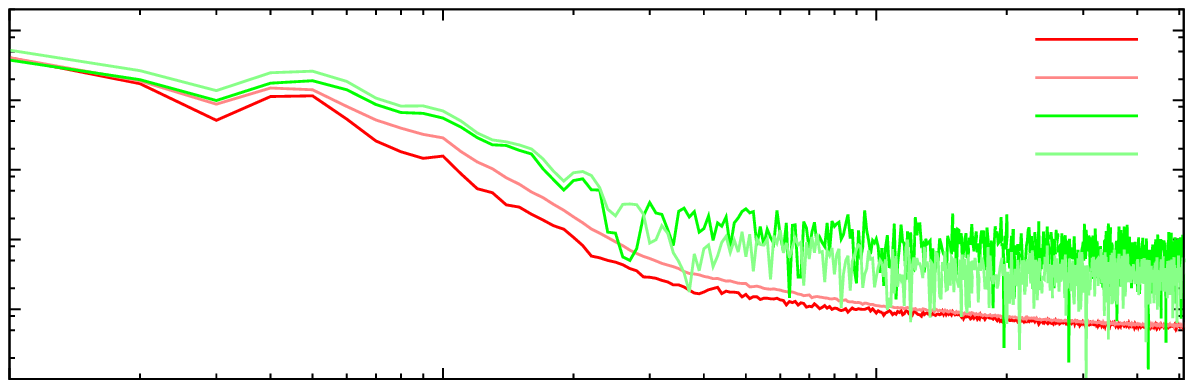}}%
    \gplfronttext
  \end{picture}%
\endgroup

%% file: graphics/reconstruction10.tex
\begingroup
  \makeatletter
  \providecommand\color[2][]{%
    \GenericError{(gnuplot) \space\space\space\@spaces}{%
      Package color not loaded in conjunction with
      terminal option `colourtext'%
    }{See the gnuplot documentation for explanation.%
    }{Either use 'blacktext' in gnuplot or load the package
      color.sty in LaTeX.}%
    \renewcommand\color[2][]{}%
  }%
  \providecommand\includegraphics[2][]{%
    \GenericError{(gnuplot) \space\space\space\@spaces}{%
      Package graphicx or graphics not loaded%
    }{See the gnuplot documentation for explanation.%
    }{The gnuplot epslatex terminal needs graphicx.sty or graphics.sty.}%
    \renewcommand\includegraphics[2][]{}%
  }%
  \providecommand\rotatebox[2]{#2}%
  \@ifundefined{ifGPcolor}{%
    \newif\ifGPcolor
    \GPcolortrue
  }{}%
  \@ifundefined{ifGPblacktext}{%
    \newif\ifGPblacktext
    \GPblacktexttrue
  }{}%
  \let\gplgaddtomacro\g@addto@macro
  \gdef\gplbacktext{}%
  \gdef\gplfronttext{}%
  \makeatother
  \ifGPblacktext
    \def\colorrgb#1{}%
    \def\colorgray#1{}%
  \else
    \ifGPcolor
      \def\colorrgb#1{\color[rgb]{#1}}%
      \def\colorgray#1{\color[gray]{#1}}%
      \expandafter\def\csname LTw\endcsname{\color{white}}%
      \expandafter\def\csname LTb\endcsname{\color{black}}%
      \expandafter\def\csname LTa\endcsname{\color{black}}%
      \expandafter\def\csname LT0\endcsname{\color[rgb]{1,0,0}}%
      \expandafter\def\csname LT1\endcsname{\color[rgb]{0,1,0}}%
      \expandafter\def\csname LT2\endcsname{\color[rgb]{0,0,1}}%
      \expandafter\def\csname LT3\endcsname{\color[rgb]{1,0,1}}%
      \expandafter\def\csname LT4\endcsname{\color[rgb]{0,1,1}}%
      \expandafter\def\csname LT5\endcsname{\color[rgb]{1,1,0}}%
      \expandafter\def\csname LT6\endcsname{\color[rgb]{0,0,0}}%
      \expandafter\def\csname LT7\endcsname{\color[rgb]{1,0.3,0}}%
      \expandafter\def\csname LT8\endcsname{\color[rgb]{0.5,0.5,0.5}}%
    \else
      \def\colorrgb#1{\color{black}}%
      \def\colorgray#1{\color[gray]{#1}}%
      \expandafter\def\csname LTw\endcsname{\color{white}}%
      \expandafter\def\csname LTb\endcsname{\color{black}}%
      \expandafter\def\csname LTa\endcsname{\color{black}}%
      \expandafter\def\csname LT0\endcsname{\color{black}}%
      \expandafter\def\csname LT1\endcsname{\color{black}}%
      \expandafter\def\csname LT2\endcsname{\color{black}}%
      \expandafter\def\csname LT3\endcsname{\color{black}}%
      \expandafter\def\csname LT4\endcsname{\color{black}}%
      \expandafter\def\csname LT5\endcsname{\color{black}}%
      \expandafter\def\csname LT6\endcsname{\color{black}}%
      \expandafter\def\csname LT7\endcsname{\color{black}}%
      \expandafter\def\csname LT8\endcsname{\color{black}}%
    \fi
  \fi
  \setlength{\unitlength}{0.0500bp}%
  \begin{picture}(10204.00,4534.00)%
    \gplgaddtomacro\gplbacktext{%
      \csname LTb\endcsname%
      \put(946,704){\makebox(0,0)[r]{\strut{} 0}}%
      \put(946,1100){\makebox(0,0)[r]{\strut{} 0.5}}%
      \put(946,1496){\makebox(0,0)[r]{\strut{} 1}}%
      \put(946,1892){\makebox(0,0)[r]{\strut{} 1.5}}%
      \put(946,2289){\makebox(0,0)[r]{\strut{} 2}}%
      \put(946,2685){\makebox(0,0)[r]{\strut{} 2.5}}%
      \put(946,3081){\makebox(0,0)[r]{\strut{} 3}}%
      \put(946,3477){\makebox(0,0)[r]{\strut{} 3.5}}%
      \put(946,3873){\makebox(0,0)[r]{\strut{} 4}}%
      \put(1581,484){\makebox(0,0){\strut{} 150}}%
      \put(2587,484){\makebox(0,0){\strut{} 200}}%
      \put(3592,484){\makebox(0,0){\strut{} 250}}%
      \put(4598,484){\makebox(0,0){\strut{} 300}}%
      \put(176,2288){\rotatebox{-270}{\makebox(0,0){\strut{}Signal / noise}}}%
      \put(3089,154){\makebox(0,0){\strut{}Pulsar phase in degree}}%
      \put(3089,4203){\makebox(0,0){\strut{}After 10 pulses}}%
    }%
    \gplgaddtomacro\gplfronttext{%
      \csname LTb\endcsname%
      \put(4114,3700){\makebox(0,0)[r]{\strut{}Bayesian reconstruction}}%
      \csname LTb\endcsname%
      \put(4114,3480){\makebox(0,0)[r]{\strut{}Classical average}}%
    }%
    \gplgaddtomacro\gplbacktext{%
      \csname LTb\endcsname%
      \put(5612,484){\makebox(0,0){\strut{} 150}}%
      \put(6631,484){\makebox(0,0){\strut{} 200}}%
      \put(7650,484){\makebox(0,0){\strut{} 250}}%
      \put(8669,484){\makebox(0,0){\strut{} 300}}%
      \put(7140,154){\makebox(0,0){\strut{}Pulsar phase in degree}}%
      \put(7140,4203){\makebox(0,0){\strut{}After 1000 pulses}}%
    }%
    \gplgaddtomacro\gplfronttext{%
      \csname LTb\endcsname%
      \put(8191,3700){\makebox(0,0)[r]{\strut{}Bayesian reconstruction}}%
      \csname LTb\endcsname%
      \put(8191,3480){\makebox(0,0)[r]{\strut{}Classical average}}%
      \csname LTb\endcsname%
      \put(9615,703){\makebox(0,0)[l]{\strut{} 0.0001}}%
      \put(9615,1496){\makebox(0,0)[l]{\strut{} 0.001}}%
      \put(9615,2288){\makebox(0,0)[l]{\strut{} 0.01}}%
      \put(9615,3080){\makebox(0,0)[l]{\strut{} 0.1}}%
      \put(9615,3873){\makebox(0,0)[l]{\strut{} 1}}%
      \put(10605,2288){\rotatebox{-270}{\makebox(0,0){\strut{}Relative frequency for single pulse}}}%
    }%
    \gplbacktext
    \put(0,0){\includegraphics{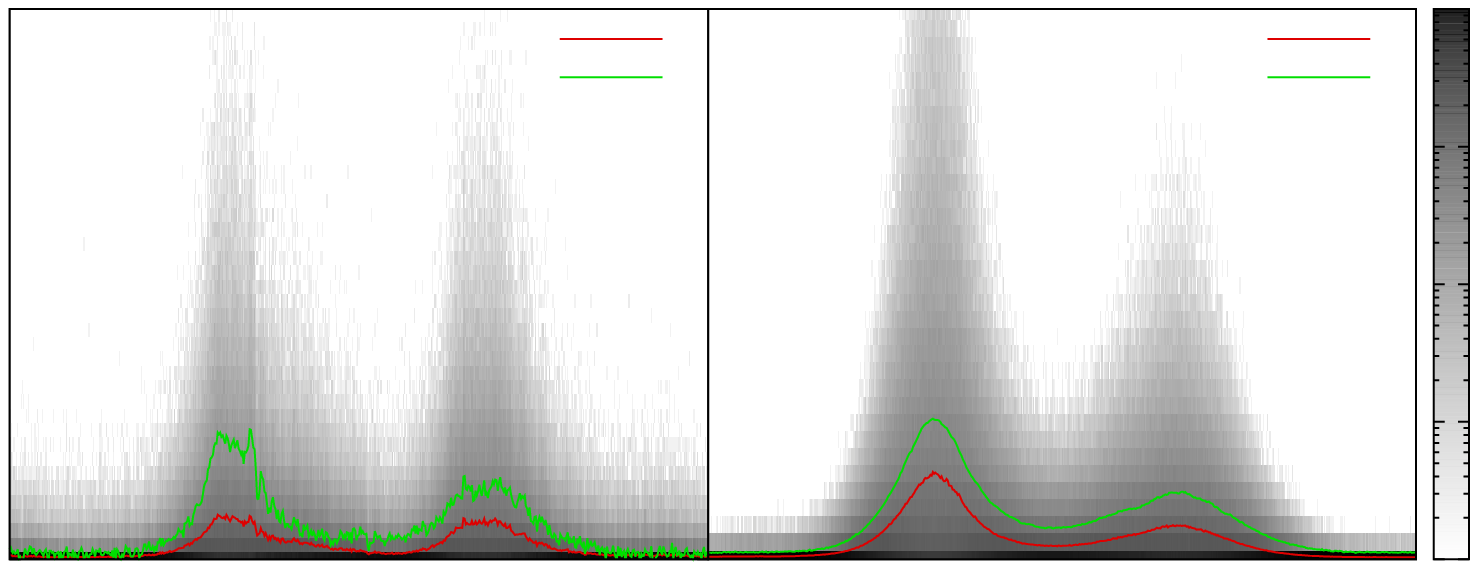}}%
    \gplfronttext
  \end{picture}%
\endgroup

%% file: graphics/movingframe.tex
\begingroup
  \makeatletter
  \providecommand\color[2][]{%
    \GenericError{(gnuplot) \space\space\space\@spaces}{%
      Package color not loaded in conjunction with
      terminal option `colourtext'%
    }{See the gnuplot documentation for explanation.%
    }{Either use 'blacktext' in gnuplot or load the package
      color.sty in LaTeX.}%
    \renewcommand\color[2][]{}%
  }%
  \providecommand\includegraphics[2][]{%
    \GenericError{(gnuplot) \space\space\space\@spaces}{%
      Package graphicx or graphics not loaded%
    }{See the gnuplot documentation for explanation.%
    }{The gnuplot epslatex terminal needs graphicx.sty or graphics.sty.}%
    \renewcommand\includegraphics[2][]{}%
  }%
  \providecommand\rotatebox[2]{#2}%
  \@ifundefined{ifGPcolor}{%
    \newif\ifGPcolor
    \GPcolortrue
  }{}%
  \@ifundefined{ifGPblacktext}{%
    \newif\ifGPblacktext
    \GPblacktexttrue
  }{}%
  \let\gplgaddtomacro\g@addto@macro
  \gdef\gplbacktext{}%
  \gdef\gplfronttext{}%
  \makeatother
  \ifGPblacktext
    \def\colorrgb#1{}%
    \def\colorgray#1{}%
  \else
    \ifGPcolor
      \def\colorrgb#1{\color[rgb]{#1}}%
      \def\colorgray#1{\color[gray]{#1}}%
      \expandafter\def\csname LTw\endcsname{\color{white}}%
      \expandafter\def\csname LTb\endcsname{\color{black}}%
      \expandafter\def\csname LTa\endcsname{\color{black}}%
      \expandafter\def\csname LT0\endcsname{\color[rgb]{1,0,0}}%
      \expandafter\def\csname LT1\endcsname{\color[rgb]{0,1,0}}%
      \expandafter\def\csname LT2\endcsname{\color[rgb]{0,0,1}}%
      \expandafter\def\csname LT3\endcsname{\color[rgb]{1,0,1}}%
      \expandafter\def\csname LT4\endcsname{\color[rgb]{0,1,1}}%
      \expandafter\def\csname LT5\endcsname{\color[rgb]{1,1,0}}%
      \expandafter\def\csname LT6\endcsname{\color[rgb]{0,0,0}}%
      \expandafter\def\csname LT7\endcsname{\color[rgb]{1,0.3,0}}%
      \expandafter\def\csname LT8\endcsname{\color[rgb]{0.5,0.5,0.5}}%
    \else
      \def\colorrgb#1{\color{black}}%
      \def\colorgray#1{\color[gray]{#1}}%
      \expandafter\def\csname LTw\endcsname{\color{white}}%
      \expandafter\def\csname LTb\endcsname{\color{black}}%
      \expandafter\def\csname LTa\endcsname{\color{black}}%
      \expandafter\def\csname LT0\endcsname{\color{black}}%
      \expandafter\def\csname LT1\endcsname{\color{black}}%
      \expandafter\def\csname LT2\endcsname{\color{black}}%
      \expandafter\def\csname LT3\endcsname{\color{black}}%
      \expandafter\def\csname LT4\endcsname{\color{black}}%
      \expandafter\def\csname LT5\endcsname{\color{black}}%
      \expandafter\def\csname LT6\endcsname{\color{black}}%
      \expandafter\def\csname LT7\endcsname{\color{black}}%
      \expandafter\def\csname LT8\endcsname{\color{black}}%
    \fi
  \fi
  \setlength{\unitlength}{0.0500bp}%
  \begin{picture}(7936.00,5668.00)%
    \gplgaddtomacro\gplbacktext{%
      \csname LTb\endcsname%
      \put(92,704){\makebox(0,0)[r]{\strut{} 0}}%
      \put(92,1421){\makebox(0,0)[r]{\strut{} 0.2}}%
      \put(92,2138){\makebox(0,0)[r]{\strut{} 0.4}}%
      \put(92,2856){\makebox(0,0)[r]{\strut{} 0.6}}%
      \put(92,3573){\makebox(0,0)[r]{\strut{} 0.8}}%
      \put(92,4290){\makebox(0,0)[r]{\strut{} 1}}%
      \put(92,5007){\makebox(0,0)[r]{\strut{} 1.2}}%
      \put(866,484){\makebox(0,0){\strut{} 140}}%
      \put(1721,484){\makebox(0,0){\strut{} 160}}%
      \put(2577,484){\makebox(0,0){\strut{} 180}}%
      \put(3432,484){\makebox(0,0){\strut{} 200}}%
      \put(4288,484){\makebox(0,0){\strut{} 220}}%
      \put(5143,484){\makebox(0,0){\strut{} 240}}%
      \put(5999,484){\makebox(0,0){\strut{} 260}}%
      \put(6854,484){\makebox(0,0){\strut{} 280}}%
      \put(7710,484){\makebox(0,0){\strut{} 300}}%
      \put(7929,2855){\rotatebox{-270}{\makebox(0,0){\strut{}Signal / noise}}}%
      \put(3967,154){\makebox(0,0){\strut{}Template phase in degree}}%
      \put(3967,5337){\makebox(0,0){\strut{}Moving frame reconstruction}}%
    }%
    \gplgaddtomacro\gplfronttext{%
      \csname LTb\endcsname%
      \put(6723,4834){\makebox(0,0)[r]{\strut{}bayesian}}%
      \csname LTb\endcsname%
      \put(6723,4614){\makebox(0,0)[r]{\strut{}bayesian, shifted window}}%
      \csname LTb\endcsname%
      \put(6723,4394){\makebox(0,0)[r]{\strut{}classical average}}%
      \csname LTb\endcsname%
      \put(6723,4174){\makebox(0,0)[r]{\strut{}classical average, shifted window}}%
    }%
    \gplbacktext
    \put(0,0){\includegraphics{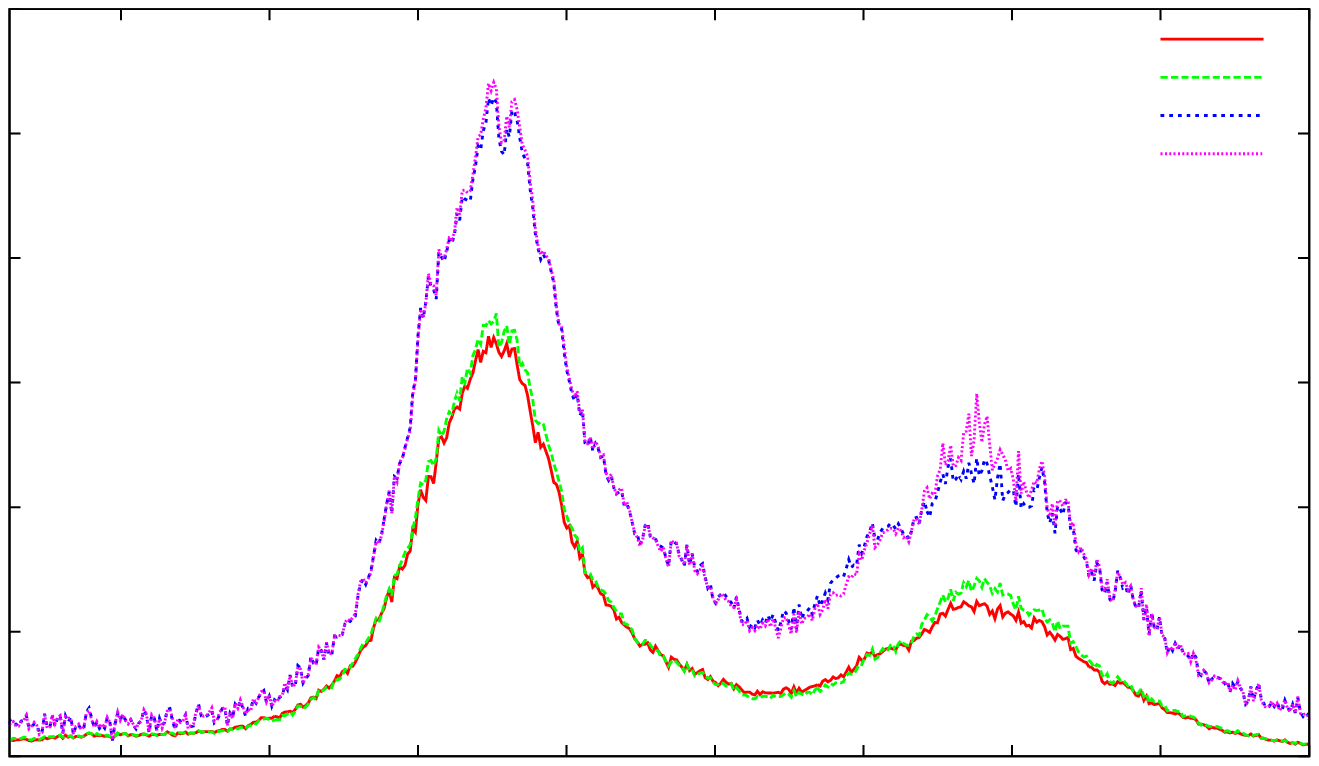}}%
    \gplfronttext
  \end{picture}%
\endgroup

%% file: graphics/nulling_simple.tex
\begingroup
  \makeatletter
  \providecommand\color[2][]{%
    \GenericError{(gnuplot) \space\space\space\@spaces}{%
      Package color not loaded in conjunction with
      terminal option `colourtext'%
    }{See the gnuplot documentation for explanation.%
    }{Either use 'blacktext' in gnuplot or load the package
      color.sty in LaTeX.}%
    \renewcommand\color[2][]{}%
  }%
  \providecommand\includegraphics[2][]{%
    \GenericError{(gnuplot) \space\space\space\@spaces}{%
      Package graphicx or graphics not loaded%
    }{See the gnuplot documentation for explanation.%
    }{The gnuplot epslatex terminal needs graphicx.sty or graphics.sty.}%
    \renewcommand\includegraphics[2][]{}%
  }%
  \providecommand\rotatebox[2]{#2}%
  \@ifundefined{ifGPcolor}{%
    \newif\ifGPcolor
    \GPcolortrue
  }{}%
  \@ifundefined{ifGPblacktext}{%
    \newif\ifGPblacktext
    \GPblacktexttrue
  }{}%
  \let\gplgaddtomacro\g@addto@macro
  \gdef\gplbacktext{}%
  \gdef\gplfronttext{}%
  \makeatother
  \ifGPblacktext
    \def\colorrgb#1{}%
    \def\colorgray#1{}%
  \else
    \ifGPcolor
      \def\colorrgb#1{\color[rgb]{#1}}%
      \def\colorgray#1{\color[gray]{#1}}%
      \expandafter\def\csname LTw\endcsname{\color{white}}%
      \expandafter\def\csname LTb\endcsname{\color{black}}%
      \expandafter\def\csname LTa\endcsname{\color{black}}%
      \expandafter\def\csname LT0\endcsname{\color[rgb]{1,0,0}}%
      \expandafter\def\csname LT1\endcsname{\color[rgb]{0,1,0}}%
      \expandafter\def\csname LT2\endcsname{\color[rgb]{0,0,1}}%
      \expandafter\def\csname LT3\endcsname{\color[rgb]{1,0,1}}%
      \expandafter\def\csname LT4\endcsname{\color[rgb]{0,1,1}}%
      \expandafter\def\csname LT5\endcsname{\color[rgb]{1,1,0}}%
      \expandafter\def\csname LT6\endcsname{\color[rgb]{0,0,0}}%
      \expandafter\def\csname LT7\endcsname{\color[rgb]{1,0.3,0}}%
      \expandafter\def\csname LT8\endcsname{\color[rgb]{0.5,0.5,0.5}}%
    \else
      \def\colorrgb#1{\color{black}}%
      \def\colorgray#1{\color[gray]{#1}}%
      \expandafter\def\csname LTw\endcsname{\color{white}}%
      \expandafter\def\csname LTb\endcsname{\color{black}}%
      \expandafter\def\csname LTa\endcsname{\color{black}}%
      \expandafter\def\csname LT0\endcsname{\color{black}}%
      \expandafter\def\csname LT1\endcsname{\color{black}}%
      \expandafter\def\csname LT2\endcsname{\color{black}}%
      \expandafter\def\csname LT3\endcsname{\color{black}}%
      \expandafter\def\csname LT4\endcsname{\color{black}}%
      \expandafter\def\csname LT5\endcsname{\color{black}}%
      \expandafter\def\csname LT6\endcsname{\color{black}}%
      \expandafter\def\csname LT7\endcsname{\color{black}}%
      \expandafter\def\csname LT8\endcsname{\color{black}}%
    \fi
  \fi
  \setlength{\unitlength}{0.0500bp}%
  \begin{picture}(4250.00,8502.00)%
    \gplgaddtomacro\gplbacktext{%
      \csname LTb\endcsname%
      \put(396,2060){\makebox(0,0)[r]{\strut{} 10}}%
      \put(396,3487){\makebox(0,0)[r]{\strut{} 20}}%
      \put(396,4915){\makebox(0,0)[r]{\strut{} 30}}%
      \put(396,6342){\makebox(0,0)[r]{\strut{} 40}}%
      \put(396,7770){\makebox(0,0)[r]{\strut{} 50}}%
      \put(680,484){\makebox(0,0){\strut{} 160}}%
      \put(1286,484){\makebox(0,0){\strut{} 200}}%
      \put(1893,484){\makebox(0,0){\strut{} 240}}%
      \put(2500,484){\makebox(0,0){\strut{} 280}}%
      \put(-242,4272){\rotatebox{-270}{\makebox(0,0){\strut{}Pulse number}}}%
      \put(1665,154){\makebox(0,0){\strut{}Phase}}%
      \put(1665,8171){\makebox(0,0){\strut{}Time domain}}%
    }%
    \gplgaddtomacro\gplfronttext{%
    }%
    \gplgaddtomacro\gplbacktext{%
      \csname LTb\endcsname%
      \put(2686,2103){\makebox(0,0)[r]{\strut{} 10}}%
      \put(2686,3530){\makebox(0,0)[r]{\strut{} 20}}%
      \put(2686,4958){\makebox(0,0)[r]{\strut{} 30}}%
      \put(2686,6385){\makebox(0,0)[r]{\strut{} 40}}%
      \put(2686,7812){\makebox(0,0)[r]{\strut{} 50}}%
      \put(2818,484){\makebox(0,0){\strut{} 0}}%
      \put(4117,484){\makebox(0,0){\strut{} 1}}%
      \put(3467,154){\makebox(0,0){\strut{}probability}}%
      \put(3467,8171){\makebox(0,0){\strut{}probability for nulling}}%
    }%
    \gplgaddtomacro\gplfronttext{%
    }%
    \gplbacktext
    \put(0,0){\includegraphics{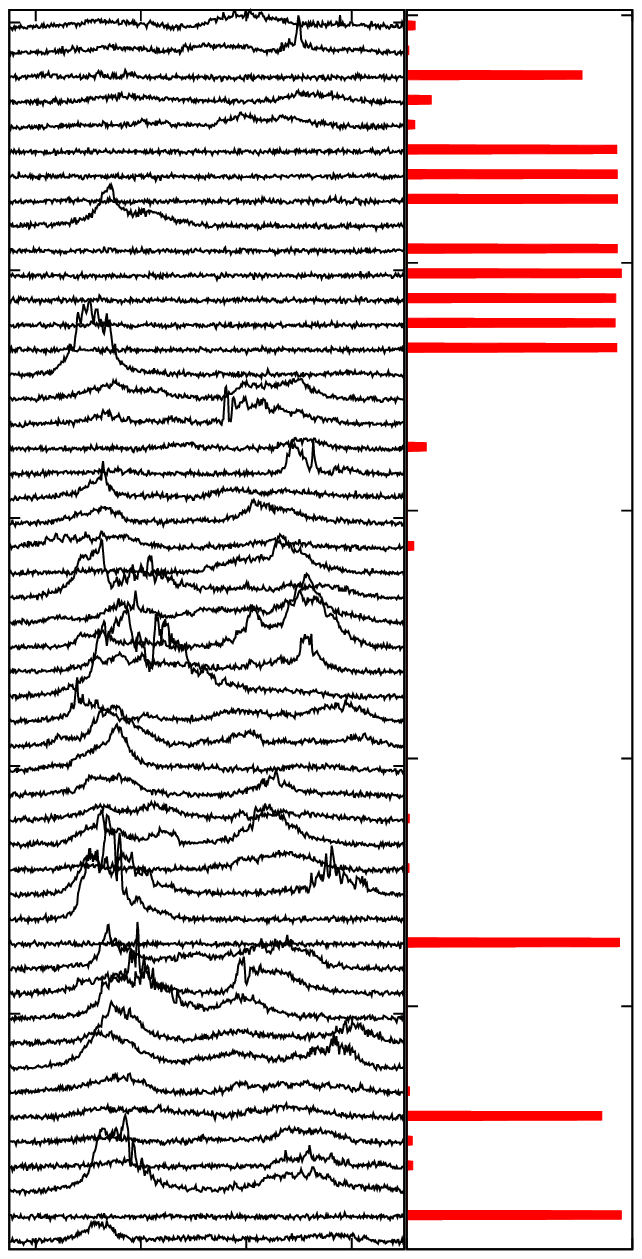}}%
    \gplfronttext
  \end{picture}%
\endgroup

%% file: graphics/toa_fixedref.tex
\begingroup
  \makeatletter
  \providecommand\color[2][]{%
    \GenericError{(gnuplot) \space\space\space\@spaces}{%
      Package color not loaded in conjunction with
      terminal option `colourtext'%
    }{See the gnuplot documentation for explanation.%
    }{Either use 'blacktext' in gnuplot or load the package
      color.sty in LaTeX.}%
    \renewcommand\color[2][]{}%
  }%
  \providecommand\includegraphics[2][]{%
    \GenericError{(gnuplot) \space\space\space\@spaces}{%
      Package graphicx or graphics not loaded%
    }{See the gnuplot documentation for explanation.%
    }{The gnuplot epslatex terminal needs graphicx.sty or graphics.sty.}%
    \renewcommand\includegraphics[2][]{}%
  }%
  \providecommand\rotatebox[2]{#2}%
  \@ifundefined{ifGPcolor}{%
    \newif\ifGPcolor
    \GPcolortrue
  }{}%
  \@ifundefined{ifGPblacktext}{%
    \newif\ifGPblacktext
    \GPblacktexttrue
  }{}%
  \let\gplgaddtomacro\g@addto@macro
  \gdef\gplbacktext{}%
  \gdef\gplfronttext{}%
  \makeatother
  \ifGPblacktext
    \def\colorrgb#1{}%
    \def\colorgray#1{}%
  \else
    \ifGPcolor
      \def\colorrgb#1{\color[rgb]{#1}}%
      \def\colorgray#1{\color[gray]{#1}}%
      \expandafter\def\csname LTw\endcsname{\color{white}}%
      \expandafter\def\csname LTb\endcsname{\color{black}}%
      \expandafter\def\csname LTa\endcsname{\color{black}}%
      \expandafter\def\csname LT0\endcsname{\color[rgb]{1,0,0}}%
      \expandafter\def\csname LT1\endcsname{\color[rgb]{0,1,0}}%
      \expandafter\def\csname LT2\endcsname{\color[rgb]{0,0,1}}%
      \expandafter\def\csname LT3\endcsname{\color[rgb]{1,0,1}}%
      \expandafter\def\csname LT4\endcsname{\color[rgb]{0,1,1}}%
      \expandafter\def\csname LT5\endcsname{\color[rgb]{1,1,0}}%
      \expandafter\def\csname LT6\endcsname{\color[rgb]{0,0,0}}%
      \expandafter\def\csname LT7\endcsname{\color[rgb]{1,0.3,0}}%
      \expandafter\def\csname LT8\endcsname{\color[rgb]{0.5,0.5,0.5}}%
    \else
      \def\colorrgb#1{\color{black}}%
      \def\colorgray#1{\color[gray]{#1}}%
      \expandafter\def\csname LTw\endcsname{\color{white}}%
      \expandafter\def\csname LTb\endcsname{\color{black}}%
      \expandafter\def\csname LTa\endcsname{\color{black}}%
      \expandafter\def\csname LT0\endcsname{\color{black}}%
      \expandafter\def\csname LT1\endcsname{\color{black}}%
      \expandafter\def\csname LT2\endcsname{\color{black}}%
      \expandafter\def\csname LT3\endcsname{\color{black}}%
      \expandafter\def\csname LT4\endcsname{\color{black}}%
      \expandafter\def\csname LT5\endcsname{\color{black}}%
      \expandafter\def\csname LT6\endcsname{\color{black}}%
      \expandafter\def\csname LT7\endcsname{\color{black}}%
      \expandafter\def\csname LT8\endcsname{\color{black}}%
    \fi
  \fi
  \setlength{\unitlength}{0.0500bp}%
  \begin{picture}(4534.00,4534.00)%
    \gplgaddtomacro\gplbacktext{%
      \csname LTb\endcsname%
      \put(171,704){\makebox(0,0)[r]{\strut{} 1e-05}}%
      \put(171,1760){\makebox(0,0)[r]{\strut{} 0.0001}}%
      \put(171,2817){\makebox(0,0)[r]{\strut{} 0.001}}%
      \put(171,3873){\makebox(0,0)[r]{\strut{} 0.01}}%
      \put(303,484){\makebox(0,0){\strut{} 10}}%
      \put(1344,484){\makebox(0,0){\strut{} 100}}%
      \put(2385,484){\makebox(0,0){\strut{} 1000}}%
      \put(3426,484){\makebox(0,0){\strut{} 10000}}%
      \put(4467,484){\makebox(0,0){\strut{} 100000}}%
      \put(4686,2288){\rotatebox{-270}{\makebox(0,0){\strut{}Phase error in radians}}}%
      \put(2385,154){\makebox(0,0){\strut{}Number of pulses}}%
      \put(2385,4203){\makebox(0,0){\strut{}Error on ToAs measured by a template from 10.000 pulses}}%
    }%
    \gplgaddtomacro\gplfronttext{%
      \csname LTb\endcsname%
      \put(3480,3700){\makebox(0,0)[r]{\strut{}$\sigma_\mathrm{err}$}}%
      \csname LTb\endcsname%
      \put(3480,3480){\makebox(0,0)[r]{\strut{}$|\langle \Phi_\mathrm{err}\rangle|$}}%
      \csname LTb\endcsname%
      \put(3480,3260){\makebox(0,0)[r]{\strut{}Estimation of classical precission}}%
    }%
    \gplbacktext
    \put(0,0){\includegraphics{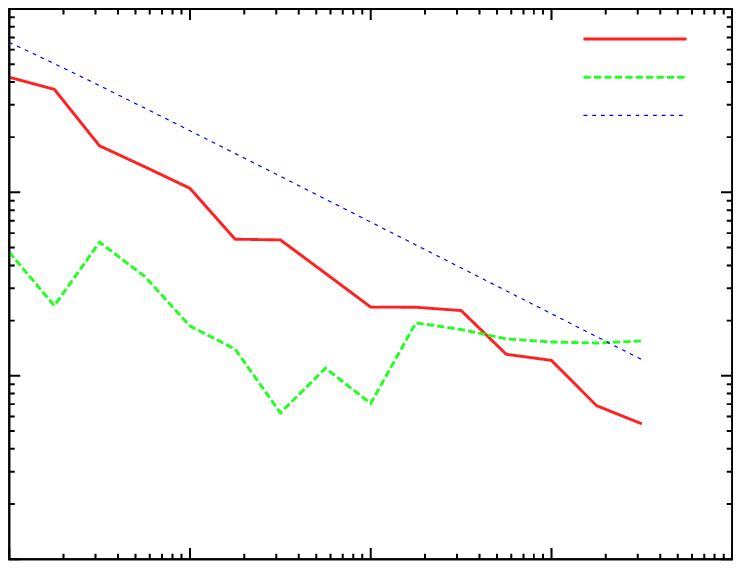}}%
    \gplfronttext
  \end{picture}%
\endgroup

%% file: graphics/toa_snr.tex
\begingroup
  \makeatletter
  \providecommand\color[2][]{%
    \GenericError{(gnuplot) \space\space\space\@spaces}{%
      Package color not loaded in conjunction with
      terminal option `colourtext'%
    }{See the gnuplot documentation for explanation.%
    }{Either use 'blacktext' in gnuplot or load the package
      color.sty in LaTeX.}%
    \renewcommand\color[2][]{}%
  }%
  \providecommand\includegraphics[2][]{%
    \GenericError{(gnuplot) \space\space\space\@spaces}{%
      Package graphicx or graphics not loaded%
    }{See the gnuplot documentation for explanation.%
    }{The gnuplot epslatex terminal needs graphicx.sty or graphics.sty.}%
    \renewcommand\includegraphics[2][]{}%
  }%
  \providecommand\rotatebox[2]{#2}%
  \@ifundefined{ifGPcolor}{%
    \newif\ifGPcolor
    \GPcolortrue
  }{}%
  \@ifundefined{ifGPblacktext}{%
    \newif\ifGPblacktext
    \GPblacktexttrue
  }{}%
  \let\gplgaddtomacro\g@addto@macro
  \gdef\gplbacktext{}%
  \gdef\gplfronttext{}%
  \makeatother
  \ifGPblacktext
    \def\colorrgb#1{}%
    \def\colorgray#1{}%
  \else
    \ifGPcolor
      \def\colorrgb#1{\color[rgb]{#1}}%
      \def\colorgray#1{\color[gray]{#1}}%
      \expandafter\def\csname LTw\endcsname{\color{white}}%
      \expandafter\def\csname LTb\endcsname{\color{black}}%
      \expandafter\def\csname LTa\endcsname{\color{black}}%
      \expandafter\def\csname LT0\endcsname{\color[rgb]{1,0,0}}%
      \expandafter\def\csname LT1\endcsname{\color[rgb]{0,1,0}}%
      \expandafter\def\csname LT2\endcsname{\color[rgb]{0,0,1}}%
      \expandafter\def\csname LT3\endcsname{\color[rgb]{1,0,1}}%
      \expandafter\def\csname LT4\endcsname{\color[rgb]{0,1,1}}%
      \expandafter\def\csname LT5\endcsname{\color[rgb]{1,1,0}}%
      \expandafter\def\csname LT6\endcsname{\color[rgb]{0,0,0}}%
      \expandafter\def\csname LT7\endcsname{\color[rgb]{1,0.3,0}}%
      \expandafter\def\csname LT8\endcsname{\color[rgb]{0.5,0.5,0.5}}%
    \else
      \def\colorrgb#1{\color{black}}%
      \def\colorgray#1{\color[gray]{#1}}%
      \expandafter\def\csname LTw\endcsname{\color{white}}%
      \expandafter\def\csname LTb\endcsname{\color{black}}%
      \expandafter\def\csname LTa\endcsname{\color{black}}%
      \expandafter\def\csname LT0\endcsname{\color{black}}%
      \expandafter\def\csname LT1\endcsname{\color{black}}%
      \expandafter\def\csname LT2\endcsname{\color{black}}%
      \expandafter\def\csname LT3\endcsname{\color{black}}%
      \expandafter\def\csname LT4\endcsname{\color{black}}%
      \expandafter\def\csname LT5\endcsname{\color{black}}%
      \expandafter\def\csname LT6\endcsname{\color{black}}%
      \expandafter\def\csname LT7\endcsname{\color{black}}%
      \expandafter\def\csname LT8\endcsname{\color{black}}%
    \fi
  \fi
  \setlength{\unitlength}{0.0500bp}%
  \begin{picture}(4534.00,4534.00)%
    \gplgaddtomacro\gplbacktext{%
      \csname LTb\endcsname%
      \put(171,1258){\makebox(0,0)[r]{\strut{} 0.001}}%
      \put(171,2050){\makebox(0,0)[r]{\strut{} 0.01}}%
      \put(171,2842){\makebox(0,0)[r]{\strut{} 0.1}}%
      \put(171,3635){\makebox(0,0)[r]{\strut{} 1}}%
      \put(303,484){\makebox(0,0){\strut{} 10}}%
      \put(2273,484){\makebox(0,0){\strut{} 100}}%
      \put(4243,484){\makebox(0,0){\strut{} 1000}}%
      \put(4686,2288){\rotatebox{-270}{\makebox(0,0){\strut{}Phase error in radiants}}}%
      \put(2385,154){\makebox(0,0){\strut{}Number of pulses}}%
      \put(2385,4203){\makebox(0,0){\strut{}Accuracy of ToAs for different SNRs}}%
    }%
    \gplgaddtomacro\gplfronttext{%
      \csname LTb\endcsname%
      \put(3480,3700){\makebox(0,0)[r]{\strut{}-26.6dB}}%
      \csname LTb\endcsname%
      \put(3480,3480){\makebox(0,0)[r]{\strut{}-21.6dB}}%
      \csname LTb\endcsname%
      \put(3480,3260){\makebox(0,0)[r]{\strut{}-16.6dB}}%
      \csname LTb\endcsname%
      \put(3480,3040){\makebox(0,0)[r]{\strut{}-11.6dB}}%
    }%
    \gplbacktext
    \put(0,0){\includegraphics{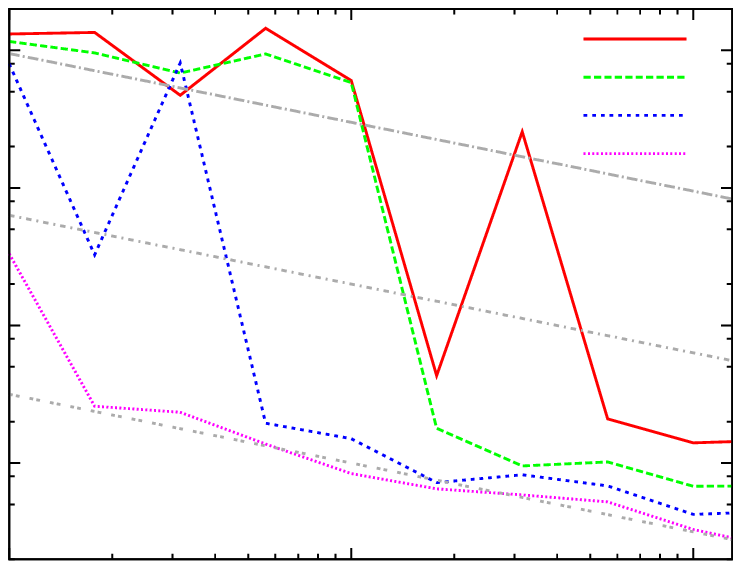}}%
    \gplfronttext
  \end{picture}%
\endgroup

%% file: graphics/toa_coordfree_20.tex
\begingroup
  \makeatletter
  \providecommand\color[2][]{%
    \GenericError{(gnuplot) \space\space\space\@spaces}{%
      Package color not loaded in conjunction with
      terminal option `colourtext'%
    }{See the gnuplot documentation for explanation.%
    }{Either use 'blacktext' in gnuplot or load the package
      color.sty in LaTeX.}%
    \renewcommand\color[2][]{}%
  }%
  \providecommand\includegraphics[2][]{%
    \GenericError{(gnuplot) \space\space\space\@spaces}{%
      Package graphicx or graphics not loaded%
    }{See the gnuplot documentation for explanation.%
    }{The gnuplot epslatex terminal needs graphicx.sty or graphics.sty.}%
    \renewcommand\includegraphics[2][]{}%
  }%
  \providecommand\rotatebox[2]{#2}%
  \@ifundefined{ifGPcolor}{%
    \newif\ifGPcolor
    \GPcolortrue
  }{}%
  \@ifundefined{ifGPblacktext}{%
    \newif\ifGPblacktext
    \GPblacktexttrue
  }{}%
  \let\gplgaddtomacro\g@addto@macro
  \gdef\gplbacktext{}%
  \gdef\gplfronttext{}%
  \makeatother
  \ifGPblacktext
    \def\colorrgb#1{}%
    \def\colorgray#1{}%
  \else
    \ifGPcolor
      \def\colorrgb#1{\color[rgb]{#1}}%
      \def\colorgray#1{\color[gray]{#1}}%
      \expandafter\def\csname LTw\endcsname{\color{white}}%
      \expandafter\def\csname LTb\endcsname{\color{black}}%
      \expandafter\def\csname LTa\endcsname{\color{black}}%
      \expandafter\def\csname LT0\endcsname{\color[rgb]{1,0,0}}%
      \expandafter\def\csname LT1\endcsname{\color[rgb]{0,1,0}}%
      \expandafter\def\csname LT2\endcsname{\color[rgb]{0,0,1}}%
      \expandafter\def\csname LT3\endcsname{\color[rgb]{1,0,1}}%
      \expandafter\def\csname LT4\endcsname{\color[rgb]{0,1,1}}%
      \expandafter\def\csname LT5\endcsname{\color[rgb]{1,1,0}}%
      \expandafter\def\csname LT6\endcsname{\color[rgb]{0,0,0}}%
      \expandafter\def\csname LT7\endcsname{\color[rgb]{1,0.3,0}}%
      \expandafter\def\csname LT8\endcsname{\color[rgb]{0.5,0.5,0.5}}%
    \else
      \def\colorrgb#1{\color{black}}%
      \def\colorgray#1{\color[gray]{#1}}%
      \expandafter\def\csname LTw\endcsname{\color{white}}%
      \expandafter\def\csname LTb\endcsname{\color{black}}%
      \expandafter\def\csname LTa\endcsname{\color{black}}%
      \expandafter\def\csname LT0\endcsname{\color{black}}%
      \expandafter\def\csname LT1\endcsname{\color{black}}%
      \expandafter\def\csname LT2\endcsname{\color{black}}%
      \expandafter\def\csname LT3\endcsname{\color{black}}%
      \expandafter\def\csname LT4\endcsname{\color{black}}%
      \expandafter\def\csname LT5\endcsname{\color{black}}%
      \expandafter\def\csname LT6\endcsname{\color{black}}%
      \expandafter\def\csname LT7\endcsname{\color{black}}%
      \expandafter\def\csname LT8\endcsname{\color{black}}%
    \fi
  \fi
  \setlength{\unitlength}{0.0500bp}%
  \begin{picture}(5668.00,6802.00)%
    \gplgaddtomacro\gplbacktext{%
      \csname LTb\endcsname%
      \put(66,2117){\makebox(0,0)[r]{\strut{} 0}}%
      \put(66,3087){\makebox(0,0)[r]{\strut{} 5}}%
      \put(66,4058){\makebox(0,0)[r]{\strut{} 10}}%
      \put(66,5028){\makebox(0,0)[r]{\strut{} 15}}%
      \put(301,1800){\makebox(0,0){\strut{} 0}}%
      \put(1328,1800){\makebox(0,0){\strut{} 5}}%
      \put(2356,1800){\makebox(0,0){\strut{} 10}}%
      \put(3383,1800){\makebox(0,0){\strut{} 15}}%
      \put(2253,6231){\makebox(0,0){\strut{}sigma measured from posterior probability distribution}}%
    }%
    \gplgaddtomacro\gplfronttext{%
      \csname LTb\endcsname%
      \put(4748,2019){\makebox(0,0)[l]{\strut{} 0.0001}}%
      \put(4748,5900){\makebox(0,0)[l]{\strut{} 0.001}}%
    }%
    \gplgaddtomacro\gplbacktext{%
      \csname LTb\endcsname%
      \put(66,660){\makebox(0,0)[r]{\strut{}-4}}%
      \csname LTb\endcsname%
      \put(66,895){\makebox(0,0)[r]{\strut{}-2}}%
      \csname LTb\endcsname%
      \put(66,1130){\makebox(0,0)[r]{\strut{} 0}}%
      \csname LTb\endcsname%
      \put(66,1365){\makebox(0,0)[r]{\strut{} 2}}%
      \csname LTb\endcsname%
      \put(66,1600){\makebox(0,0)[r]{\strut{} 4}}%
      \csname LTb\endcsname%
      \put(300,440){\makebox(0,0){\strut{} 0}}%
      \csname LTb\endcsname%
      \put(1321,440){\makebox(0,0){\strut{} 5}}%
      \csname LTb\endcsname%
      \put(2342,440){\makebox(0,0){\strut{} 10}}%
      \csname LTb\endcsname%
      \put(3362,440){\makebox(0,0){\strut{} 15}}%
    }%
    \gplgaddtomacro\gplfronttext{%
      \csname LTb\endcsname%
      \put(3294,1427){\makebox(0,0)[r]{\strut{}absolute error $\cdot 10^3$}}%
    }%
    \gplbacktext
    \put(0,0){\includegraphics{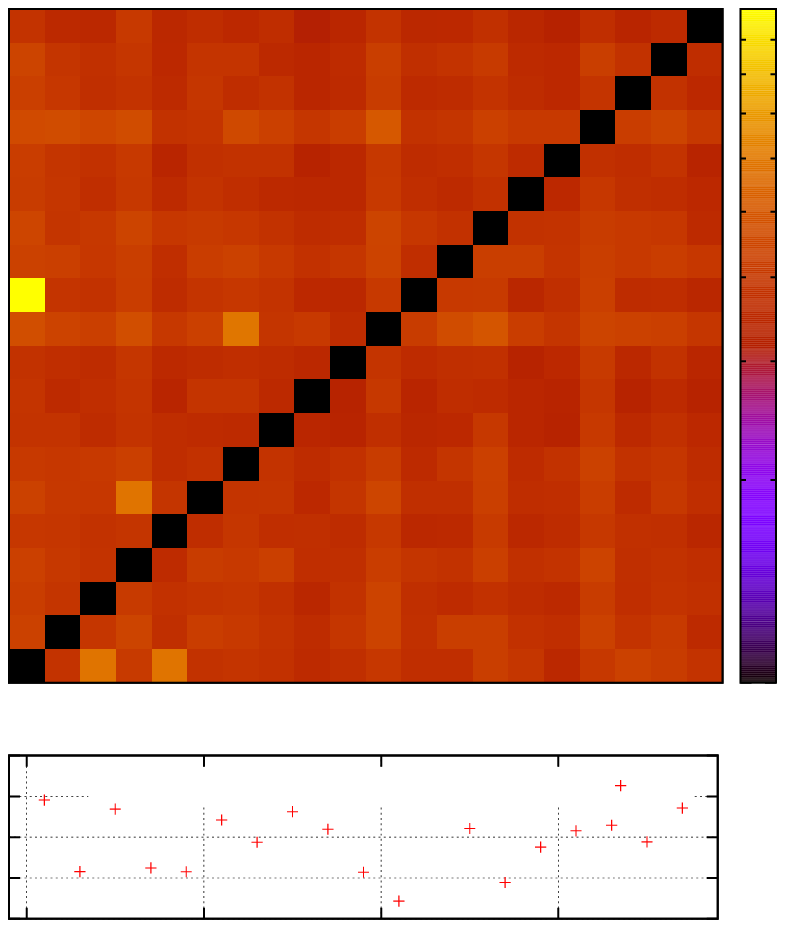}}%
    \gplfronttext
  \end{picture}%
\endgroup

%% file: graphics/scatter.tex
\begingroup
  \makeatletter
  \providecommand\color[2][]{%
    \GenericError{(gnuplot) \space\space\space\@spaces}{%
      Package color not loaded in conjunction with
      terminal option `colourtext'%
    }{See the gnuplot documentation for explanation.%
    }{Either use 'blacktext' in gnuplot or load the package
      color.sty in LaTeX.}%
    \renewcommand\color[2][]{}%
  }%
  \providecommand\includegraphics[2][]{%
    \GenericError{(gnuplot) \space\space\space\@spaces}{%
      Package graphicx or graphics not loaded%
    }{See the gnuplot documentation for explanation.%
    }{The gnuplot epslatex terminal needs graphicx.sty or graphics.sty.}%
    \renewcommand\includegraphics[2][]{}%
  }%
  \providecommand\rotatebox[2]{#2}%
  \@ifundefined{ifGPcolor}{%
    \newif\ifGPcolor
    \GPcolortrue
  }{}%
  \@ifundefined{ifGPblacktext}{%
    \newif\ifGPblacktext
    \GPblacktexttrue
  }{}%
  \let\gplgaddtomacro\g@addto@macro
  \gdef\gplbacktext{}%
  \gdef\gplfronttext{}%
  \makeatother
  \ifGPblacktext
    \def\colorrgb#1{}%
    \def\colorgray#1{}%
  \else
    \ifGPcolor
      \def\colorrgb#1{\color[rgb]{#1}}%
      \def\colorgray#1{\color[gray]{#1}}%
      \expandafter\def\csname LTw\endcsname{\color{white}}%
      \expandafter\def\csname LTb\endcsname{\color{black}}%
      \expandafter\def\csname LTa\endcsname{\color{black}}%
      \expandafter\def\csname LT0\endcsname{\color[rgb]{1,0,0}}%
      \expandafter\def\csname LT1\endcsname{\color[rgb]{0,1,0}}%
      \expandafter\def\csname LT2\endcsname{\color[rgb]{0,0,1}}%
      \expandafter\def\csname LT3\endcsname{\color[rgb]{1,0,1}}%
      \expandafter\def\csname LT4\endcsname{\color[rgb]{0,1,1}}%
      \expandafter\def\csname LT5\endcsname{\color[rgb]{1,1,0}}%
      \expandafter\def\csname LT6\endcsname{\color[rgb]{0,0,0}}%
      \expandafter\def\csname LT7\endcsname{\color[rgb]{1,0.3,0}}%
      \expandafter\def\csname LT8\endcsname{\color[rgb]{0.5,0.5,0.5}}%
    \else
      \def\colorrgb#1{\color{black}}%
      \def\colorgray#1{\color[gray]{#1}}%
      \expandafter\def\csname LTw\endcsname{\color{white}}%
      \expandafter\def\csname LTb\endcsname{\color{black}}%
      \expandafter\def\csname LTa\endcsname{\color{black}}%
      \expandafter\def\csname LT0\endcsname{\color{black}}%
      \expandafter\def\csname LT1\endcsname{\color{black}}%
      \expandafter\def\csname LT2\endcsname{\color{black}}%
      \expandafter\def\csname LT3\endcsname{\color{black}}%
      \expandafter\def\csname LT4\endcsname{\color{black}}%
      \expandafter\def\csname LT5\endcsname{\color{black}}%
      \expandafter\def\csname LT6\endcsname{\color{black}}%
      \expandafter\def\csname LT7\endcsname{\color{black}}%
      \expandafter\def\csname LT8\endcsname{\color{black}}%
    \fi
  \fi
  \setlength{\unitlength}{0.0500bp}%
  \begin{picture}(5668.00,4250.00)%
    \gplgaddtomacro\gplbacktext{%
      \csname LTb\endcsname%
      \put(760,484){\makebox(0,0){\strut{} 220}}%
      \put(1454,484){\makebox(0,0){\strut{} 230}}%
      \put(2147,484){\makebox(0,0){\strut{} 240}}%
      \put(2840,484){\makebox(0,0){\strut{} 250}}%
      \put(3534,484){\makebox(0,0){\strut{} 260}}%
      \put(4227,484){\makebox(0,0){\strut{} 270}}%
      \put(4920,484){\makebox(0,0){\strut{} 280}}%
      \put(-88,2344){\rotatebox{-270}{\makebox(0,0){\strut{}receiver voltage in arbitrary units}}}%
      \put(2569,154){\makebox(0,0){\strut{}pulsar phase in degree}}%
    }%
    \gplgaddtomacro\gplfronttext{%
      \csname LTb\endcsname%
      \put(4020,3812){\makebox(0,0)[r]{\strut{}Mode 1 - 44.5\%}}%
      \csname LTb\endcsname%
      \put(4020,3592){\makebox(0,0)[r]{\strut{}Mode 2 - 33.9\%}}%
      \csname LTb\endcsname%
      \put(4020,3372){\makebox(0,0)[r]{\strut{}Mode 3 - 21.6\%}}%
    }%
    \gplbacktext
    \put(0,0){\includegraphics{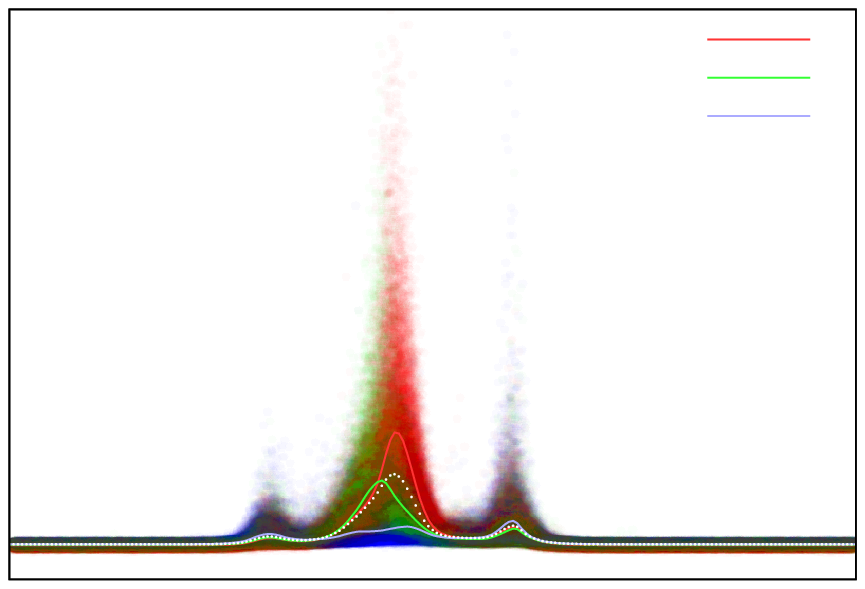}}%
    \gplfronttext
  \end{picture}%
\endgroup

%% file: graphics/histo_5000_10_200.tex
\begingroup
  \makeatletter
  \providecommand\color[2][]{%
    \GenericError{(gnuplot) \space\space\space\@spaces}{%
      Package color not loaded in conjunction with
      terminal option `colourtext'%
    }{See the gnuplot documentation for explanation.%
    }{Either use 'blacktext' in gnuplot or load the package
      color.sty in LaTeX.}%
    \renewcommand\color[2][]{}%
  }%
  \providecommand\includegraphics[2][]{%
    \GenericError{(gnuplot) \space\space\space\@spaces}{%
      Package graphicx or graphics not loaded%
    }{See the gnuplot documentation for explanation.%
    }{The gnuplot epslatex terminal needs graphicx.sty or graphics.sty.}%
    \renewcommand\includegraphics[2][]{}%
  }%
  \providecommand\rotatebox[2]{#2}%
  \@ifundefined{ifGPcolor}{%
    \newif\ifGPcolor
    \GPcolortrue
  }{}%
  \@ifundefined{ifGPblacktext}{%
    \newif\ifGPblacktext
    \GPblacktexttrue
  }{}%
  \let\gplgaddtomacro\g@addto@macro
  \gdef\gplbacktext{}%
  \gdef\gplfronttext{}%
  \makeatother
  \ifGPblacktext
    \def\colorrgb#1{}%
    \def\colorgray#1{}%
  \else
    \ifGPcolor
      \def\colorrgb#1{\color[rgb]{#1}}%
      \def\colorgray#1{\color[gray]{#1}}%
      \expandafter\def\csname LTw\endcsname{\color{white}}%
      \expandafter\def\csname LTb\endcsname{\color{black}}%
      \expandafter\def\csname LTa\endcsname{\color{black}}%
      \expandafter\def\csname LT0\endcsname{\color[rgb]{1,0,0}}%
      \expandafter\def\csname LT1\endcsname{\color[rgb]{0,1,0}}%
      \expandafter\def\csname LT2\endcsname{\color[rgb]{0,0,1}}%
      \expandafter\def\csname LT3\endcsname{\color[rgb]{1,0,1}}%
      \expandafter\def\csname LT4\endcsname{\color[rgb]{0,1,1}}%
      \expandafter\def\csname LT5\endcsname{\color[rgb]{1,1,0}}%
      \expandafter\def\csname LT6\endcsname{\color[rgb]{0,0,0}}%
      \expandafter\def\csname LT7\endcsname{\color[rgb]{1,0.3,0}}%
      \expandafter\def\csname LT8\endcsname{\color[rgb]{0.5,0.5,0.5}}%
    \else
      \def\colorrgb#1{\color{black}}%
      \def\colorgray#1{\color[gray]{#1}}%
      \expandafter\def\csname LTw\endcsname{\color{white}}%
      \expandafter\def\csname LTb\endcsname{\color{black}}%
      \expandafter\def\csname LTa\endcsname{\color{black}}%
      \expandafter\def\csname LT0\endcsname{\color{black}}%
      \expandafter\def\csname LT1\endcsname{\color{black}}%
      \expandafter\def\csname LT2\endcsname{\color{black}}%
      \expandafter\def\csname LT3\endcsname{\color{black}}%
      \expandafter\def\csname LT4\endcsname{\color{black}}%
      \expandafter\def\csname LT5\endcsname{\color{black}}%
      \expandafter\def\csname LT6\endcsname{\color{black}}%
      \expandafter\def\csname LT7\endcsname{\color{black}}%
      \expandafter\def\csname LT8\endcsname{\color{black}}%
    \fi
  \fi
  \setlength{\unitlength}{0.0500bp}%
  \begin{picture}(10200.00,4520.00)%
    \gplgaddtomacro\gplbacktext{%
      \csname LTb\endcsname%
      \put(645,2260){\makebox(0,0)[r]{\strut{} 0}}%
      \csname LTb\endcsname%
      \put(645,2600){\makebox(0,0)[r]{\strut{} 10}}%
      \csname LTb\endcsname%
      \put(645,2940){\makebox(0,0)[r]{\strut{} 20}}%
      \csname LTb\endcsname%
      \put(645,3281){\makebox(0,0)[r]{\strut{} 30}}%
      \csname LTb\endcsname%
      \put(645,3621){\makebox(0,0)[r]{\strut{} 40}}%
      \csname LTb\endcsname%
      \put(144,3110){\rotatebox{-270}{\makebox(0,0){\strut{}observed frequency}}}%
      \put(5320,4240){\makebox(0,0){\strut{}25 vs. 50 pulse integration window}}%
    }%
    \gplgaddtomacro\gplfronttext{%
      \csname LTb\endcsname%
      \put(2379,3794){\makebox(0,0)[r]{\strut{}single template}}%
      \csname LTb\endcsname%
      \put(2379,3608){\makebox(0,0)[r]{\strut{}10 modes}}%
    }%
    \gplgaddtomacro\gplbacktext{%
      \csname LTb\endcsname%
      \put(645,595){\makebox(0,0)[r]{\strut{} 0}}%
      \csname LTb\endcsname%
      \put(645,833){\makebox(0,0)[r]{\strut{} 5}}%
      \csname LTb\endcsname%
      \put(645,1071){\makebox(0,0)[r]{\strut{} 10}}%
      \csname LTb\endcsname%
      \put(645,1309){\makebox(0,0)[r]{\strut{} 15}}%
      \csname LTb\endcsname%
      \put(645,1546){\makebox(0,0)[r]{\strut{} 20}}%
      \csname LTb\endcsname%
      \put(645,1784){\makebox(0,0)[r]{\strut{} 25}}%
      \csname LTb\endcsname%
      \put(747,409){\makebox(0,0){\strut{}-0.2}}%
      \csname LTb\endcsname%
      \put(2576,409){\makebox(0,0){\strut{}-0.15}}%
      \csname LTb\endcsname%
      \put(4405,409){\makebox(0,0){\strut{}-0.1}}%
      \csname LTb\endcsname%
      \put(6235,409){\makebox(0,0){\strut{}-0.05}}%
      \csname LTb\endcsname%
      \put(8064,409){\makebox(0,0){\strut{} 0}}%
      \csname LTb\endcsname%
      \put(9893,409){\makebox(0,0){\strut{} 0.05}}%
      \csname LTb\endcsname%
      \put(144,1427){\rotatebox{-270}{\makebox(0,0){\strut{}observed frequency}}}%
      \csname LTb\endcsname%
      \put(5320,130){\makebox(0,0){\strut{}ToA phase in radians}}%
    }%
    \gplgaddtomacro\gplfronttext{%
      \csname LTb\endcsname%
      \put(2379,2093){\makebox(0,0)[r]{\strut{}single template}}%
      \csname LTb\endcsname%
      \put(2379,1907){\makebox(0,0)[r]{\strut{}5 modes}}%
    }%
    \gplbacktext
    \put(0,0){\includegraphics{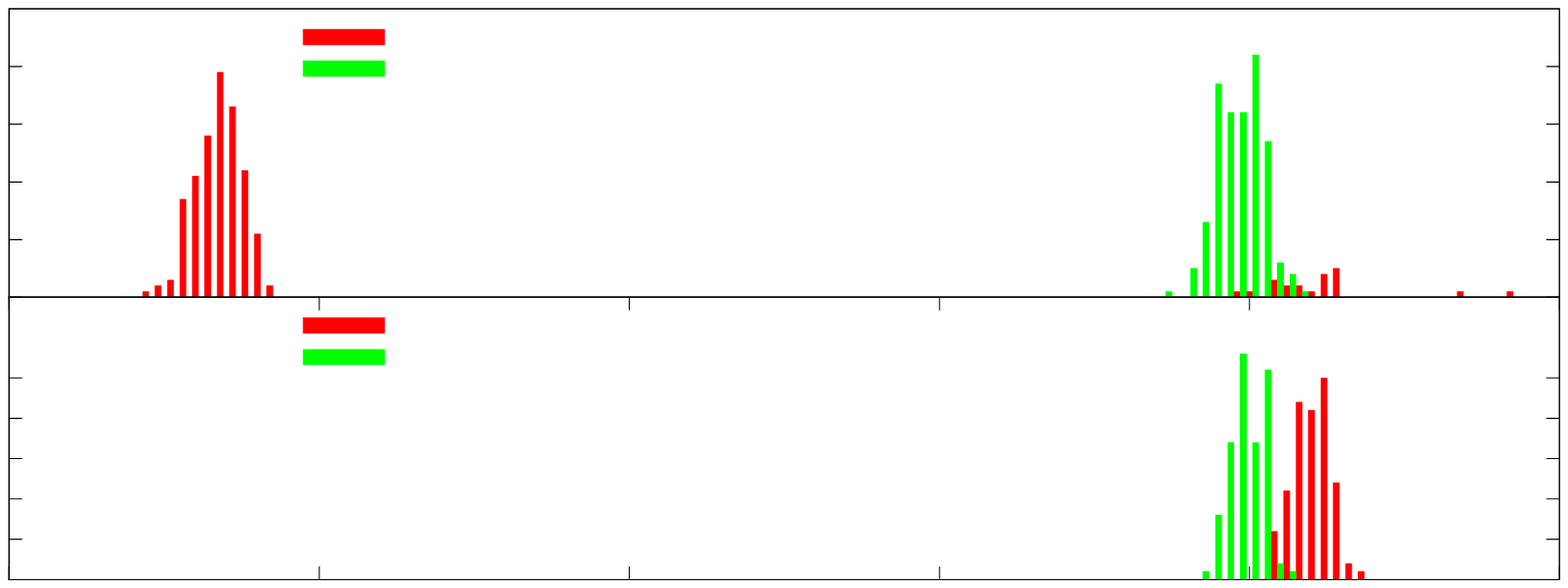}}%
    \gplfronttext
  \end{picture}%
\endgroup